\documentclass[useAMS]{aa}
\usepackage[varg]{txfonts}
\usepackage{natbib}
\usepackage{lscape}
\bibpunct{(}{)}{;}{a}{}{,} 
\usepackage{supertabular}
\usepackage{color}
\usepackage[normalem]{ulem}

\usepackage{float}
\restylefloat{figure}
\usepackage{graphicx}
\usepackage{caption}
\usepackage{subcaption}

\begin{document}

\title{Distant clusters of galaxies\thanks{The catalogue, similar to Table~A.1, is available in electronic
            form at the CDS via anonymous ftp to cdsarc.u-strasbg.fr (130.79.128.5) or via
            http://cdsweb.u-strasbg.fr/cgi-bin/qcat?J/A+A/}
            in the 2XMM\thanks{Based on observations obtained with
            XMM-Newton, an ESA science mission with instruments and contributions directly funded by ESA Member
            States and NASA}/SDSS footprint: follow-up observations with the LBT\thanks{The LBT is an
            international collaboration among institutions in the United States, Italy and Germany. LBT Corporation
            partners are: The University of Arizona on behalf of the Arizona Board of Regents; Istituto Nazionale
            di Astrofisica, Italy; LBT Beteiligungsgesellschaft, Germany, representing the Max-Planck Society,
            The Leibniz Institute for Astrophysics Potsdam, and Heidelberg University; The Ohio State University,
            and The Research Corporation, on behalf of The University of Notre Dame, University of Minnesota and
            University of Virginia. --- \emph{http://www.lbto.org/for-investigators.html}}}
\author{A.~Rabitz\inst{\ref{inst1}}\and G.~Lamer\inst{\ref{inst1}}\and A.~Schwope\inst{\ref{inst1}}\and
        A.~Takey\inst{\ref{inst2}}}
\institute{Leibniz-Institut für Astrophysik Potsdam (AIP), An der Sternwarte 16, 14482 Potsdam,
           Germany\label{inst1}
           \and National Research Institute of Astronomy and Geophysics (NRIAG), 11421 Helwan, Cairo,
           Egypt\label{inst2}}
\abstract{Galaxy clusters at high redshift are important to test cosmological models 
         and models for the growth of structure. They are difficult to find in
         wide-angle optical surveys, however, leaving dedicated follow-up of X-ray selected
         candidates as one promising identification route.}
         {We aim to increase the number of galaxy clusters beyond the SDSS-limit, \(z \sim 0.75\).}
         {We compiled a list of extended X-ray sources from the 2XMMp catalogue within
         the footprint of the Sloan Digital Sky Survey. Fields without optical counterpart
         were selected for further investigation. Deep optical imaging and follow-up
         spectroscopy were obtained with the Large Binocular
Telescope, Arizona (LBT), of those candidates not known to the literature.}
         {From initially 19 candidates, selected by visually screening X-ray images of
         478 \emph{XMM-Newton} observations and the corresponding SDSS images, 
         6 clusters were found in the literature. 
         Imaging data through r,z filters were obtained for the remaining candidates, and 7 were 
         chosen for multi-object (MOS) spectroscopy. Spectroscopic redshifts, optical magnitudes, and X-ray
         parameters (flux, temperature, and luminosity) are presented for the clusters with 
         spectroscopic redshifts. The distant clusters studied here constitute one additional 
         redshift bin for studies of the \(L_{X}-T\) relation, which does not seem to evolve 
         from high to low redshifts.}
         {The selection method of distant galaxy clusters presented here was highly 
         successful. It is based solely on archival optical (SDSS) and X-ray (XMM-Newton) data. 
         Out of 19 selected candidates, 6 of the 7 candidates selected for spectroscopic follow-up
         were verified as distant clusters, a further candidate is most likely a group of
         galaxies at \(z\sim1.21\).
         Out of the remaining 12 candidates, 6 were known previously as galaxy clusters, one object is a likely X-ray emission
         from an AGN radio jet, and for 5 we see no clear evidence for them to be high-redshift galaxy clusters.}
\titlerunning{Distant clusters of galaxies in the 2XMM/SDSS footprint}
\date{Received 08/05/2017 / Accepted 11/07/2017}
\maketitle

\section{Introduction}
\label{Introduction}

Clusters of galaxies are considered the largest gravitationally bound structures in the Universe, and
they emit radiation on a wide wavelength range. The multi-component nature of clusters gives rise to various
detection methods, since distinct physical effects contribute to different parts of the spectrum they emit.
Beneath the main component of galaxy clusters, which is thought to be composed of dark matter and is hence
not directly observable, the next major fraction is the hot ionized intra-cluster medium (ICM).
The ICM can be observed through thermal emission of its gravitationally heated gas, which mainly emits
in the soft X-ray regime. Inverse-Compton scattering of the ICM and photons of the cosmic microwave
background, the Sunyaev-Zel'dovich \citep[SZ effect; ][]{Sunyaev1972}, can be investigated at millimeter
wavelengths. Optical and near-infrared (near-IR) observations, however, allow redshift measurements
with moderate effort, and for instance reveal the dynamical state of the galaxy population of the cluster, as well
as the morphology and star formation history of the individual members.
The suitability of different wavelength regimes for detecting galaxy clusters is proven by various
campaigns, for example \citet{Boehringer2001}, \citet{Gladders2005}, and \citet{Vale2006} for the selection using
X-ray, optical, and millimeter wavelength, respectively.

Important properties of galaxy clusters such as their number density and masses deliver important constraints
for cosmological models.
Large-area X-ray surveys like the \emph{ROSAT} All Sky Survey \citep{Voges1999} detected large numbers of
clusters, see for example the Meta-Catalog of X-ray detected Clusters of galaxies \citep[MCXC; ][]{Piffaretti2011}.
Observations with \emph{XMM-Newton} increase the number of deeply exposed regions of the X-ray sky, allowing
for different luminosity constraints of potential clusters. Many cluster surveys are based on X-ray selection
and optical verification of sources, either working on complete sets of archival data for the lower redshift
regime \citep[e.g. \emph{XMM-Newton} and SDSS; ][]{Takey2011,Takey2013,Takey2016} or on fields that require dedicated
deep follow-up imaging and spectroscopy for discovering low-luminosity and high-redshift clusters
\citep[e.g. ][]{Fassbender2011_XDCP,Pacaud2016}.

\begin{table*}
 \caption{List of all X-ray selected galaxy cluster candidates within SDSS blank fields}
 \label{tab:survey}
 \centering
 \begin{small}\begin{tabular}{c c c c c c c c}
  \hline\hline
  Name & RA(J2000) & Dec(J2000) & LBT/LBC & LBT/MODS & $z$ & ref. \\
  \hline
  2XMMp~J030212.0-000133 & 03:02:12.1 & -00:01:32 & --- & --- & 1.185 & \citet{Suhada2011} \\
  2XMMp~J083026.2+524133 & 08:30:26.2 & +52:41:33 & r, z (0.3h) & 1.5h & \(0.9855 \pm 0.0037\) & \emph{this work} \\
  2XMMp~J084836.4+445345 & 08:48:36.1 & +44:53:43 & --- & --- & 1.273 & \citet{Stanford1997} \\
  2XMMp~J084858.3+445158 & 08:48:58.4 & +44:51:58 & --- & --- & 1.261 & \citet{Rosati1999} \\
  2XMMp~J092120.2+371735 & 09:21:20.2 & +37:17:35 & r, z (0.6h) & 2.5h & \(1.2134 \pm 0.0014\) & \emph{this work} \\
  2XMMp~J093437.4+551340 & 09:34:37.5 & +55:13:42 & r, z (0.48h) & 1.5h & \(0.83855 \pm 0.00094\) & \emph{this work} \\
  2XMMp~J093607.2+613245 & 09:36:07.3 & +61:32:45 & r, z (0.73h) & --- & --- & \\
  2XMMp~J100451.6+411626 & 10:04:51.6 & +41:16:27 & --- & --- & 0.82 & \citet{Hoeft2008} \\
  2XMMp~J105320.0+440816 & 10:53:19.8 & +44:08:17 & r, z (0.62h) & 2.0h & \(0.8955 \pm 0.0032\) & \emph{this work} \\
  2XMMp~J105344.2+573517 & 10:53:43.5 & +57:35:18 & z (0.3h) & --- & 1.134 & \citet{Hashimoto2005} \\
  2XMMp~J120735.1+250538 & 12:07:35.2 & +25:05:37 & r, z (0.3h) & --- & --- & \\
  2XMMp~J120815.5+250001 & 12:08:16.0 & +25:00:02 & r, z (0.3h) & 2.0h & \(0.9912 \pm 0.0027\) & \emph{this work} \\
  2XMMp~J123113.1+154550 & 12:31:13.1 & +15:45:48 & --- & --- & 0.893 & Rabitz et al. (in preparation) \\
  2XMMp~J123759.3+180332 & 12:37:59.2 & +18:03:35 & r, z (0.3h) & 2.0h & \(0.8874 \pm 0.0024\) & \emph{this work} \\
  2XMMp~J133038.6-013832 & 13:30:38.7 & -01:38:32 & r, z (0.3h) & --- & --- & \\
  2XMMp~J133853.9+482033 & 13:38:54.1 & +48:20:34 & r, z (0.85h) & 2.0h & \(0.7517 \pm 0.0016\) & \emph{this work} \\
  2XMMp~J144854.8+085400 & 14:48:54.7 & +08:53:59 & r, z (0.3h) & --- & --- & \citet{Lamer2011} \\
  2XMMp~J145220.8+165458 & 14:52:20.6 & +16:54:57 & r, z (0.77h) & --- & --- & \\
  2XMMp~J151716.8+001302 & 15:17:16.9 & +00:12:58 & r, z (0.3h) & --- & --- & \\
  \hline
 \end{tabular}\end{small}
 \tablefoot{Summary of the observational state of our sample. The first column lists the extended
            X-ray emission in the 2XMMp catalogue, from which this sample is built. The coordinates
            based on a cross-correlation with the 3XMM-DR6 catalogue are given in Cols. 2~and~3.
            The next two columns state the exposure time spent at imaging in LBT/LBC r-SLOAN, z-SLOAN,
            and at spectroscopy with LBT/MODS, respectively, followed by the spectroscopic redshift
            (Col.~6) and its reference (Col.~7).}
\end{table*}

Large samples of clusters were confirmed using the red-sequence method from photometric data of optical
sky surveys, for example the MaxBCG Catalog based on SDSS photometry \citep{Koester2007} and redMaPPer using data from
SDSS and the Dark Energy Survey (DES) \citep{Rykoff2016}. Other cluster surveys make use of spectroscopically confirmed
galaxies as verification, for instance spectra of BCGs from the SDSS \citep{Takey2011}. 
However, redshifts from SDSS data alone (either photometric or spectroscopic) are limited to redshifts below \(z\sim0.75\).
\citet{Buddendiek2015} cross-matched ROSAT All Sky Surveys sources with red galaxies from the SDSS, and using
follow-up observations with the William Herschel Telescope (WHT) and the Large Binocular Telescope, Arizona (LBT), they were able to identify clusters at \(0.6 \lesssim z \lesssim 1.0\).
\citet{Bleem2015} conducted a study using the {SZ effect} 
to discover galaxy clusters. Since the dimming of the observed SZ flux caused by the luminosity distance to the cluster
is partly compensated for by the higher energy density of the cosmic microwave background (CMB) at higher redshifts, current SZ surveys (e.g. from the SPT)
allow probing high-mass galaxy clusters out to \(z>1\). \citet{Bleem2015} used IR imaging and spectroscopy of
\emph{SPITZER} and gound-based facilities to verify overdensities of galaxies and to determine their redshift.
Another approach is to select clusters purely on their near-IR and IR colour, where the \(z\)-band and the
\(3.6\mu m\) band of \emph{SPITZER}, for instance, allow for tracing the 4000\AA~break of galaxies out to \(z>1\)
\citep[e.g.][]{Muzzin2009,Webb2015,Wilson2009}.

Here we investigate galaxy clusters that were serendipitously found by \emph{XMM-Newton} observations
and are listed in the 2XMMp catalogue, the predecessor of the public 2XMM catalogue \citep{Watson2009}.
By explicitly considering SDSS blank fields in our selection, we intend to increase the number of high-redshift
(\(z\sim1\)) galaxy clusters. This is the redshift regime where existing large-area optical surveys lack sensitivity to unambiguously confirm these clusters. 
We carried out follow-up observations  at the LBT using the 
LBC prime focus camera for  imaging and MODS for multi-object spectroscopy, resulting in identifications of six new clusters and one
high-redshift group.
Furthermore, six of the selected X-ray sources were identified with clusters of galaxies that have been published previously.

This new sample complements the cluster samples published by \citet{Takey2013} and \citet{Takey2014}, which were
also selected from the 2XMM catalogue, adding 13 high-redshift objects to the cluster identifications in the 
2XMM/SDSS footprint.

The paper is organized as follows: In Section~\ref{sec:sample} we explain the sample definition based on archival
\emph{XMM-Newton} data. Details on data reduction and the analysis of optical pre-imaging and spectroscopic follow-up
are outlined in Section~\ref{Sec:imaging} and Section~\ref{Sec:spectroscopy}, respectively.
The treatment of archival X-ray data is described in Section~\ref{sec:Xray_reduction}. Section~\ref{sec:results}
 presents the results of this work. We describe how we derive the spectroscopic properties and cluster mass
(Sect.~\ref{sec:dynamics}), the generation and treatment of X-ray spectra to measure fluxes and ICM temperatures
(Sect.~\ref{sec:Xray_analysis}), and detail the properties of the individual clusters in the following subsections.
An analysis of the \(L-T\) relation (Sect.~\ref{sec:results_L-T}), using our sample in comparison to data from the
literature, concludes the results section.
We summarize our findings in Section~\ref{sec:conclusions}.
In the appendix we list photometric and spectroscopic properties of the fields observed with LBC and MODS at the LBT
(Appendix~A), and present optical and IR images of the previously known clusters (Appendix~B),
as well as of the rejected or unclassified fields (Appendix~C).

Throughout the paper we assume a standard cosmology with \(\Omega_{M}=0.3\), \(\Omega_{\Lambda}=0.7\) and
\(H_{0}=70~\textnormal{km~s}^{-1}~\textnormal{Mpc}^{-1}\). The critical density of the Universe for the respective
mean cluster redshifts can be derived using \(\rho_{c}=3H(z)^{2}\cdot(8\pi~\textnormal{G})^{-1}\), with the gravitational
constant G and the Hubble parameter \(H(z)\).
All given optical magnitudes refer to the AB system.

\begin{table*}\caption{Cluster properties from optical and X-ray spectroscopy.}
\label{tab:resulttab}
\centering\begin{small}
\begin{tabular}{c c c c c c c c c}
\hline
\hline
name & n & \(z_{cl}\) & \(\sigma\) &  \(r_{200}\) & \(M_{200}\) & \(F_{0.5-2}(300\rm kpc)\) & \(L_{bol}(r_{500})\) & \(k_{\rm B}T\) \\
 \hline
2XMMp~J083026.2+524133 & 8  & \(0.9856(27)\) & \(930_{-230}^{+300}\) & \(2.44_{-0.6}^{+0.79}\) & \(5.1_{-2.9}^{+6.7}\) & \(6.477\pm0.095\) & \(1168\pm4\) & \(7.82^{+0.4}_{-0.39}\) \\
2XMMp~J093437.4+551340 & 11 & \(0.83858(73)\) & \(430_{-96}^{+120}\) & \(1.24_{-0.28}^{+0.35}\) & \(0.56_{-0.3}^{+0.62}\) & \(7.4\pm0.83\) & \(56\pm6.1\) & \(2.96^{+0.86}_{-0.65}\) \\
2XMMp~J105319.8+440817 & 8  & \(0.8955(32)\) & \(780_{-360}^{+660}\) & \(2.14_{-0.97}^{+1.82}\) & \(3.1_{-2.6}^{+16.6}\) & \(2.37\pm0.27\) & \(240\pm25\) & \(3.6^{+1.2}_{-0.76}\) \\
2XMMp~J120815.5+250001 & 16 & \(0.9929(26)\) & \(1200_{-240}^{+300}\) & \(3.13_{-0.62}^{+0.78}\) & \(10.9_{-5.3}^{+10.3}\) & \(2.79\pm0.42\) & \(363\pm44\) & \(4.1^{+2.2}_{-1.1}\) \\
2XMMp~J123759.3+180332 & 16  & \(0.8874(24)\) & \(1200_{-250}^{+310}\) & \(3.31_{-0.7}^{+0.87}\) & \(11.4_{-5.8}^{+11.6}\) & \(4.75\pm0.47\) & \(536\pm40\) & \(5.0^{+1.9}_{-1.1}\) \\
2XMMp~J133853.9+482033 & 8  & \(0.74969(30)\) & \(122_{-26}^{+32}\) & \(0.37_{-0.07}^{+0.1}\) & \(0.013_{-0.006}^{+0.014}\) & \(1.1\pm0.16\) & \(148\pm22\) & --- \\
\hline
\hline
\end{tabular}\end{small}
\tablefoot{The dynamical mass \(M_{200}\) is given in \(10^{14}M_{\odot}\), \(r_{200}\) in Mpc. The errors of \(\sigma\) correspond
to a confidence interval of 90\%, \(M_{200}\) and \(r_{200}\) furthermore include the uncertainty in \(A_{\rm 1D}\). The X-ray
properties flux and ICM temperature are based on measurements within a radius of 300kpc, the bolometric luminosity was extrapolated
to \(r_{500}\) (details see Sect.~\ref{sec:Xray_analysis}). The X-ray flux is in units of \(10^{-14}~\rm erg~cm^{-2}~s^{-1}\),
the luminosity in \(10^{42}~\rm erg~s^{-1}\).
Note that two clusters (2XMMp~J120815.5+250001 and 2XMMp~J123759.3+180332) have extremely large dynamical mass estimates
(\(>10^{15}M_{\odot}\)) but, for this sample, an average ICM temperature. The dynamical mass estimates might be overestimated due to
velocity measurements from galaxies not gravitationally bound to the cluster (see Sect.~\ref{sec:LBT_013} and ~\ref{sec:LBT_015}).}
\end{table*}

\section{Sample definition}
\label{sec:sample}
The catalogue 2XMMp, a pre-release of the second \emph{XMM-Newton} catalogue \citep[2XMM;][]{Watson2009},
lists all sources detected with the XMM-EPIC cameras 
in \(\sim2400\) observations before April 2006, covering an area
on the sky of \(\sim360~\textnormal{deg}^2\).
When this project started, the 2XMMp catalogue was the current and most complete catalogue of
\emph{XMM-Newton} serendipitous sources. Since then, the \emph{XMM-Newton} catalogue was updated continuously
through incremented observations and changes in the data reduction pipeline. In Sect.~\ref{sec:Xray_reduction}
we compare the initial sample parameters to the more recent 3XMM-DR6 catalogue \citep{Rosen2016}.

For our cluster sample we selected only observations within the footprint of the Sloan Digital Sky Survey
DR6 \citep{Adelman2008} and at Galactic latitudes \(|bII| > 20^{\circ}\). 
Observations not suitable for the detection of extended sources (e.g. because of very extended or very bright
targets, or observations with severe background contamination) were removed by means of visual screening of
the EPIC images. This reduced the number of EPIC pointings used for the compilation of our input catalogue
to 478. When we take repeated or overlapping pointings in the same sky area into account, the total survey
area is \(\sim60~\textnormal{deg}^2\).

For the 2XMM data processing, the source detection pipeline of XMM-SAS version 7.1 was used in a
configuration enabling the classification of sources as point-like or extended. Extended sources were
fitted with radially symmetric \(\beta\)-profiles, convolved with the corresponding EPIC point-spread function (PSF), in order to
determine the source parameters. Since the spurious detection rate of extended sources is relatively high
(e.g. near bright point-sources or because of blending of multiple point-sources), another visual screening
step was necessary to clean the sample. The final list of extended sources comprises 412 entries.

In order to identify candidates for distant galaxy clusters, we correlated the source positions with the
SDSS DR6 photometric and spectroscopic catalogues and generated SDSS finding charts for each of the
extended X-ray sources. This correlation yielded 393 extended X-ray sources with either a galaxy or cluster
candidate detected in the SDSS.

Only 19 sources of our selection (see Table~\ref{tab:survey}) have no plausible counterpart in
the SDSS imaging and therefore are considered as candidates for distant clusters of galaxies, where the
member galaxies are beyond the detection limit of the SDSS.

\section{Optical pre-imaging}
\label{Sec:imaging}
\subsection{Imaging data - data description and reduction}
\label{imaging_reduction}

\begin{figure}
 \includegraphics[angle=0,width=9.0cm]{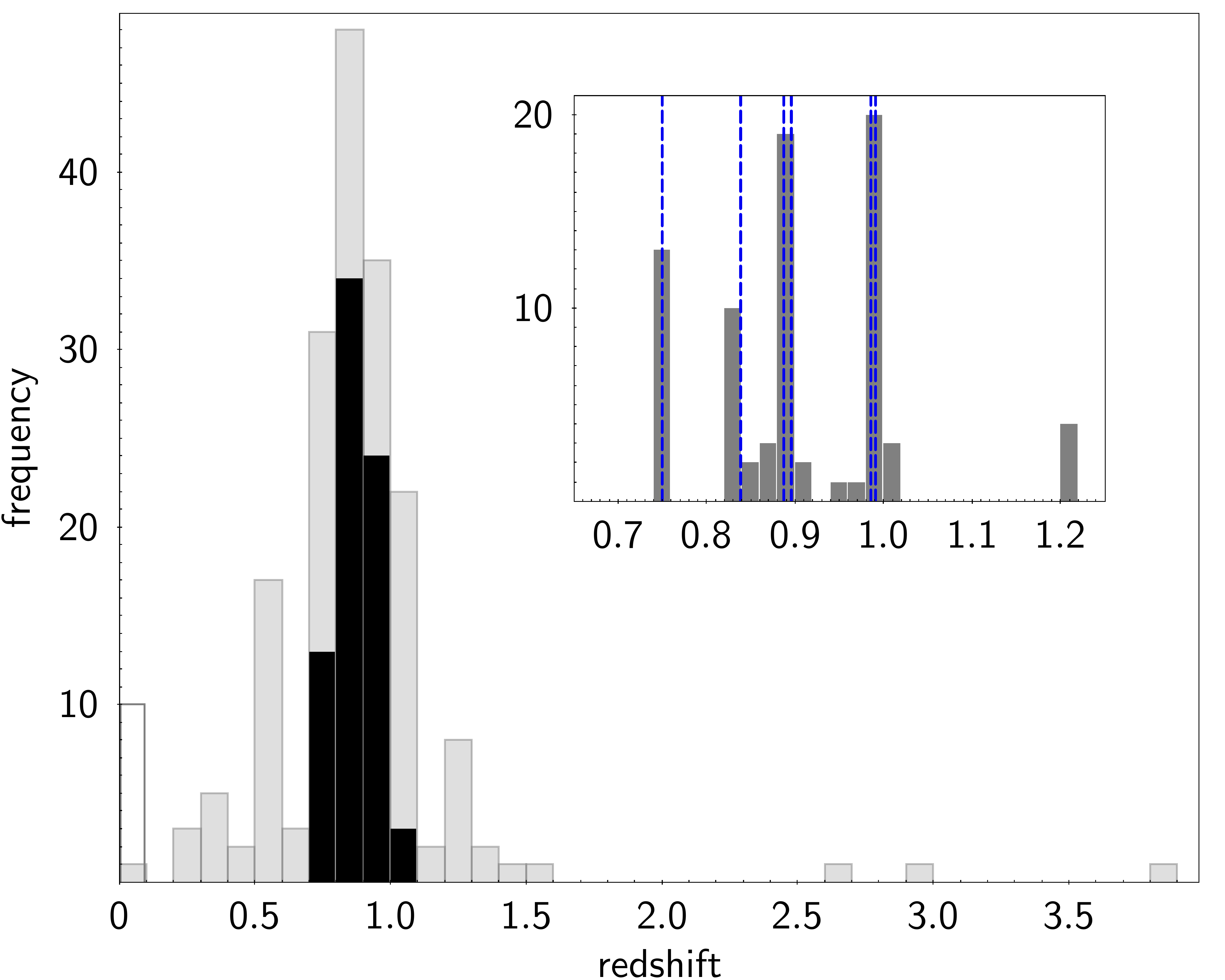}
 \caption{Histogram built from all spectra of our spectroscopic observations. The transparent region in the main plot
 marks the content of stars identified in the sample, the light shaded regions reflect the content of galaxies, where the fraction of all identified cluster member galaxies is plotted black. The inset details the mean cluster
 redshift as vertical blue dashed lines overplotted on the histogram of the identified member galaxies that match our selection
 criteria.}
 \label{fig:redshift_histo}
\end{figure}

\begin{figure*}
  \begin{subfigure}{0.5\textwidth}
    \centering\includegraphics[angle=0,width=0.75\linewidth]{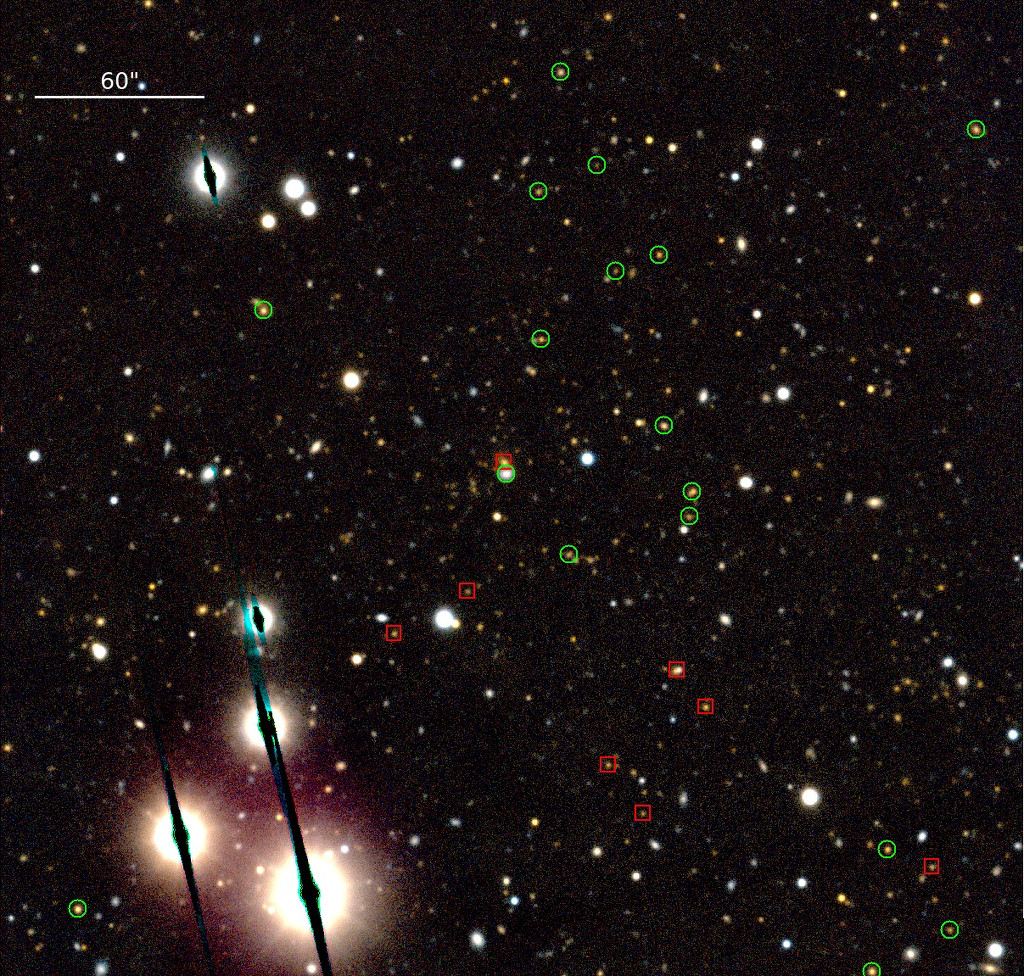}
    \subcaption{2XMMp~J083026.2+524133}
    \label{fig:LBT_003_RGB}
  \end{subfigure}\hfill
  \begin{subfigure}{0.5\textwidth}
    \centering\includegraphics[angle=0,width=0.75\linewidth]{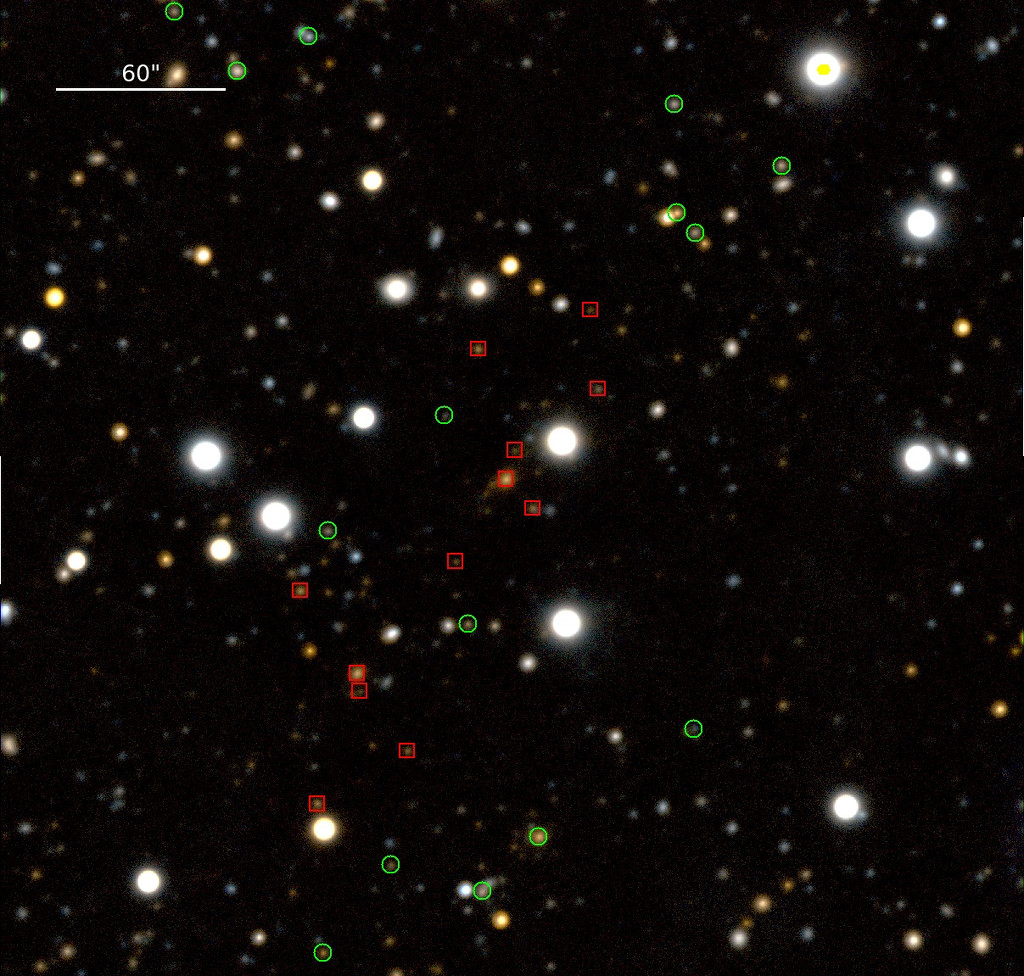}
    \subcaption{2XMMp~J093437.4+551340}
    \label{fig:LBT_007_RGB}
  \end{subfigure}\vskip0.25cm

  \begin{subfigure}{0.5\textwidth}
     \centering\includegraphics[angle=0,width=0.75\linewidth]{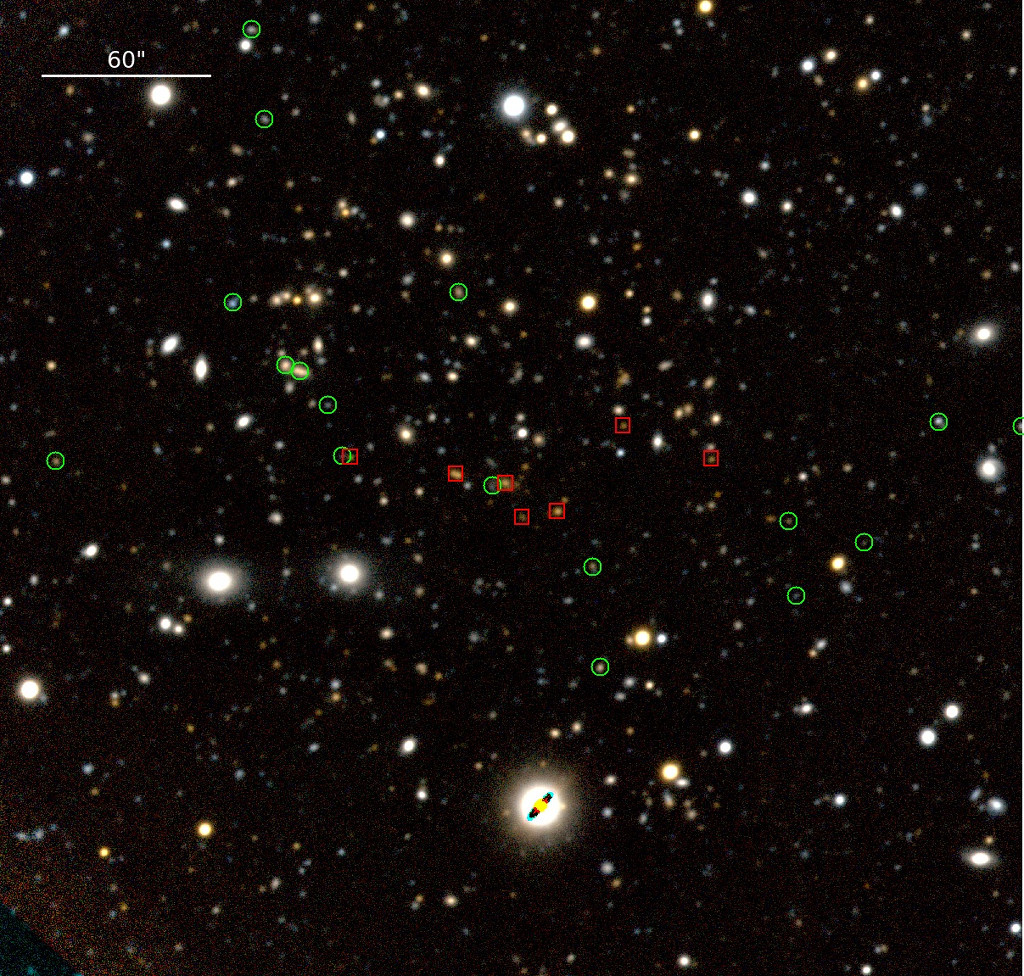}
     \subcaption{2XMMp~J105319.8+440817}
     \label{fig:LBT_010_RGB}
  \end{subfigure}
  \begin{subfigure}{0.5\textwidth}
     \centering\includegraphics[angle=0,width=0.7425\linewidth]{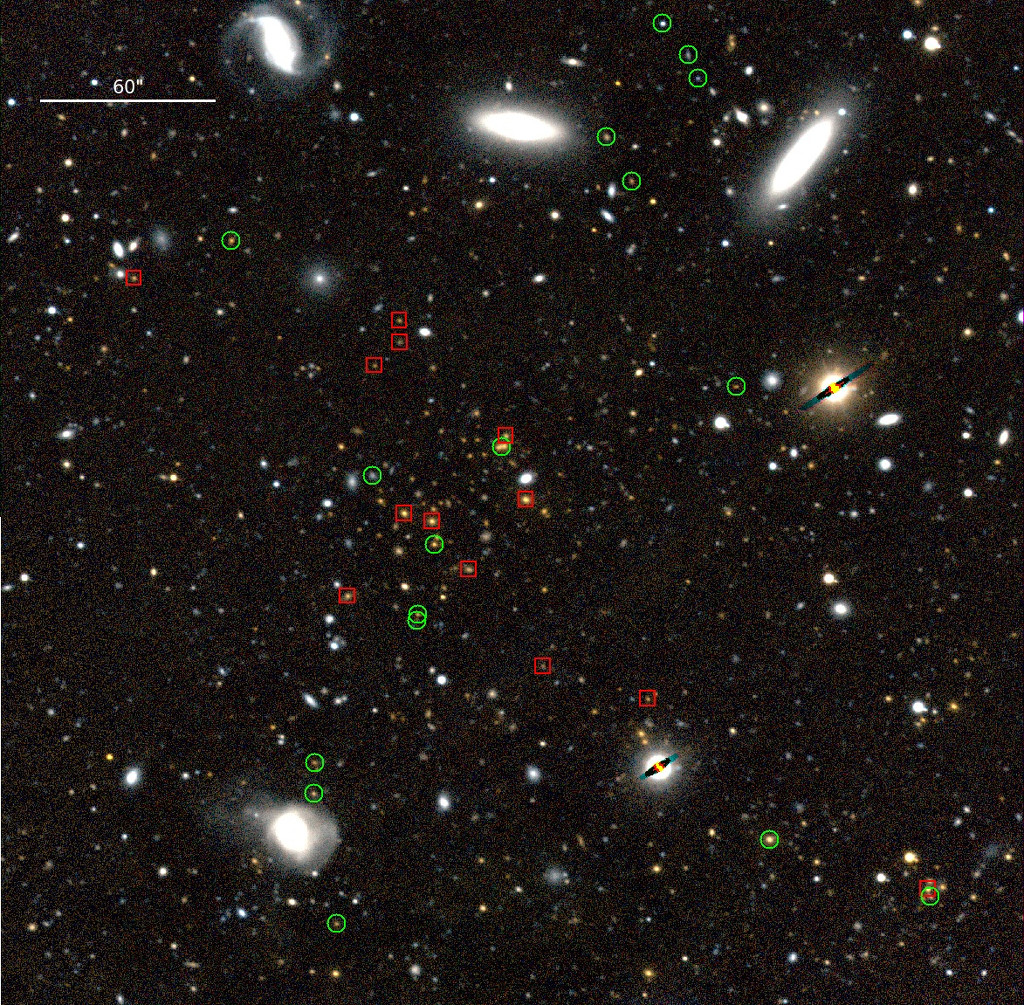}
     \subcaption{2XMMp~J120815.5+250001}\label{fig:LBT_013_RGB}
  \end{subfigure}\vskip0.25cm

  \begin{subfigure}{0.5\textwidth}
    \centering\includegraphics[angle=0,width=0.75\linewidth]{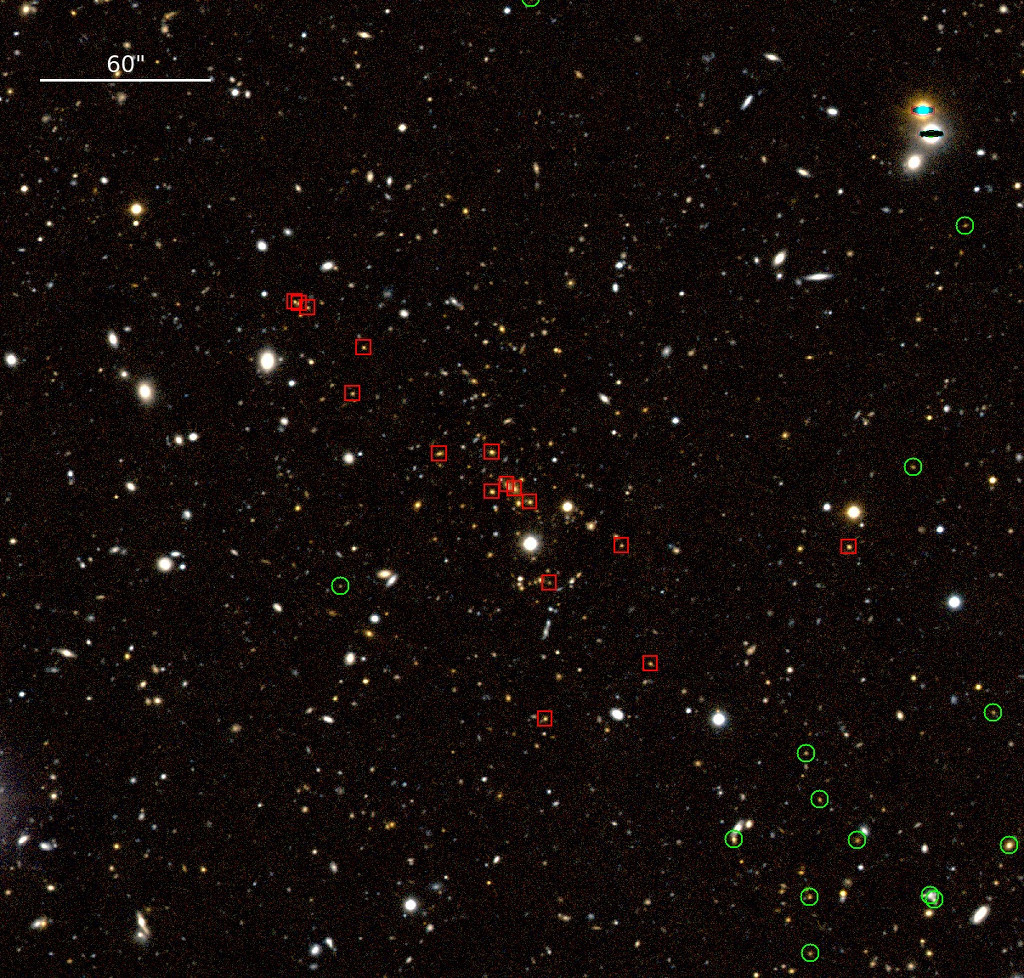}
    \subcaption{2XMMp~J123759.3+180332}
    \label{fig:LBT_015_RGB}
  \end{subfigure}\hfill
  \begin{subfigure}{0.5\textwidth}
    \centering\includegraphics[angle=0,width=0.75\linewidth]{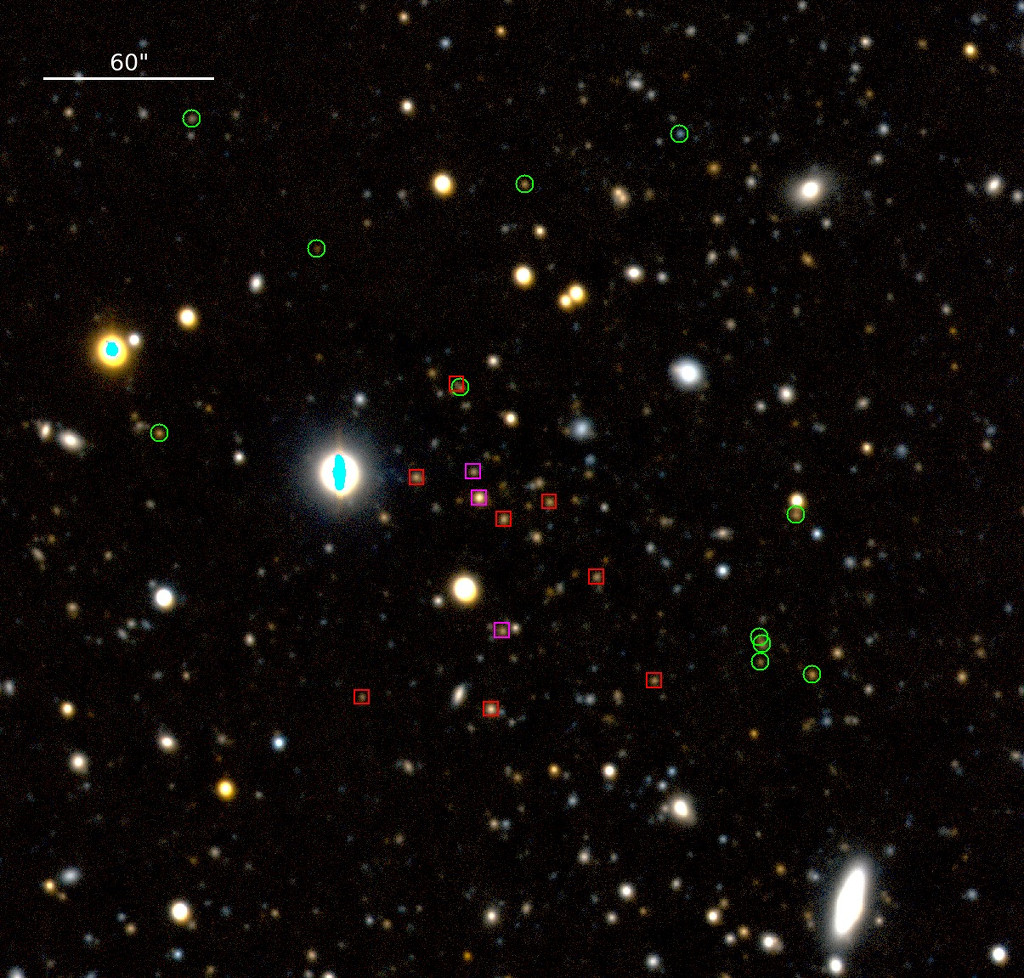}
    \subcaption{2XMMp~J133853.9+482033}
    \label{fig:LBT_018_RGB}
  \end{subfigure}\hfill
  \caption{\label{fig:LBT_RGB_cutouts}
           LBC colour images of the confirmed cluster fields of Section~\ref{sec:LBT_clusters}. We assigned r-SLOAN and z-SLOAN images from
           the LBC to the blue and red channel of the colour image. The green channel was created from the mean of both bands. 
           Overplotted red squares and green circles show cluster galaxies and spectroscopic sources outside our membership criteria, respectively.
           In panel (\subref{fig:LBT_018_RGB}), which shows the cluster 2XMMp~J133853.9+482033, we additionally highlight with
purple squares the group of galaxies that was ruled out by
           the iterative sigma-clipping method (compare Sect.~\ref{sec:LBT_018}).
           North is up, east is left, and the images are centred on the respective X-ray position.}
\end{figure*}

Deep optical imaging for 13 of the initial 19 fields was carried out between March 2008 and November
2011 at the LBT.
Our imaging strategy included binocular observation to efficiently use the telescope capabilities.
LBC-blue and LBC-red were equipped with r-SLOAN and z-SLOAN filters, respectively. Dither patterns of at
least nine positions with single-exposure times of 90s or 120s were used to cleanly bridge the chip gaps of
the instruments in the reduced images. For details on the individual target coordinates
and the exposure times, we refer to Table \ref{tab:survey}.

All imaging data were reduced with the code {\tt THELI} \citep{Schirmer2013,Erben2005}. 
The reduction included subtracting a master-bias and flat-fielding the raw data using sky-flats, applying
a background-model correction to reduce the imprinted fringing pattern, weighting, astrometric
calibration, and coaddition of the individual science exposures. The photometric zero-point 
(AB system) of the mosaic image was calculated using the SDSS-DR9 catalogue data as reference.

We extracted AB magnitudes using {\tt SExtractor} \citep{Bertin1996} with critical settings
that affect background estimation and object detection fixed for all fields and filters.
The dual-image mode of {\tt SExtractor} was used to simultaneously
measure isophotal object magnitudes in r-SLOAN and
z-SLOAN images on fixed positions that were determined on the z-band images.
The limiting magnitude of the coadded mosaics, taking the flux from the sky background
within a given aperture into account, is given by
\begin{equation}
 \label{eq:maglim}
 m_{lim}=ZP-2.5\rm log\left(5\sqrt{\rm N_{pix}}\sigma_{sky}\right)~,
\end{equation}
where we use the zero-point (ZP), the number of pixels (\(\rm N_{pix}\)) within the aperture
with 2\arcsec~radius, and the variation of the sky background as measured by {\tt SExtractor}.
The resulting 5\(\sigma\) detection limits for both filters are \(25.42<r_{lim}<26.49\) and
\(24.41<z_{lim}<25.49\), depending on the total exposure times of 0.3 to 0.85 hours.

False-colour images, created from the reduced r-SLOAN and z-SLOAN exposures, are shown in Section~\ref{sec:results}
and the respective sub-paragraphs for each of the seven spectroscopically confirmed galaxy clusters.

\subsection{Target selection for spectroscopic follow-up}
\label{target_selection}
We inspected the colour-magnitude diagrams of all fields in order to select the tentative cluster
galaxies using the red-sequence method \citep[see][]{Gladders2000}.
In these colour-magnitude plots, the colour of passive galaxies shows a clear dependency on
redshift, which is due to the prominence of the Balmer break (\(4000\AA\) at restframe). Hence, galaxies of the
same cluster undergoing similar evolution will scatter little in colour, but vary in magnitude.
Since the fraction of early-type cluster member galaxies decreases with redshift,
the red sequence for distant clusters may be only sparsely populated. For this reason, not only the colour was decisive for
the selection of a possible BCG and cluster galaxies, but also the position with respect to the extended
X-ray emission and the actual observability, such as brightness and typical constraints of multi-object
spectroscopy (slit collision and possible spectral overlap). Wherever possible, we filled the remaining
space on the slit-masks with less likely targets and objects of opportunity (e.g. isolated X-ray sources
or possible lensing features). 
Colour-magnitude diagrams of the six spectroscopically confirmed clusters are shown in
Figures~\ref{fig:LBT_colormag_I} and \ref{fig:LBT_colormag_II}.

\section{Spectroscopic follow-up}
\label{Sec:spectroscopy}
\subsection{Spectroscopy - data description}
\label{data_description}
The spectroscopic follow-up was executed in February and March 2012. The instrument of choice for this
campaign was the Multi-Object Double Spectrograph \citep[MODS;][]{Pogge2010} at the LBT. For each of the
seven follow-up fields, one multi-slit mask (MOS-mask) was cut, intended to cover the most probable cluster
galaxy candidates (see Sect.~\ref{target_selection}) and a few objects of opportunity.
We only observed using the red spectral arm of MODS with grating G670L, resulting in a spectral coverage of
\(5200-10~000\)~\AA.
Depending on the target brightness and observing conditions, we took up to five individual exposures of
30 minutes per field; the total exposure time per field is summarized in Table~\ref{tab:survey}.

\subsection{Spectroscopy - data reduction}
Data reduction was performed as a two-stage process: pre-reduction, and pipeline reduction. During
pre-reduction we made use of
{\tt modsCCDRed}\footnote{\label{modsCCDRed}http://www.astronomy.ohio-state.edu/MODS/Software/modsCCDRed/},
a set of Python-scripts that subtracted the bias-pattern and applied a flat-field correction. The resulting
pre-reduced data were further reduced with a set of scripts written mainly in {\tt ESO-Midas}. These scripts
were initially intended to reduce VLT/FORS2 data, but are perpetually generalised to suit different multi-object
spectroscopy instruments, including MODS. 

The workflow of the pipeline reduction includes the following tasks: cosmic filtering of scientific frames,
wavelength calibration, weighted extraction of 1D spectra \citep[following][]{Horne1986},
including generation of respective error spectra, creation of 2D sky-subtracted frames, and
co-addition of single spectra. 
Spectrophotometric standard stars observed during our observation period were used to flux-calibrate
the final spectra.

\subsection{Spectroscopy - data analysis}
\label{data_analysis}
We were able to extract a total of 215 spectra from our data. All 1D flux- and error-spectra were
inspected with {\tt EZ} \citep[a tool for automatic redshift measurement, see ][]{Garilli2010}
to obtain first impressions on quality and redshift of each individual spectrum.
The first inspection revealed 194 preliminary redshifts, of which 176 were graded secure because they showed several
spectral features. In these science-grade spectra, we identified 10 late-type stars.
Spectra of obvious galaxy cluster candidates, primarily passive galaxies (see Sect.~\ref{target_selection}), were inspected
in more detail. We measured their redshift by fitting a double-Gaussian to the prominent CaII H\&K absorption lines,
or alternatively, when the signal-to-noise ratio was found to be too low, by solely ftting the [OII] (3727\AA) emission. All galaxy
redshifts were converted into the barycentre of the solar system.
In Fig.~\ref{fig:redshift_histo} we show a histogram of all science-grade redshifts from our spectroscopic program,
where the inset highlights the redshifts of the successfully identified clusters.

\begin{figure*}
 \centering
 \begin{subfigure}{0.33\textwidth}\centering
  \includegraphics[width=5.5cm]{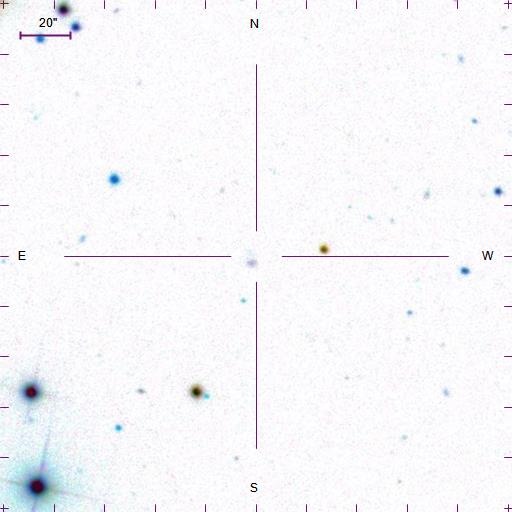}
  \caption{}
  \label{fig:LBT_003_SDSS}
 \end{subfigure}\hfill
 \begin{subfigure}{0.33\textwidth}\centering
  \includegraphics[width=5.5cm]{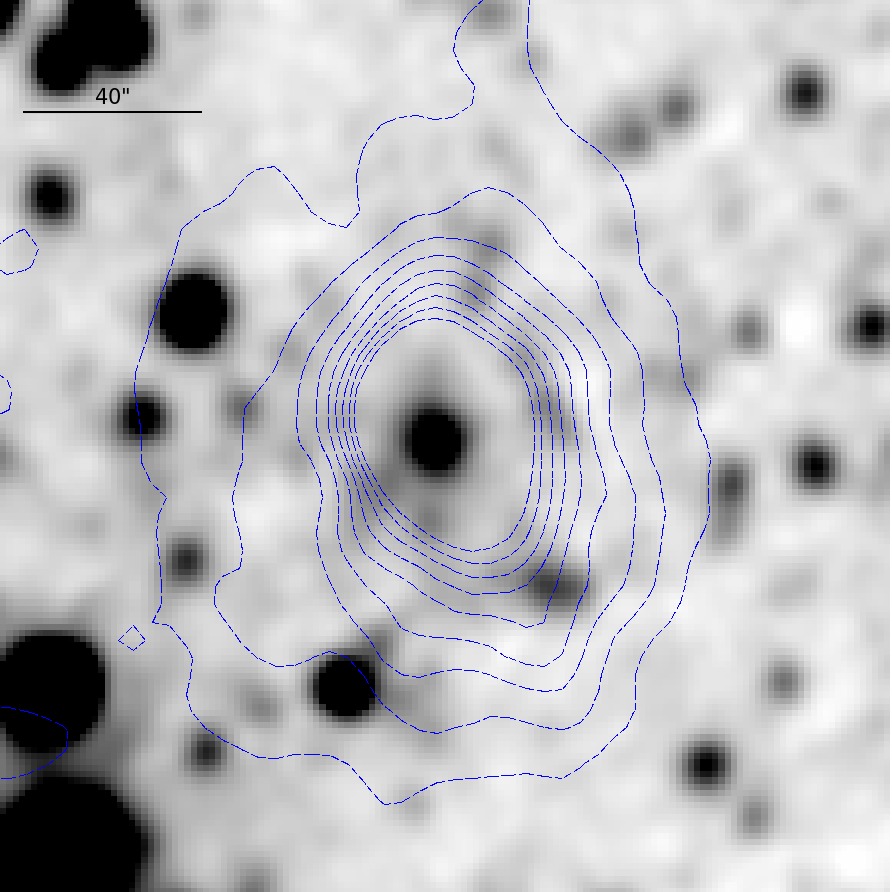}
  \caption{}
  \label{fig:LBT_003_WISE}
 \end{subfigure}\hfill
 \begin{subfigure}{0.33\textwidth}\centering
  \includegraphics[width=5.5cm]{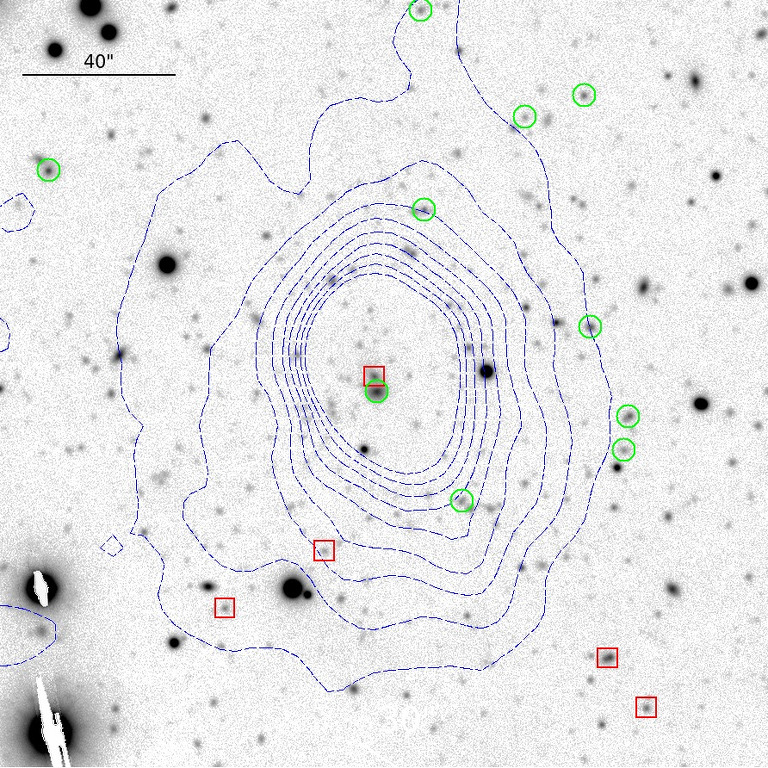}
  \caption{}
  \label{fig:LBT_003_LBC}
 \end{subfigure}\hfill
 
 \begin{subfigure}{0.33\textwidth}\centering
  \includegraphics[width=5.5cm]{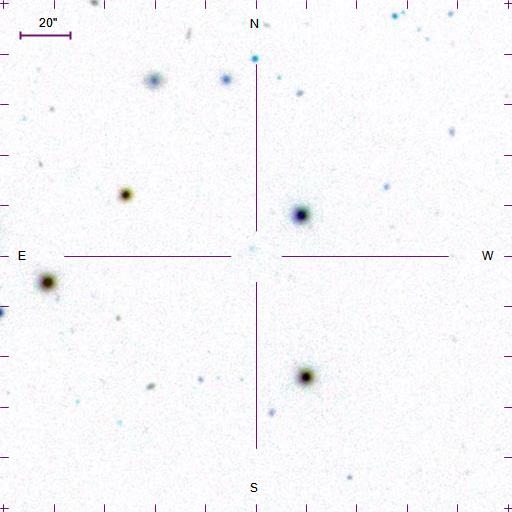}
  \caption{}
  \label{fig:LBT_007_SDSS}
 \end{subfigure}\hfill
 \begin{subfigure}{0.33\textwidth}\centering
  \includegraphics[width=5.5cm]{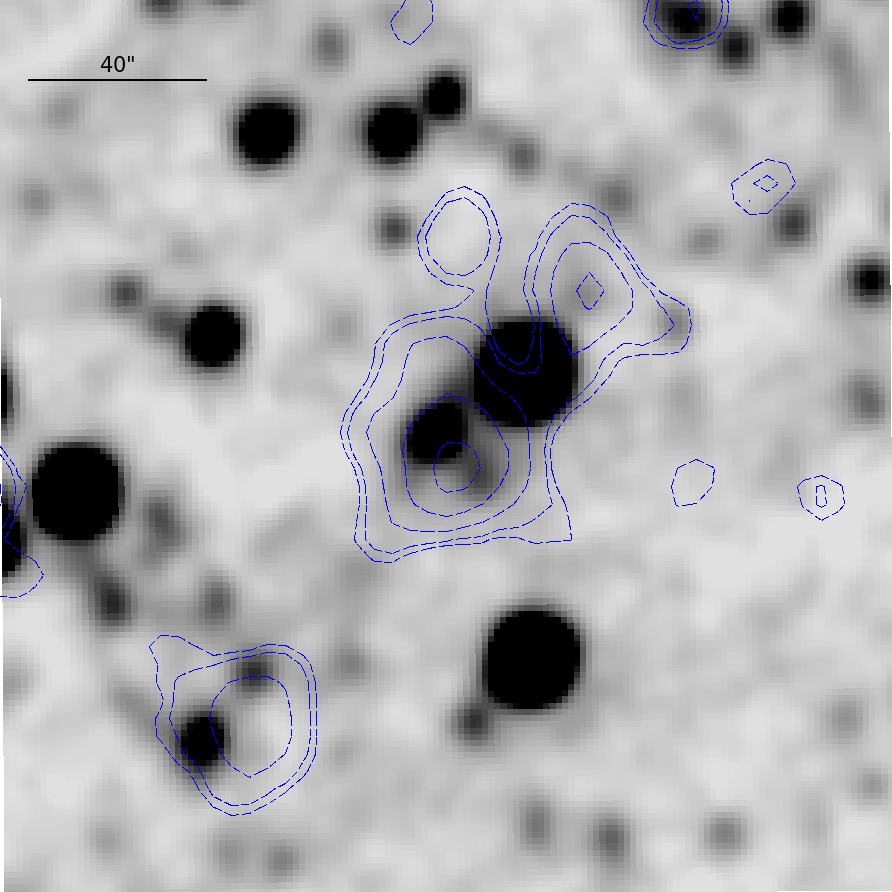}
  \caption{}
  \label{fig:LBT_007_WISE}
 \end{subfigure}\hfill
 \begin{subfigure}{0.33\textwidth}\centering
  \includegraphics[width=5.5cm]{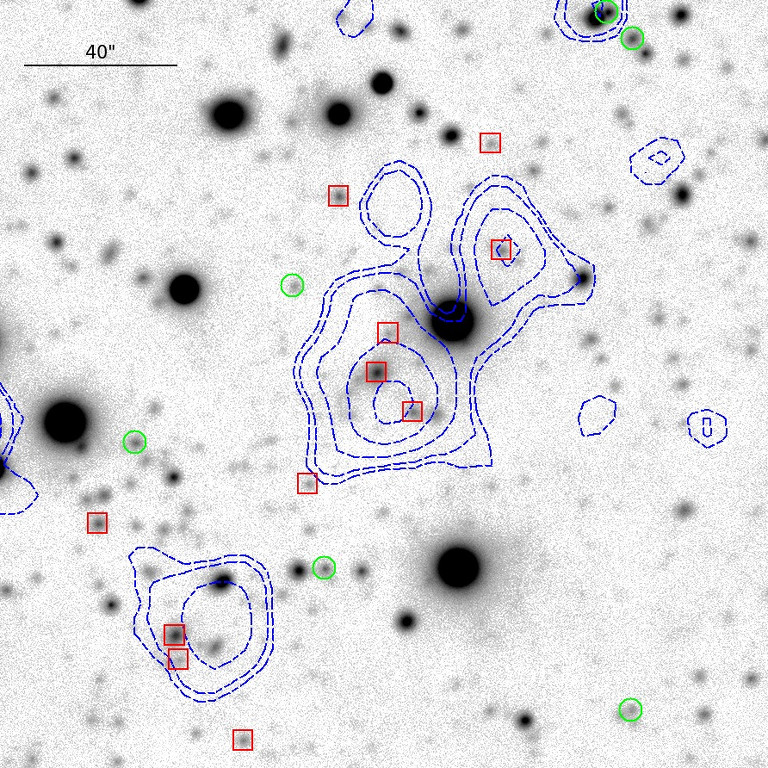}
  \caption{}
  \label{fig:LBT_007_LBC}
 \end{subfigure}
 
 \begin{subfigure}{0.33\textwidth}\centering
  \includegraphics[width=5.5cm]{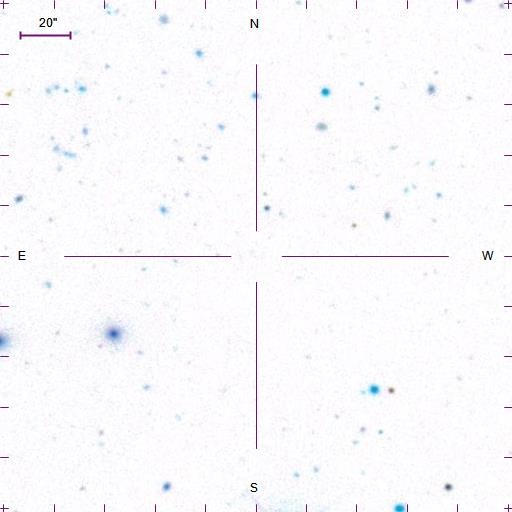}
  \caption{}
  \label{fig:LBT_010_SDSS}
 \end{subfigure}\hfill
 \begin{subfigure}{0.33\textwidth}\centering
  \includegraphics[width=5.5cm]{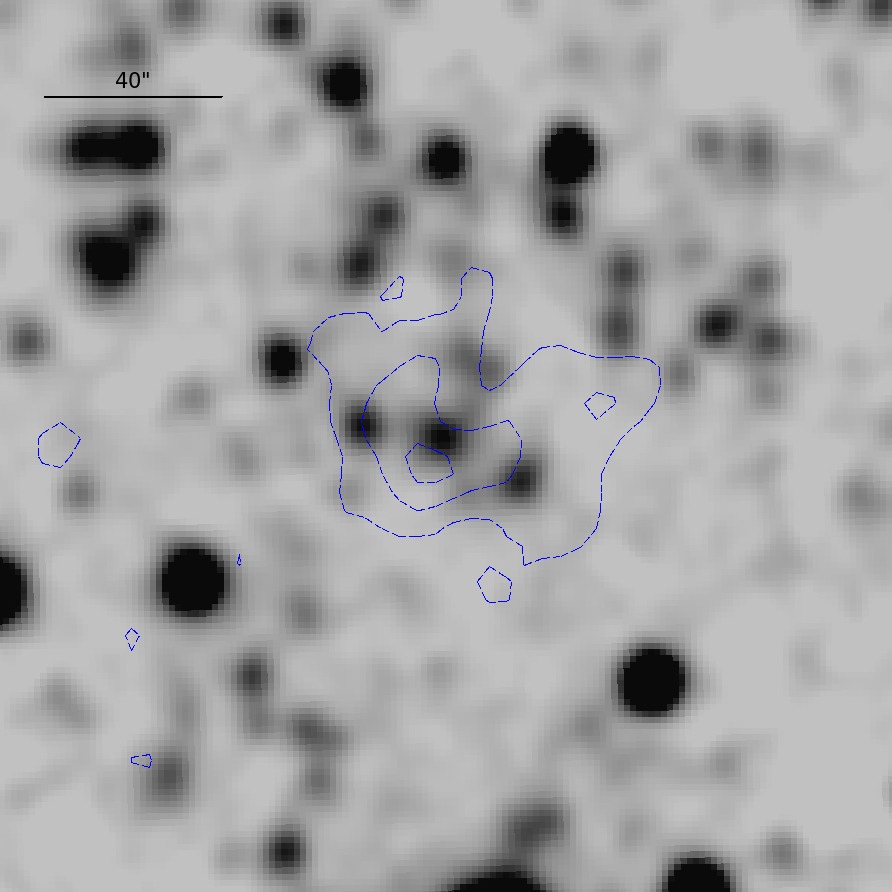}
  \caption{}
  \label{fig:LBT_010_WISE}
 \end{subfigure}\hfill
 \begin{subfigure}{0.33\textwidth}\centering
  \includegraphics[width=5.5cm]{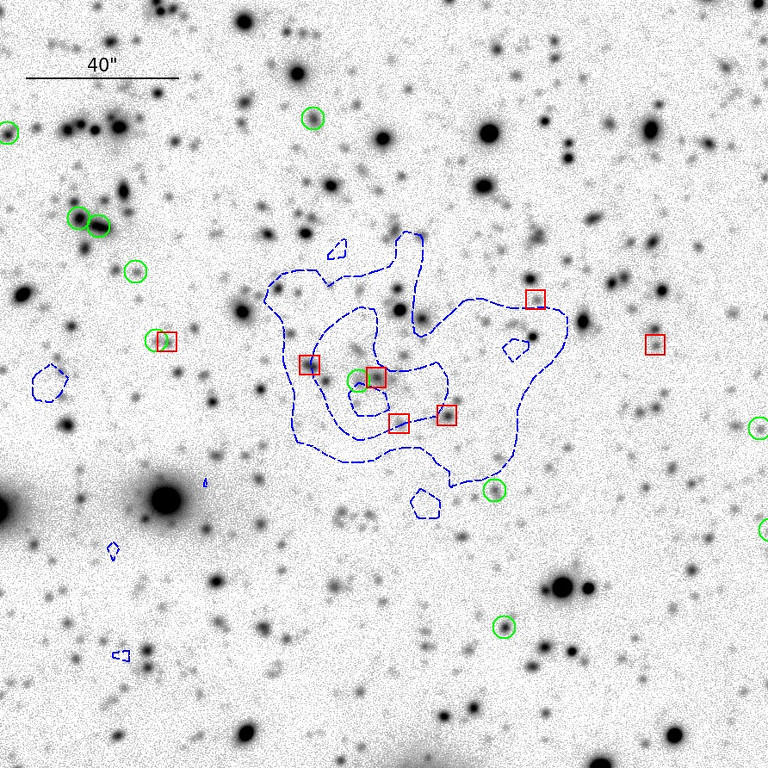}
  \caption{}
  \label{fig:LBT_010_LBC}
 \end{subfigure}\hfill
\caption{\label{fig:LBT_3cutouts_I}
         Optical and near-IR images of the fields of 2XMMp~J083026.2+524133 (top row), 2XMMp~J093437.4+551340 (middle), and 2XMMp~J105319.8+440817 (bottom).
         The first column shows SDSS cutouts of the respective fields, the central column shows the W1-band (3.4\(\mu m\)) images from the AllWISE survey, and
         on the right side, we present LBC images (mean of the r- and z-SLOAN filters). Images of each row are centred at the respective X-ray positions
         and show the same region of the sky. Blue dashed lines (middle and right images) indicate contours from the X-ray flux, where the levels are
         chosen for illustration alone. Red squares and green circles are overplotted on the LBC images, referring to cluster galaxies and spectroscopic
         sources outside our membership criteria, respectively. North is up, east is left, and the images are centred on the respective X-ray position.}
\end{figure*}

\section{Archival X-ray data and data reduction}
\label{sec:Xray_reduction}
The whole sample of extended sources was initially selected from the 2XMMp (see Sect.~\ref{sec:sample}).
Since in the meantime the number of \emph{XMM-Newton} observations (OBSIDs) and the knowledge on the instruments
have vastly increased, we compared our initial targets with the recent 3XMM-DR6 catalogue \citep{Rosen2016}
to ensure that the detections are characterized as extended sources.
For all initial targets, we found an extended detection in the 3XMM-DR6, with the exception of 2XMMp~J145220.8+165458,
which is now classified as a point source.
The extents of all other sources agreed well and were mostly within the errors of the two catalogues. The
number of observations covering our sample increased from 28 to 36 between 2XMMp and 3XMM-DR6.
A comparison between 2XMMp and 3XMM-DR6, containing source-identifiers (SRCID), extent parameters (EP\_EXTENT),
and OBSID, is listed in Table~D.1. The table also contains the flare-cleaned exposure time for the
EPIC instruments for all OBSIDs from which we extracted spectra (see Sect.~\ref{sec:Xray_analysis}).

We downloaded all publicly available \emph{XMM-Newton} observations until December 2015 that overlapped our
sample from the \emph{XMM-Newton} Science Archive \citep[XSA; ][]{Arviset2002}. The data reduction and analysis was
carried out using the \emph{XMM-Newton} Science Analysis Software (SAS) version 15.
From background-flare-cleaned event file lists, we generated X-ray images in the energy range between 0.5 and 3.0 keV
for each EPIC camera and OBSIDs individually.

\begin{figure*}
 \centering
 \begin{subfigure}{0.33\textwidth}\centering
  \includegraphics[width=5.5cm]{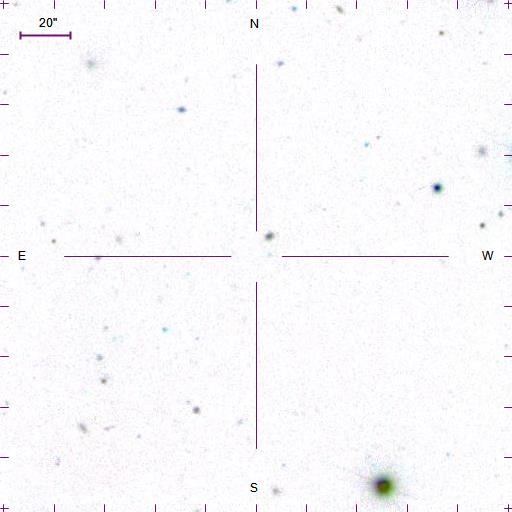}
  \caption{}
  \label{fig:LBT_013_SDSS}
 \end{subfigure}\hfill
 \begin{subfigure}{0.33\textwidth}\centering
  \includegraphics[width=5.5cm]{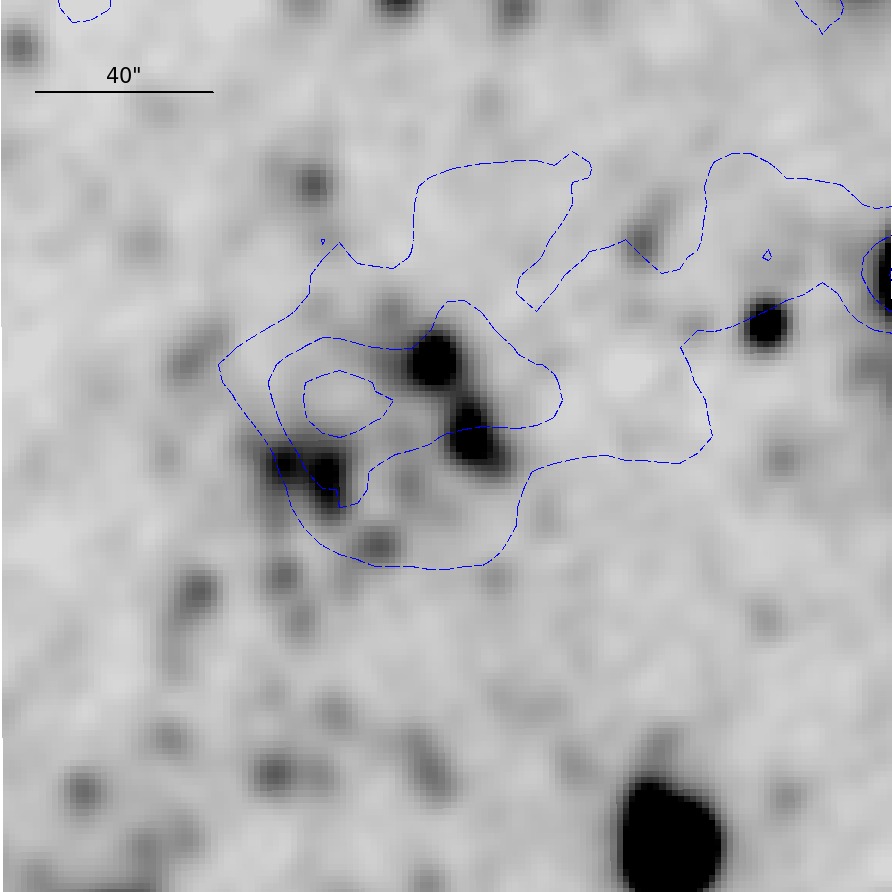}
  \caption{}
  \label{fig:LBT_013_WISE}
 \end{subfigure}\hfill
 \begin{subfigure}{0.33\textwidth}\centering
  \includegraphics[width=5.5cm]{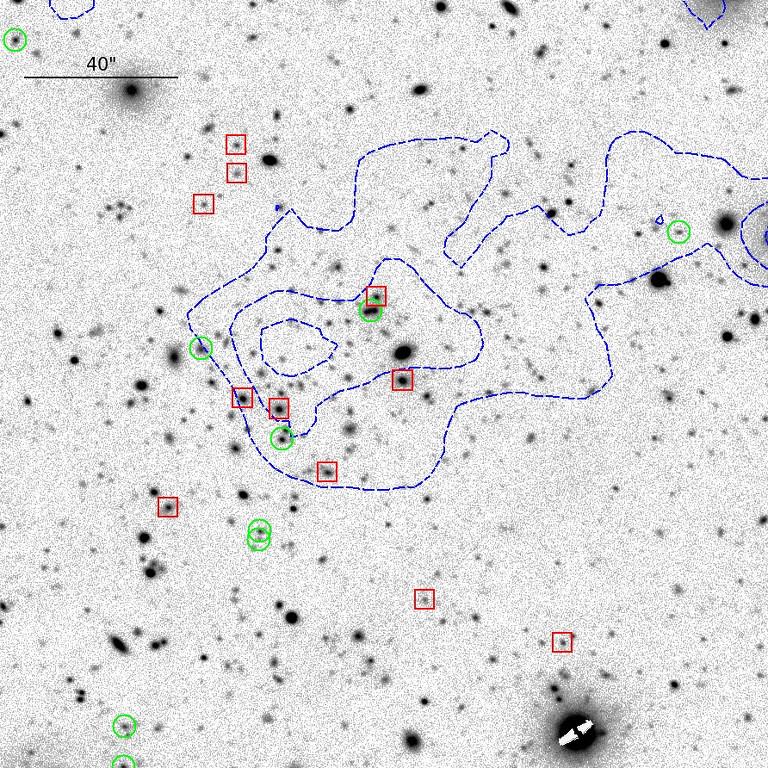}
  \caption{}
  \label{fig:LBT_013_LBC}
 \end{subfigure}\hfill
 
 \begin{subfigure}{0.33\textwidth}\centering
  \includegraphics[width=5.5cm]{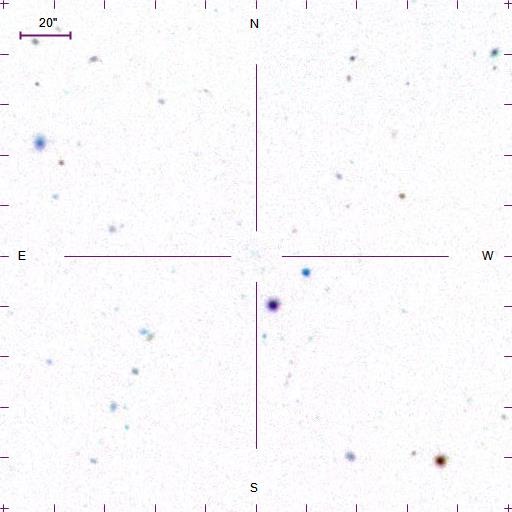}
  \caption{}
  \label{fig:LBT_015_SDSS}
 \end{subfigure}\hfill
 \begin{subfigure}{0.33\textwidth}\centering
  \includegraphics[width=5.5cm]{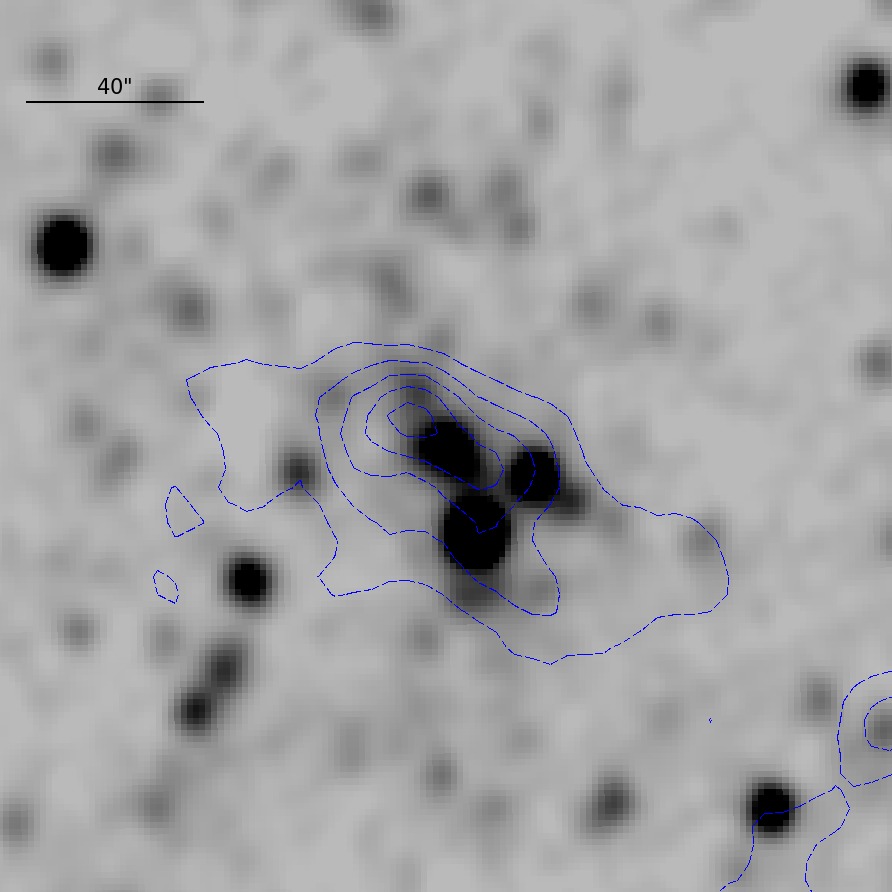}
  \caption{}
  \label{fig:LBT_015_WISE}
 \end{subfigure}\hfill
 \begin{subfigure}{0.33\textwidth}\centering
  \includegraphics[width=5.5cm]{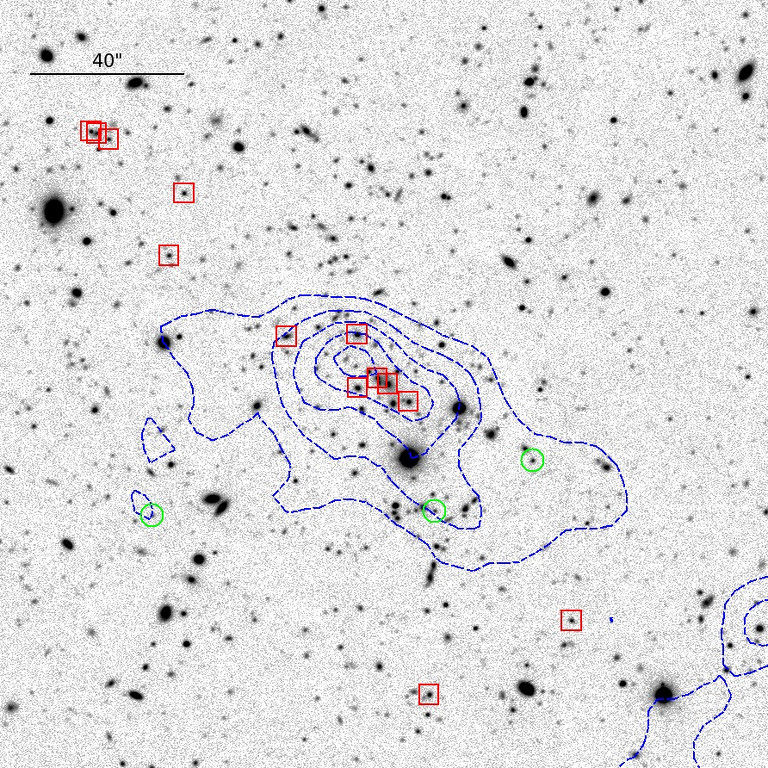}
  \caption{}
  \label{fig:LBT_015_LBC}
 \end{subfigure}
 
 \begin{subfigure}{0.33\textwidth}\centering
  \includegraphics[width=5.5cm]{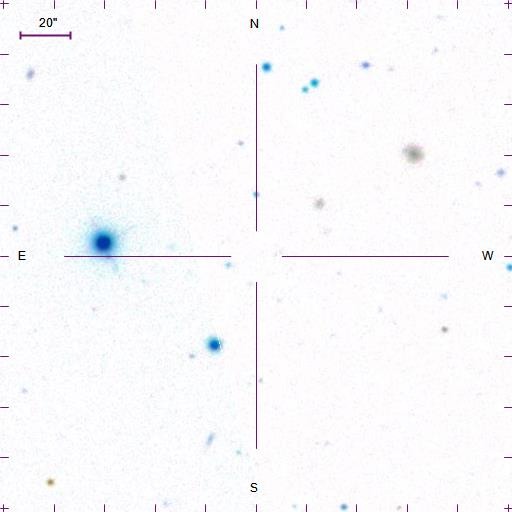}
  \caption{}
  \label{fig:LBT_018_SDSS}
 \end{subfigure}\hfill
 \begin{subfigure}{0.33\textwidth}\centering
  \includegraphics[width=5.5cm]{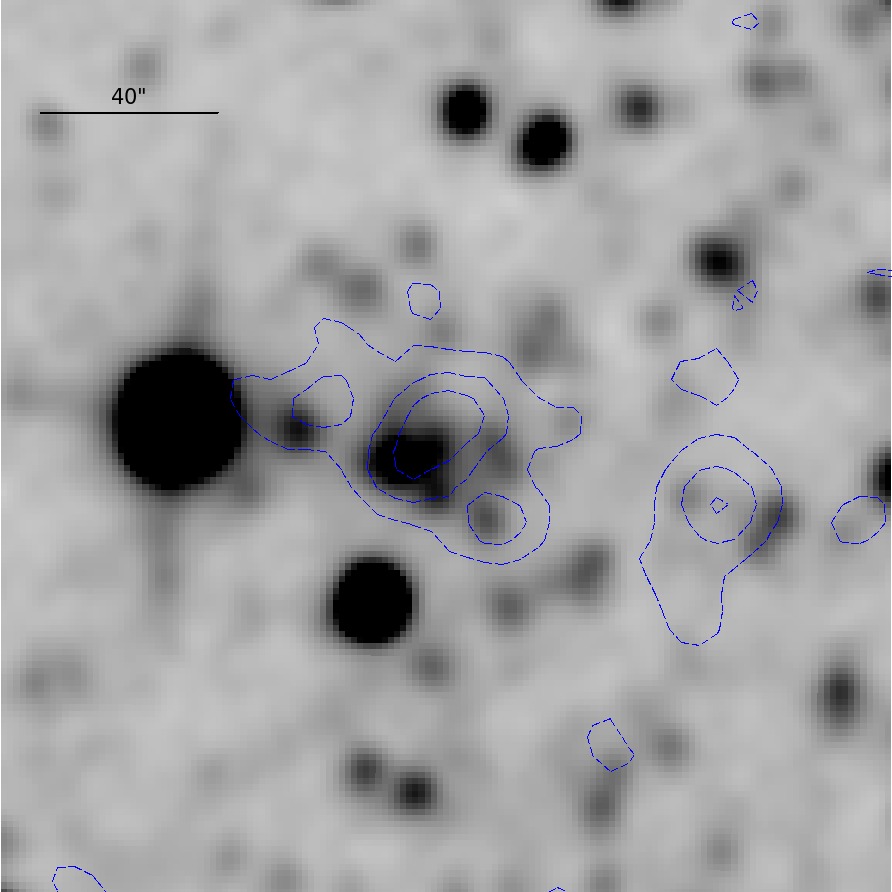}
  \caption{}
  \label{fig:LBT_018_WISE}
 \end{subfigure}\hfill
 \begin{subfigure}{0.33\textwidth}\centering
  \includegraphics[width=5.5cm]{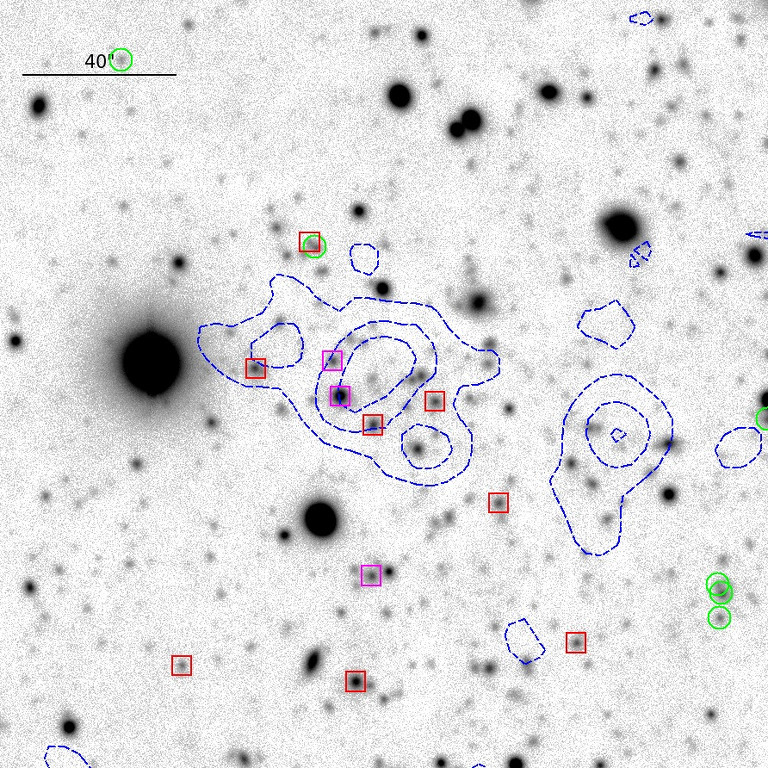}
  \caption{}
  \label{fig:LBT_018_LBC}
 \end{subfigure}\hfill
\caption{\label{fig:LBT_3cutouts_II}
         Optical and near-IR images of the fields of 2XMMp~J120815.5+250001 (top row), 2XMMp~J123759.3+180332 (middle), and 2XMMp~J133853.9+482033 (bottom).
         The first column shows SDSS cutouts of the respective fields, the central column the W1-band (3.4\(\mu m\)) images from the AllWISE survey, and
         on the right side we present LBC images (mean of the r- and z-SLOAN filters). Images of each row are centred at the respective X-ray positions
         and show the same region of the sky. Blue dashed lines (middle and right images) indicate contours from the X-ray flux, where the levels are
         chosen for illustration alone. Red squares and green circles are overplotted on the LBC images, referring to cluster galaxies and spectroscopic
         sources outside our membership criteria, respectively, while purple squares in panel (\subref{fig:LBT_018_LBC}) are explained in Sect.~\ref{sec:LBT_018}.
         North is up, east is left, and the images are centred on the respective X-ray position.}
\end{figure*}

\section{Results: galaxy and cluster properties}
\label{sec:results}

\subsection{Cluster redshifts and dynamical masses}
\label{sec:dynamics}

In this section we describe our procedure of cluster member selection and the velocity dispersion and dynamical
mass calculation for the seven fields of which we have dedicated spectroscopic data from LBT/MODS.
We proceed with the analysis of the \emph{XMM-Newton} data, from which we extract spectra and compute fluxes,
luminosities, and temperatures. We give the results and further details for all selected fields, which are grouped
according to their available data and properties: (i) the six fields of newly confirmed clusters with
LBT/MODS spectroscopy, (ii) targets with no spectroscopy from this work, but published redshifts from the literature,
and (iii) the remaining fields containing a high-redshift group of galaxies, and candidates that are not yet confimed or are
rejected.

At this point, we also introduce images from the Wide-field Infrared Survey Explorer \citep[WISE, see][]{Wright2010},
whose imaging bands are centred at 3.4, 4.6, 12, and 22 \(\mu m\) (W1-W4).
Especially the bands W1 and W2 are useful for colour-selecting galaxies between \(0.75 < z < 1.75\) because of an
apparently monotonical reddening. This method was introduced by \citet{Papovich2008} and later adopted by
\citet{Gettings2012} for their search for distant galaxy clusters using WISE data. Cutout IR images from WISE
for our cluster candidates therefore allow high-redshift galaxies to be better recognized than from optical
survey data (SDSS) alone. We did not use the IR colour as an
excluding criterion for the decision between low and high
redshift of clusters -- a qualitative examination of this criterion is part of ongoing
work comprising a larger sample of galaxy clusters, and will be presented in a future paper (Rabitz et al. in prep.).
As additional detail for the LBT/MODS confirmed clusters, we provide cutouts from the SDSS, WISE, and LBC pre-imaging
of the respective cluster fields (Figs.~\ref{fig:LBT_RGB_cutouts}, \ref{fig:LBT_3cutouts_I}, and \ref{fig:LBT_3cutouts_II})
together with photometric and spectroscopic results from our observations (Figs.~\ref{fig:LBT_colormag_I} and \ref{fig:LBT_colormag_II}).

Since LBT imaging data are also available for groups (ii) and (iii) of our data, we present the respective LBC images
and SDSS finding charts in Appendices~B and C, while additional information for the high-redshift
group of galaxies are given in Figs.~\ref{fig:LBT_006_cutouts}, \ref{fig:LBT_006_RGB}, and \ref{fig:LBT_006_colorhist}.

We summarize the properties of the LBT/MODS-confirmed clusters in Table~\ref{tab:resulttab}, and those of the
clusters with references in Table~\ref{tab:results_known}.
To evaluate cluster redshift and velocity dispersion, we adopted the bi-weight method outlined in
\citet{Beers1990}. Our strategy to determine \(z_{cl}\) was as follows. We selected all galaxies as
preliminary cluster members fulfilling \(|z_{mem}-z_{BCG}|\cdot c \le 4000~\textnormal{km~s}^{-1}\),
where we used the redshift of an obvious brightest cluster galaxy \(z_{BCG}\) and the speed of light
\(c\), and calculated the preliminary cluster redshift and its velocity dispersion. Using a conservative
clipping based on the preliminary velocity dispersion,
\(|z_{pre}-z_{mem}| \cdot c \le 3\sigma_{pre} \cdot (1+z_{pre})\), we rejected further galaxies from
the sample in order to derive updated bi-weight estimators and repeated this step when necessary.
Errors were calculated from \(10^{5}\) bootstrap simulations and the \(90\%\) confidence interval of
jackknife rebinning, for \(z_{cl}\) and \(\sigma\), respectively, as recommended in \citet{Beers1990}.

For clusters with very few member galaxies, we were unable to calculate a statistically meaningful
velocity dispersion. We therefore show the number of galaxies (\(n\)) taken into account for the
calculation in Table~\ref{tab:resulttab}. All \(\sigma\) values were scaled to the cluster redshift
(factorized with \((1+z_{cl})^{-1}\), as stated in \citet{Peebles1993}).

\citet{Munari2013} examined the relation between the velocity dispersion of different physical tracers in simulated
galaxy clusters to the mass of their respective dark matter halos, where they used 
Navarro-Frenk-White \citep[NFW; ][]{Navarro1997} profile fits.
We averaged the best-fitting parameters they found for the two corresponding simulations \citep[ceased star formation and AGN
feedback; Table~1 in][]{Munari2013} and galaxies as tracers. Accordingly, the parameters \(A_{\rm1D}=1169.75\pm11.45\) and
\(\alpha=0.3593\pm0.0068\) were used to calculate \(M_{200}\), the mass within which the mean matter density is 200 times
\(\rho_{c}(z)\), using
\begin{equation}
 \label{eq:munari}
 \frac{\sigma_{\rm 1D}}{\textnormal{km~s}^{-1}}=A_{\rm 1D}\left[\frac{h(z)~M_{200}}{10^{15}M_{\odot}}\right]^{\alpha}~.
\end{equation}
In Eq.~\ref{eq:munari}, \(\sigma_{\rm 1D}\) denotes the line-of-sight velocity dispersion, and \(h(z)\) is the Hubble parameter
at the cluster redshift \(z\) normalized by 100~km~s\(^{-1}\)~Mpc\(^{-1}\).
The uncertainties for \(M_{200}\) are based on the intrinsic error of the fitting parameters and our error of
\(\sigma\) for each galaxy cluster.

To computate \(r_{200}\), the radius within which the mean density of the halo is 200 times \(\rho_{c}(z)\),
we used
\begin{equation}
  r_{200} = \sqrt[3]{\frac{3\pi}{4}\frac{M_{200}}{200\rho_{c}(z)}} ~ ,
\end{equation}
assuming a mean density and spherical symmetry.

These equations are valid when we assume that the clusters are in hydrostatic equilibrium and are spherically symmetric. Such
estimates are therefore only considered as proxies. The main results from the spectroscopy on clusters are summarized in
Table~\ref{tab:resulttab}, while the individual galaxy redshifts are given in Appendix~A.

\subsection{X-ray fluxes and temperatures}
\label{sec:Xray_analysis}
X-ray spectroscopy was carried out for all clusters with redshift
known either from the literature or from this work.
Source extraction regions were centred on the positions of the detected extended source in the 3XMM-DR6 catalogue
(see Table~\ref{tab:survey}) and set to the common radius of \(300\rm kpc\) for the respective cluster redshift.
Using the [0.5--3.0~keV] EPIC images, we searched for suitable background regions at approximately the same off-axis
distances as the source, with a strong preference for the same chip whenever possible, and a radius sufficiently large in
order to increase the quality and statistics of the background. Obvious point sources contaminating the background region
were excluded. Taking the source radius into account, the background and source spectra for the individual instruments were
extracted and response matrix files were created using the SAS task {\tt especget}.

The photons of all spectra were binned to at least one count per bin with the FTOOL task {\tt{grppha}}. We used the XSPEC
version 12.9.0m \citep{Arnaud1996}, the abundance table of \citet{Wilms2000}, and the atomic cross-sections of \citet{Verner1996}.
In order to derive the ICM temperature of the selected clusters, we fitted the extracted EPIC spectra using the {\tt APEC} model
\citep{Smith2001}, while we accounted for the Galactic absorption with the {\tt tbabs} model
\citep[T\"ubingen-Boulder ISM absorption model;][]{Wilms2000}, taking the column density of neutral hydrogen as a parameter
(see below).
For each fit we fixed the following parameters in order to reduce the number of free
variables: (i) the metallicity, (ii) the column density of Galactic neutral hydrogen, and (iii) the redshift of
the cluster. The value for (i) was fixed to \(Z=0.23\pm 0.01 ~Z_{\odot}\) for the whole sample of X-ray data. This
value was computed by \citet{McDonald2016} in their analysis of 153 mass-selected galaxy clusters between
\(0 < z < 1.5\).
We used the nH tool provided by HEASARC\footnote{https://heasarc.gsfc.nasa.gov/cgi-bin/Tools/w3nh/w3nh.pl}, which determines
the weighted average column density of neutral Galactic hydrogen (ii) based on data by the Leiden/Argentine/Bonn (LAB) survey
\citep{Kalberla2005,Hartmann1997}.
For (iii) we queried the literature for existing redshifts of our cluster candidates in addition to the clusters with
spectroscopic redshifts based on data of this work (see Table~\ref{tab:survey}). The best-fitting parameters, cluster temperature,
the normalization, and their respective 1\(\sigma\) errors were determined by minimizing the C-statistics.
The spectral fit enabled calculating the X-ray flux and luminosity in the energy range [0.5--2.0 keV]. The luminosity
was further converted into a bolometric luminosity ([0.1 -- 50 keV]) using a dummy response matrix.
The derived fluxes, \(F_{0.5-2}(300\rm kpc)\), and (bolometric) luminosities, \(L_{0.5-2}(300\rm kpc)\) and \(L_{bol}(300\rm kpc)\),
take into account photons emitted within the extraction region of 300 kpc.
In order to derive \(L_{bol}\) for the more physically motivated radius \(r_{500}\), we used the iterative procedure based on
\(L-T\) and \(L-M\) relations published by \citet{Pratt2009}. The whole procedure is outlined in detail in \citet{Takey2011},
but can be summarized to work as follows. It calculates a \(M_{500}\) from the input bolometric luminosity, used to derive a first
guess of \(r_{500}\) based on the \(L-M\) relation. The temperature from the \(L-T\) relation is then used to calculate the core
radius of the \(\beta\) model \citep[see relations in][]{Finoguenov2007}. The flux ratio between the input radius and the aperture
extrapolated from the \(\beta\) model is finally used to correct the bolometric luminosity for the next iteration. 

We present the X-ray properties of the ICM-temperature, \(F_{0.5-2}(300\rm kpc)\) and \(L_{bol}(r_{500}),\) and their 1\(\sigma\) confidence
interval in Tables~\ref{tab:resulttab} and~\ref{tab:results_known}.

\subsection{Results of individual clusters with LBT follow-up}
\label{sec:LBT_clusters}
\subsubsection{2XMMp~J083026.2+524133}
\label{sec:LBT_003}
A redshift based on the X-ray spectrum (\(z=0.99\)) for this source has been published by \citet{Lamer2008}.

\begin{figure*}
  \begin{subfigure}{0.5\textwidth}
    \includegraphics[width=8.5cm]{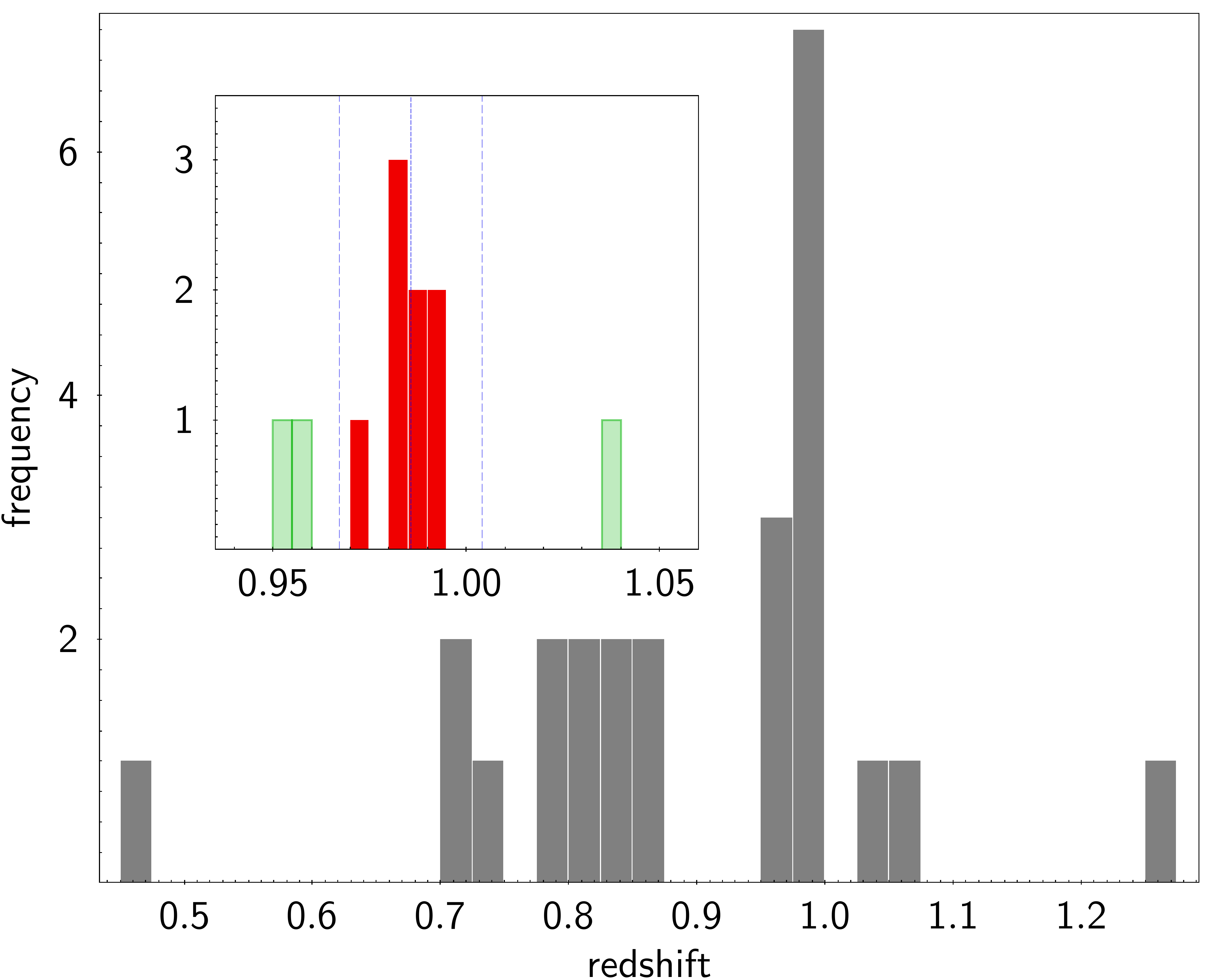}
    \subcaption{redshift histogram}
    \label{fig:LBT_003_histo}
  \end{subfigure}
  \begin{subfigure}{0.5\textwidth}
    \includegraphics[width=8.5cm]{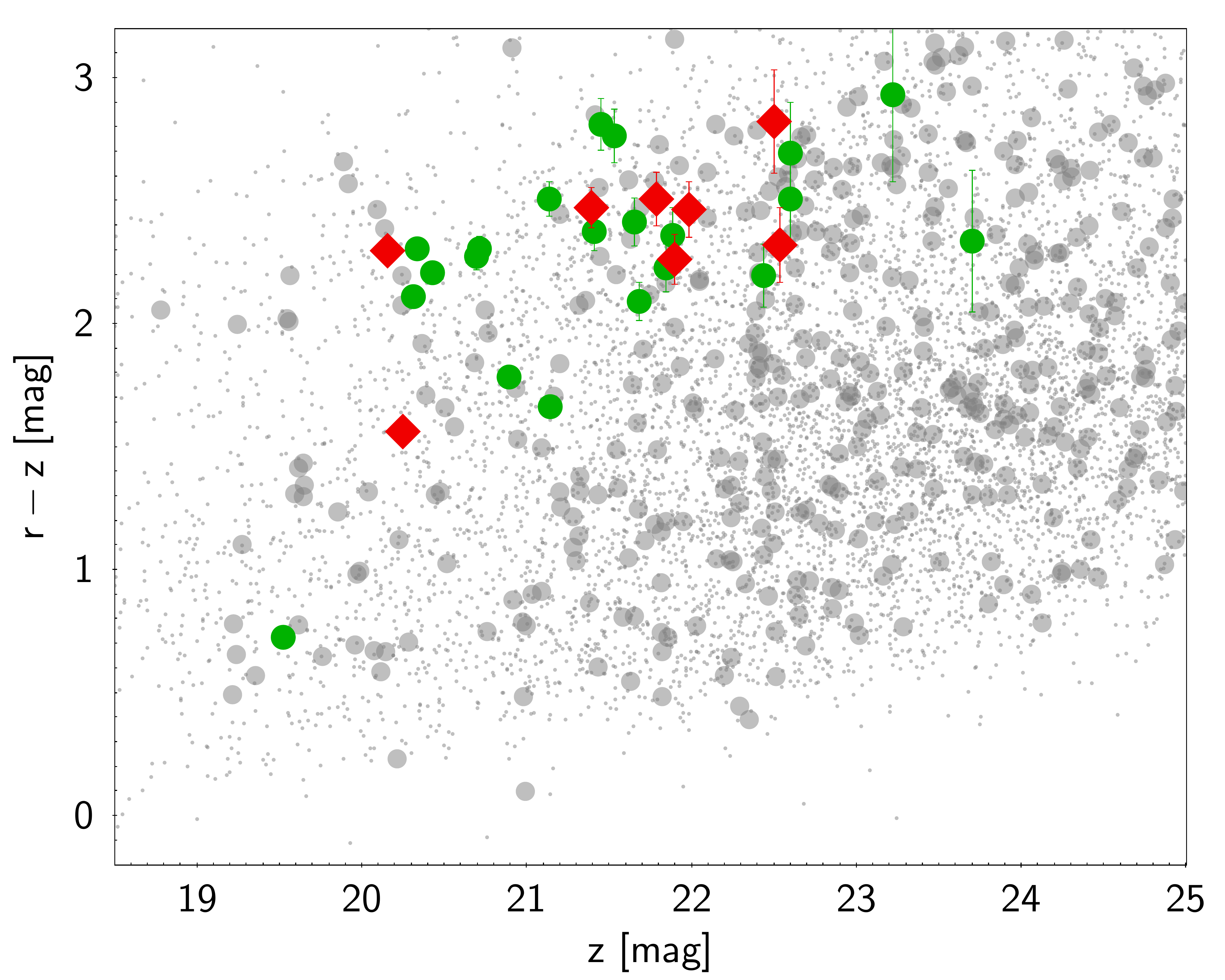}
    \subcaption{colour-magnitude diagram}
    \label{fig:LBT_003_colormag}
  \end{subfigure}
  
  \begin{subfigure}{0.5\textwidth}
    \includegraphics[width=8.5cm]{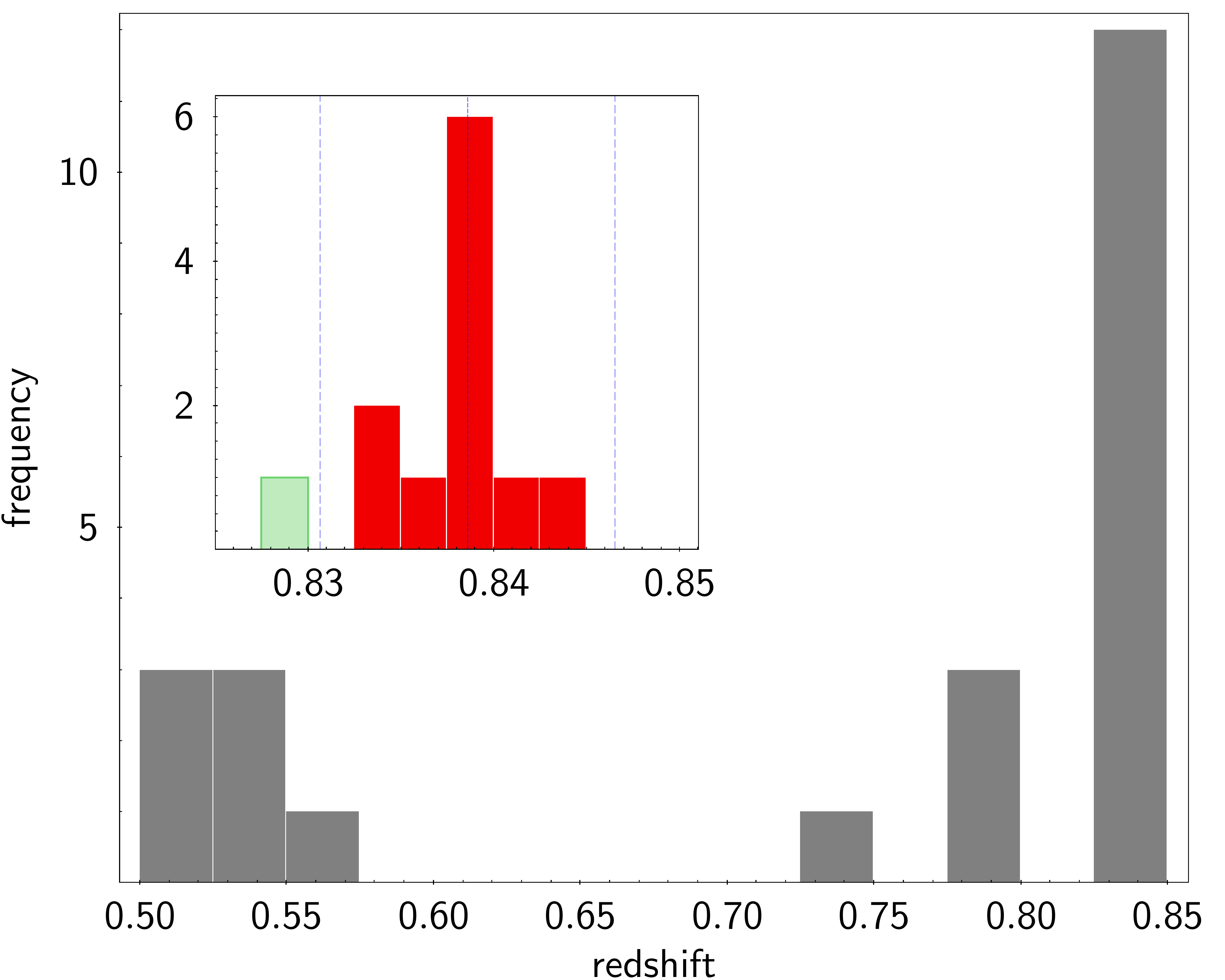}
    \subcaption{redshift histogram}
    \label{fig:LBT_007_histo}
  \end{subfigure}
  \begin{subfigure}{0.5\textwidth}
    \includegraphics[width=8.5cm]{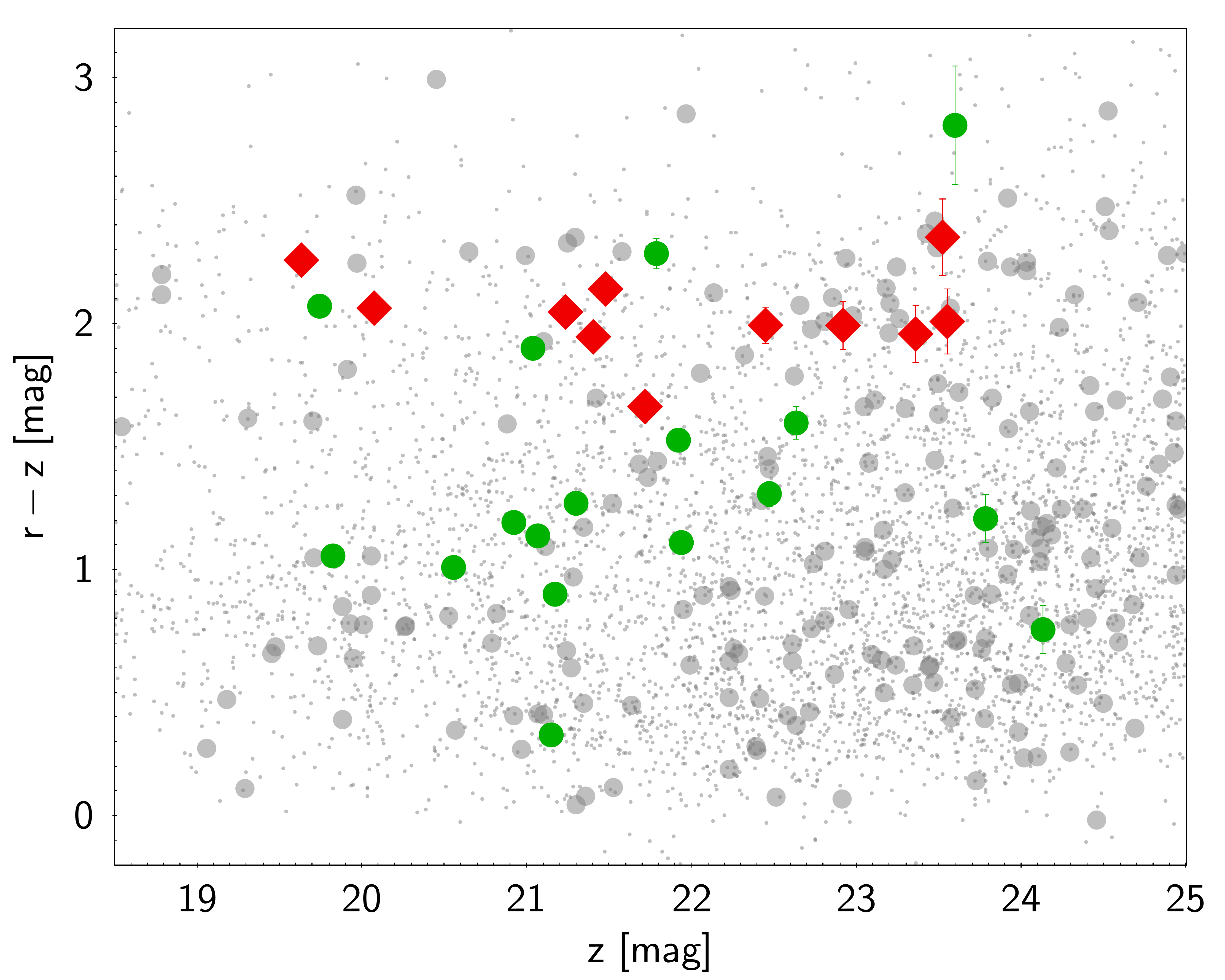}
    \subcaption{colour-magnitude diagram}
    \label{fig:LBT_007_colormag}
  \end{subfigure}
  
  \begin{subfigure}{0.5\textwidth}
    \includegraphics[width=8.5cm]{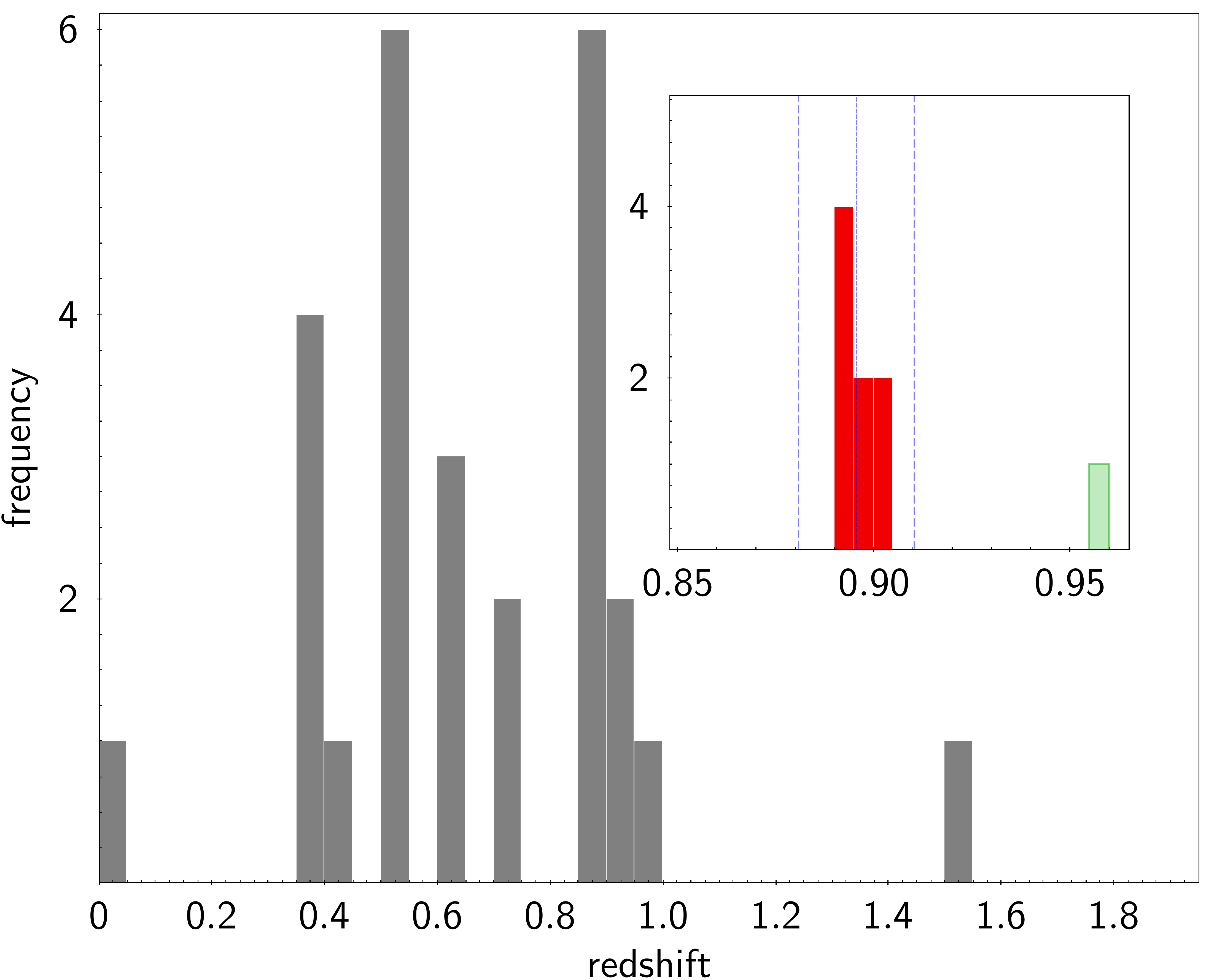}
    \subcaption{redshift histogram}
    \label{fig:LBT_010_histo}
  \end{subfigure}
  \begin{subfigure}{0.5\textwidth}
    \includegraphics[width=8.5cm]{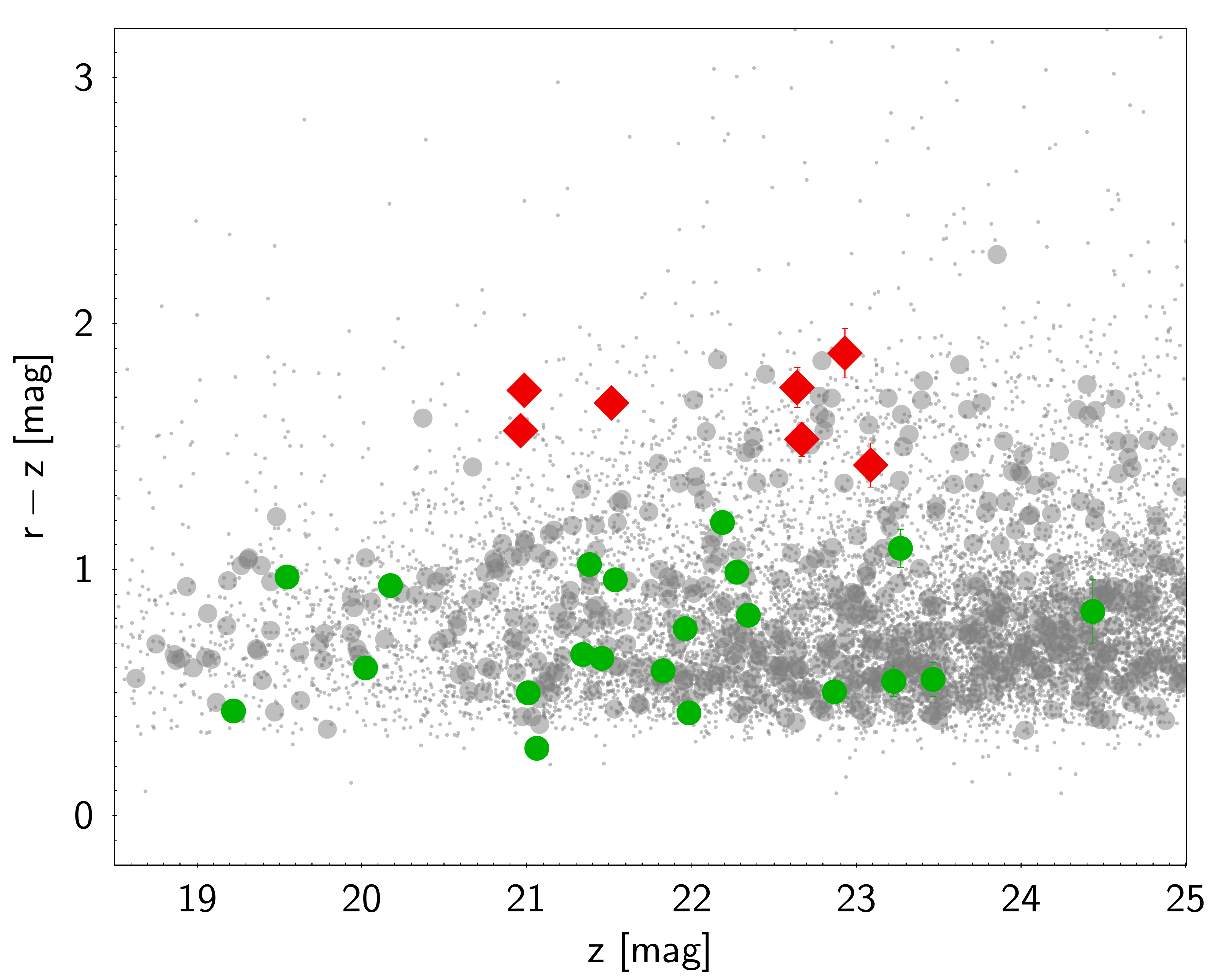}
    \subcaption{colour-magnitude diagram}
    \label{fig:LBT_010_colormag}
  \end{subfigure}
  \caption{\label{fig:LBT_colormag_I}
           Redshift histograms (left side) and colour-magnitude diagrams (right side) of 2XMMp~J083026.2+524133 (top), 2XMMp~J093437.4+551340
           (middle), and 2XMMp~J105319.8+440817 (bottom). Objects with spectra are plotted in green, while the confirmed cluster members are shown in red.
           Error bars plotted in the colour-magnitude diagrams indicate the photometric uncertainties and are visible when they exceed
           the symbol size.}
\end{figure*}

In the follow-up imaging of the field, many faint galaxies of similar colour to the apparent BCG are visible
(compare Fig.~\ref{fig:LBT_003_RGB}), the colour-magnitude diagram in Fig.~\ref{fig:LBT_003_colormag} indicates no
dominant red sequence, and the spectroscopic follow-up revealed only eight galaxies matching our criteria for cluster members
(Fig.~\ref{fig:LBT_003_histo}). Nevertheless, we were able to determine a mean cluster redshift of \(z=0.9856\), thus
confirming the initial finding from \citet{Lamer2008}, and a velocity dispersion (see Table~\ref{tab:resulttab}). The
dynamical mass of 2XMMp~J083026.2+524133 within the radius of \(r_{200}=\left(2.44_{-0.6}^{+0.79}\right)\textnormal{Mpc}\)
was calculated to be \(M_{200}=5.1_{-2.9}^{+6.7}\times10^{14}M_{\odot}\).

The source spectrum was extracted from \emph{XMM-Newton} observations with a total exposure time of 216.7, 223.4, and 170.3
ksec for MOS1, MOS2, and PN, respectively. Owing to the high luminosity of the source
(\(L_{0.5-2}(300\rm kpc)=(189.6\pm2.8)\times10^{42}~\rm erg~s^{-1}\)), we were able to
constrain the ICM temperature with good accuracy to \(k_{B}T=7.82^{+0.4}_{-0.39}~\rm{keV}\). 
The temperature is in good agreement with the result from \citet{Lamer2008}, which was based on fewer observations.

The high cluster mass from the dynamical analysis and the large virial radius agree within their errors with X-ray estimates
from \citet{Lamer2008} considering a typical scaling factor of 1.52 between \(r_{500}\) and \(r_{200}\) \citep{Piffaretti2011}.

We also note an independent cluster mass proxy from \citet{Culverhouse2010}, who reported the gas mass derived from SZ observations.
Their analysis of \emph{XMM-Newton} data results in a cluster gas mass within \(r_{2500}\) that agrees excellently with their SZ
proxy. Furthermore, \citet{Culverhouse2010} reported a cluster temperature of \(7.6\pm0.8~\rm keV\), which exactly matches our results.
Another and more recent approach using SZ measurements \citep{Schammel2013} derived a much lower cluster mass,
\(M_{200}=3.6\times10^{14}M_{\odot}~/~4.7\times10^{14}M_{\odot}\) (computed based on different models), compared to our dynamical
calculation.

\subsubsection{2XMMp~J093437.4+551340}
\label{sec:LBT_007}
The two-band photometry yields a tight red sequence within the field of view (compare Fig.~\ref{fig:LBT_007_colormag}).
While the optical counterpart to the BCG is hardly visible in the SDSS, AllWISE W1-imaging shows clear detections
of the brightest cluster galaxies that coincide with the extended X-ray contours,  see Figures
\ref{fig:LBT_007_SDSS}, \ref{fig:LBT_007_WISE}, and \ref{fig:LBT_007_LBC}.

Optical spectroscopy resulted in 23 galaxy redshifts for the observed field.
The mean cluster redshift and its velocity dispersion were calculated using an iterative clipping of redshifts from
the sample (see beginning of Sect.~\ref{sec:results}) from  a final census of 11 member galaxies to
\(z_{cl}=0.83858 \pm 0.00073\) and \(\sigma=\left(430_{-96}^{+120}\right)\textnormal{km~s}^{-1}\).
In the redshift histogram (Fig.~\ref{fig:LBT_007_histo}), the confirmed member galaxies of galaxy cluster
2XMMp~J093437.4+551340 are visible as a compact overdensity.
Based on 11 cluster galaxies, a dynamical cluster mass and radius for an overdensity of \(\Delta=200\) were computed.
For these values and for the X-ray parameters of this cluster, we refer to Table~\ref{tab:resulttab}.

The publicly available X-ray data with \(\sim 10\)ksec exposure time were sufficient to determine the ICM temperature within
reasonable accuracy, \(k_{B}T=2.96^{+0.86}_{-0.65}~\textnormal{keV}\), taking the spectroscopic redshift of the
cluster into account. X-ray flux and luminosity were calculated to be \(F_{0.5-2}(300\rm kpc)=(7.4\pm0.83)\times10^{-14}~\rm erg~cm^{-2}~s^{-1}\) and
\(L_{0.5-2}(300\rm kpc)=(20.9\pm2.3)\times10^{42}~\rm erg~s^{-1}\), respectively (see also Table~\ref{tab:resulttab}).

\begin{figure*}
  \begin{subfigure}{0.5\textwidth}
    \includegraphics[width=8.5cm]{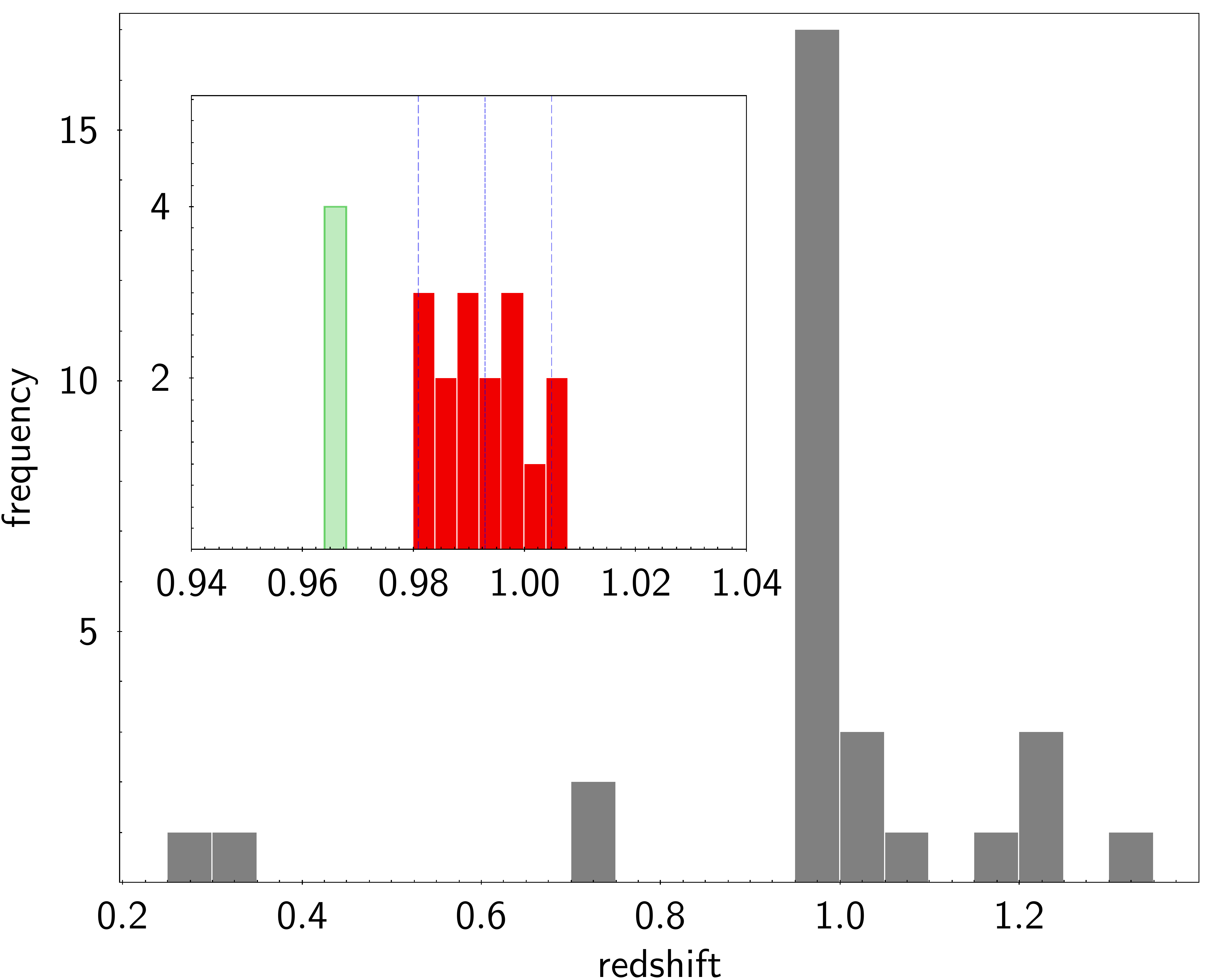}
    \subcaption{redshift histogram}
    \label{fig:LBT_013_histo}
  \end{subfigure}
  \begin{subfigure}{0.5\textwidth}
    \includegraphics[width=8.5cm]{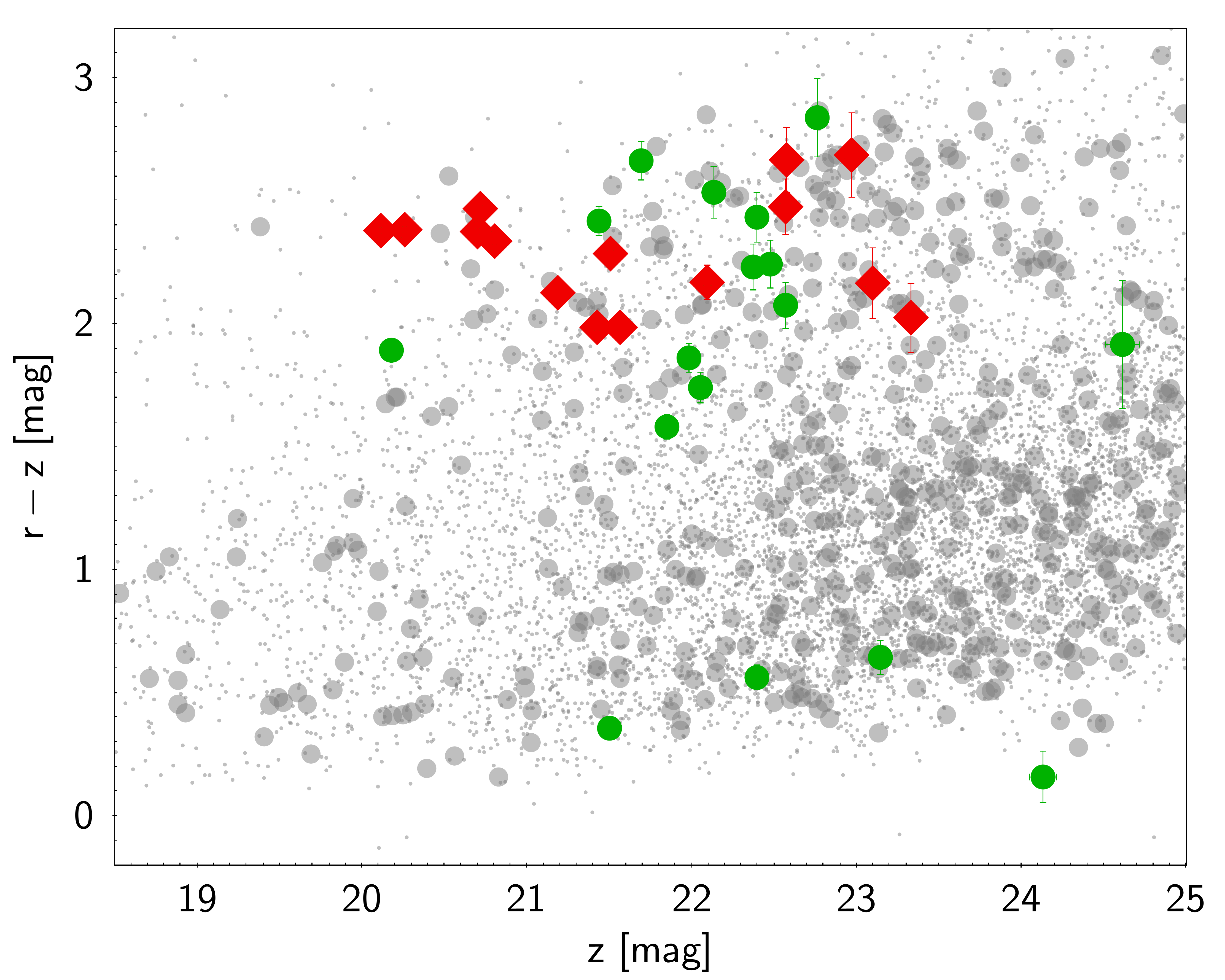}
    \subcaption{colour-magnitude diagram}
    \label{fig:LBT_013_colormag}
  \end{subfigure}
  
  \begin{subfigure}{0.5\textwidth}
    \includegraphics[width=8.5cm]{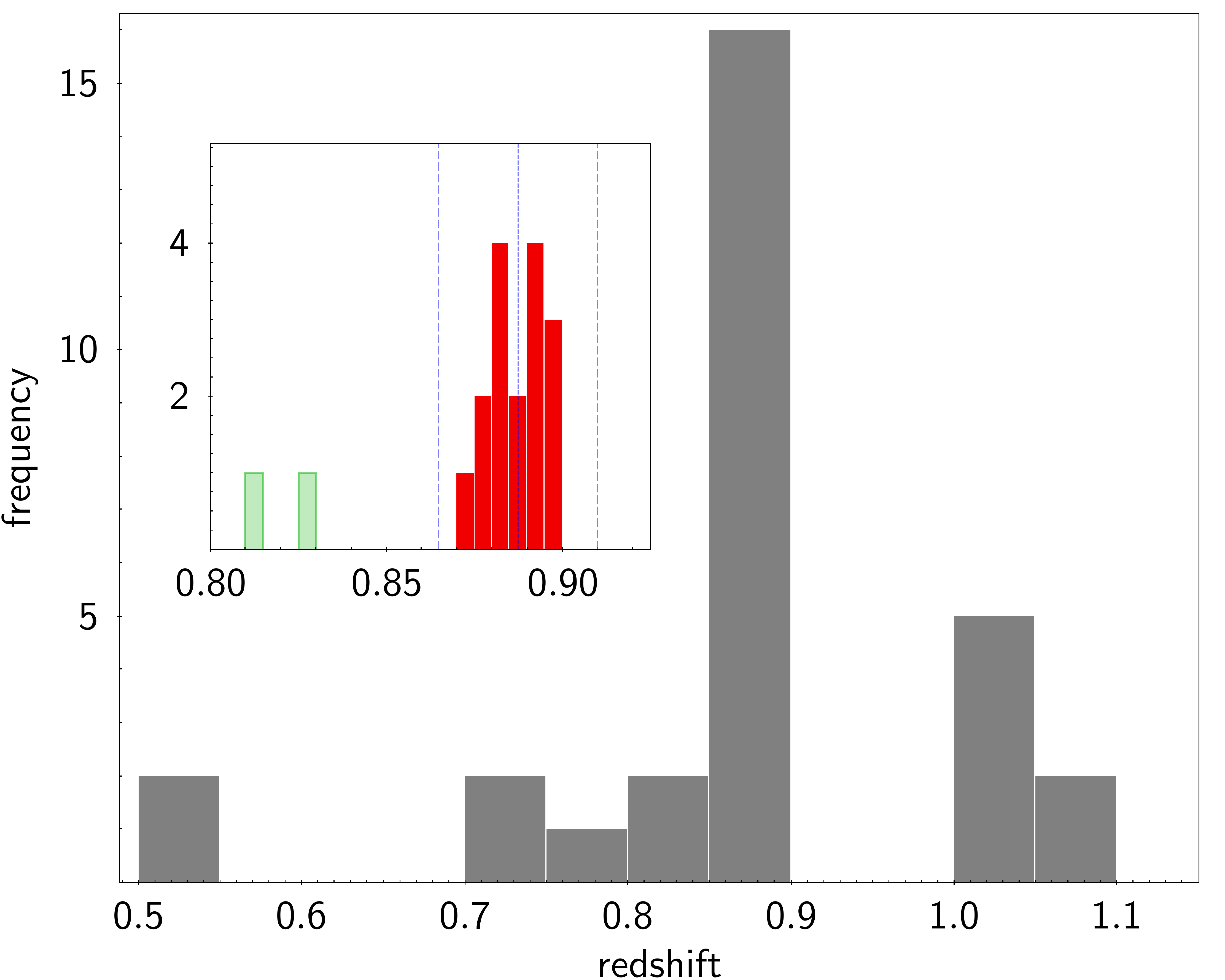}
    \subcaption{redshift histogram}
    \label{fig:LBT_015_histo}
  \end{subfigure}
  \begin{subfigure}{0.5\textwidth}
    \includegraphics[width=8.5cm]{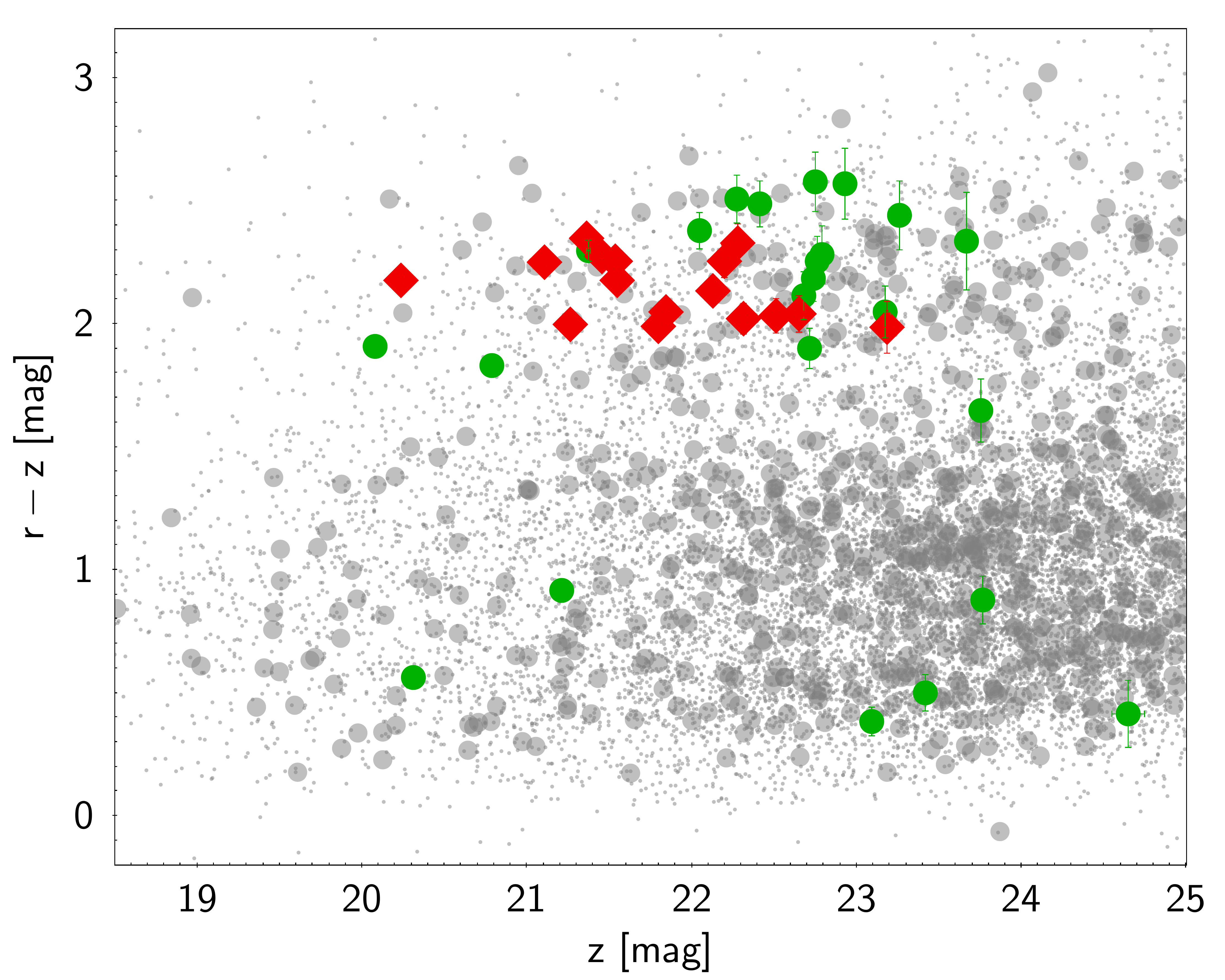}
    \subcaption{colour-magnitude diagram}
    \label{fig:LBT_015_colormag}
  \end{subfigure}
  
  \begin{subfigure}{0.5\textwidth}
    \includegraphics[width=8.5cm]{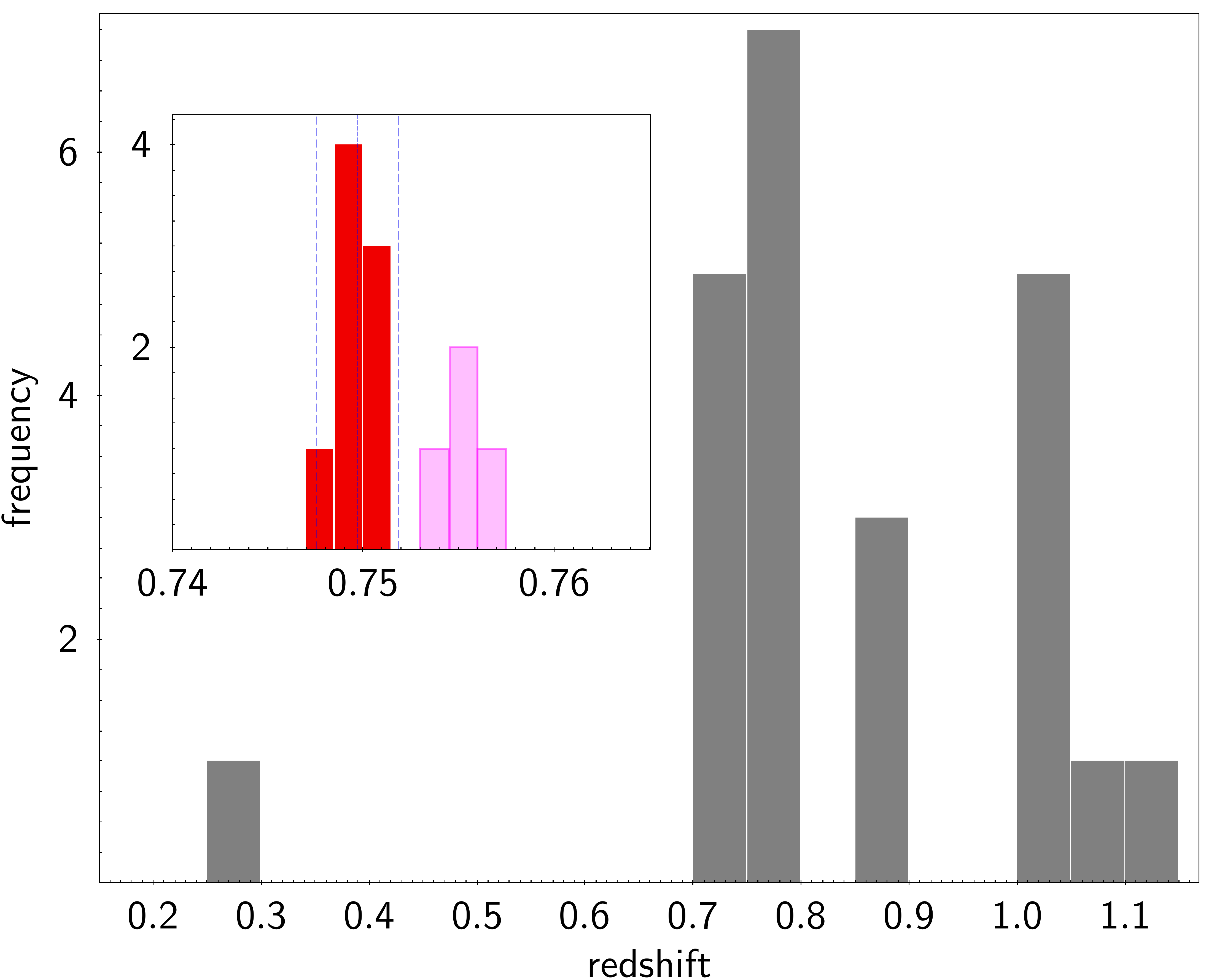}
    \subcaption{redshift histogram}
    \label{fig:LBT_018_histo}
  \end{subfigure}
  \begin{subfigure}{0.5\textwidth}
    \includegraphics[width=8.5cm]{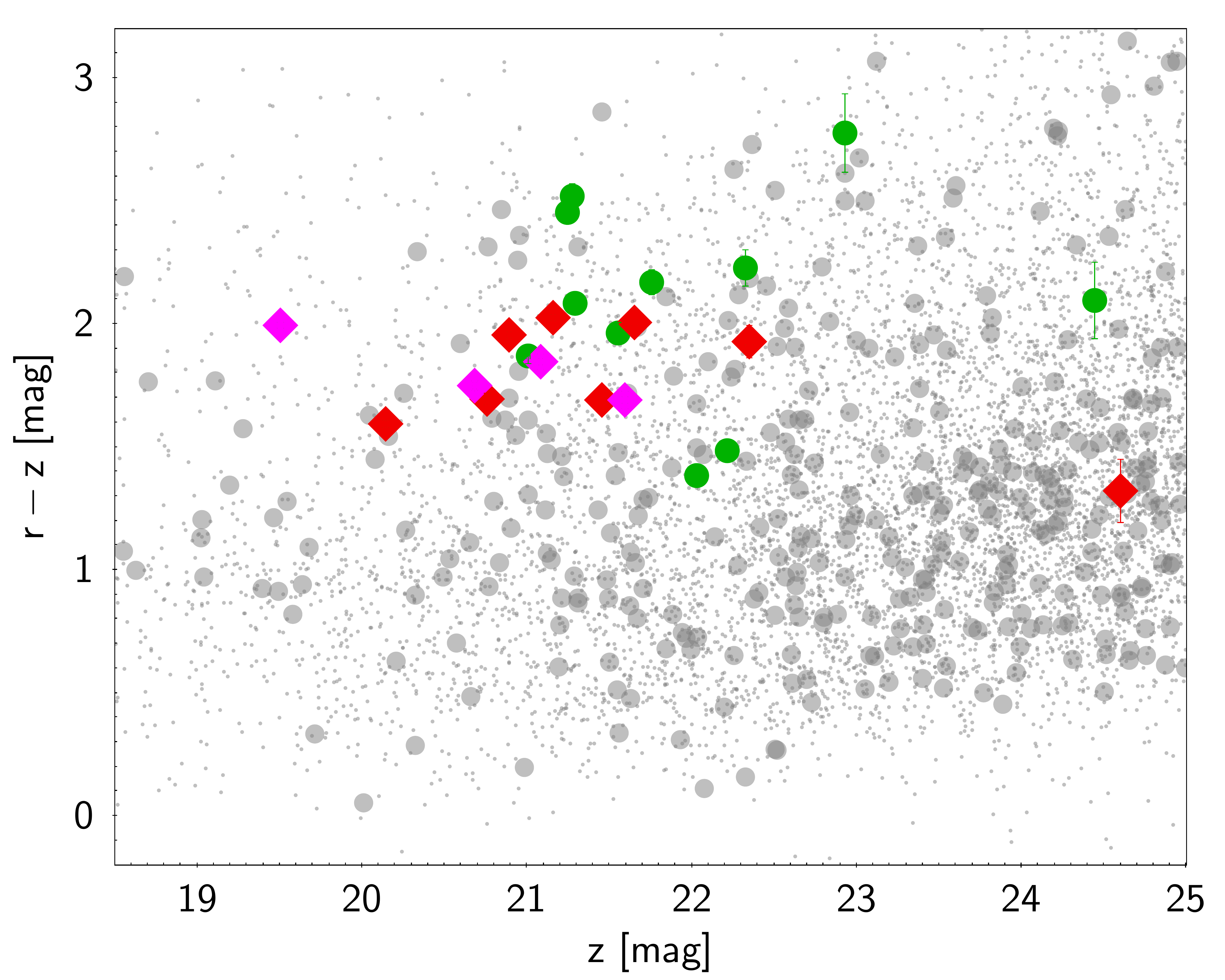}
    \subcaption{colour-magnitude diagram}
    \label{fig:LBT_018_colormag}
  \end{subfigure}
  \caption{\label{fig:LBT_colormag_II}
           Redshift histograms (left side) and colour-magnitude diagrams (right side) of 2XMMp~J120815.5+250001 (top), 2XMMp~J123759.3+180332
           (middle), and 2XMMp~J133853.9+482033 (bottom). Objects with spectra are plotted in green, while the confirmed cluster members are shown in red.
           Error bars plotted in the colour-magnitude diagrams indicate the photometric uncertainties and are visible when they exceed
           the symbol size. Galaxies excluded by the iterative sigma
           clipping method are shown in purple in panels (e) and (f).}
\end{figure*}

\subsubsection{2XMMp~J105319.8+440817}
\label{sec:LBT_010}
When we analysed the colour-magnitude diagram (Fig.~\ref{fig:LBT_010_colormag}) from the optical pre-imaging, a tight red
sequence of galaxies became visible. Using the spectroscopic follow-up, we confirm eight galaxies out of the 28 spectra
of this mask. We note that the BCG is relatively faint (20.96 mag in z-SLOAN) in comparison to the other clusters in our
sample. The BCG and cluster members are not detected by the SDSS (Fig.~\ref{fig:LBT_010_SDSS}), but have IR
counter-parts in the AllWISE survey (Fig.~\ref{fig:LBT_010_WISE}).

Based on the eight identified galaxies within the redshift range of \(0.890 \lesssim z \lesssim 0.903,\) we calculated a
low-precision velocity dispersion of \(\sigma=\left(780_{-360}^{+660}\right)\textnormal{km~s}^{-1}\) (see
Tab.~\ref{tab:resulttab}). We also note that a close galaxy pair (the double-member east of the X-ray centre;
Fig.~\ref{fig:LBT_010_LBC}) might bias the \(\sigma\) by possible interaction. Spectra of both galaxies were taken
from the same slit on the MOS-mask and appear only \(\sim1.8\arcsec\) and \(\sim 2950\textnormal{km~s}^{-1}\) apart in
projection and recessional velocity, respectively.

As a peculiarity within the cluster field, imaging revealed a slightly distorted blue object only \(\sim5\arcsec\)
away from the BCG. We identified this object as a background galaxy and were able to assign a redshift of \(z=3.8354\).
The object is not bent to an Einstein ring, nor were any possible multiple images to this galaxy identified. Based on its
magnitude (22.3~mag in z) with respect to its high redshift and the apparent magnitudes of the cluster members at
\(z=0.8955\) (compare Table~A.1), the galaxy is most likely lensed by the BCG of the galaxy cluster.

In the XSA we found two observations covering the source 2XMMp~J105319.8+440817 with cleaned exposure times
between 4 and 11 ksec (Table~D.1). Using our common procedure, we extracted an X-ray flux of
\(F_{0.5-2}(300\rm kpc)=(2.37\pm0.27)\times10^{-14}~\rm erg~cm^{-2}~s^{-1}\) and a model luminosity of
\(L_{0.5-2}(300\rm kpc)=(76.7\pm8.7)\times10^{42}~\rm erg~s^{-1}\). The best-fitting value for the gas temperature of the cluster
was calculated to be \(k_{B}T=3.6^{+1.2}_{-0.76}~\textnormal{keV}\) (Table~\ref{tab:resulttab}).

\subsubsection{2XMMp~J120815.5+250001}
\label{sec:LBT_013}  
In the LBC imaging data, very many galaxies around the centre of 2XMMp~J120815.5+250001
are visible (compare Fig.\ref{fig:LBT_013_RGB}). The colour-magnitude diagram in Fig.~\ref{fig:LBT_013_colormag}
appears to have a rich red sequence, which is not very tight in colour.
Within the MODS field of view of 2XMMp~J120815.5+250001, we spectroscopically identified 20 galaxies
in the redshift range of \(0.96 < z < 1.06\).
A cluster consisting of those 20 galaxies would exhibit line-of-sight velocities
(\(|v_{los}|>6000~\textnormal{km~s}^{-1}\)) that are too high
for this to be considered a gravitationally bound system.
We detect several bright galaxies with greatly different recessional velocities near the centre of the
elongated X-ray emission (Fig.~\ref{fig:LBT_013_LBC}), imaged as the brightest four galaxies near the
inner X-ray contour. When we consider each as a possible BCG, we obtain different samples of cluster members and velocity
dispersions, regarding our member selection algorithm.
All four bright galaxies appear related to red sources (W1-W2 between 0.22 and 0.35) in the AllWISE data,
visible as bright sources in Fig.~\ref{fig:LBT_013_WISE}, and thus give no clear indication of a single BCG.

In the following, we consider the galaxy at the position 12:08:15 +25:00:01 with a z-SLOAN magnitude of 20.262~mag
and a distance of \(~5\arcsec\) with respect to the X-ray centre, as the proper BCG. Our algorithm estimates a
cluster redshift of \(z=0.9929\pm0.0026\) and \(\sigma=\left(1200_{-240}^{+300}\right)\textnormal{km~s}^{-1}\),
taking the 16 galaxies into account as cluster members. The respective radius and mass we determine are listed in
Table~\ref{tab:resulttab}.

We extracted source spectra for all EPIC instruments using the one \emph{XMM-Newton} observation available for this
cluster. The {\tt APEC} model fit constrained the ICM temperature for 2XMMp~J120815.5+250001 to be \(kT=4.1^{+2.2}_{1.1}~\rm keV\).
X-ray flux and luminosity were calculated to be \(F_{0.5-2}(300\rm kpc)=(2.79\pm0.42)\times10^{-14}~\rm erg~cm^{-2}~s^{-1}\)
and \(L_{0.5-2}(300\rm kpc)=(106\pm16)\times10^{42}~\rm erg~s^{-1}\), respectively.

The velocity dispersion of this cluster may be contaminated by galaxies that are not bound to the cluster,
but are part of a larger structure. This is further supported by the relatively low ICM temperature found from
the analysis of EPIC spectra compared to the large \(\sigma\) from optical spectroscopy.

\subsubsection{2XMMp~J123759.3+180332}
\label{sec:LBT_015}  
The source 2XMMp~J123759.3+180332 appears very rich in red galaxies in the optical pre-imaging (Fig.~\ref{fig:LBT_015_RGB}),
with an apparent concentration in the central part of the X-ray emission (Fig.~\ref{fig:LBT_015_LBC}), and bright
and possibly blended IR emission in the AllWISE imaging data (Fig.~\ref{fig:LBT_015_WISE}). The optical richness
is also recovered in the corresponding colour-magnitude plot (Fig.~\ref{fig:LBT_015_colormag}) as a clearly visible
red sequence, of which a considerable fraction of objects were confirmed as cluster members.
Applying our clipping algorithm, we were able to identify 16 members within the redshift range of
\(0.865 \lesssim z \lesssim 0.9\), and consequently fixed the cluster redshift to \(z=0.8874\pm0.0024\) with
\(\sigma=\left(1200_{-250}^{+310}\right)\textnormal{km~s}^{-1}\). The dynamical cluster mass computed based on
\(\sigma\) results in a high value and confidence interval. A high cluster mass, however, slightly
contradicts the average X-ray temperature (compare Table~\ref{tab:resulttab}). The population of cluster member
galaxies might contain an unidentified fraction of interloping galaxies, which we cannot identify for statistical reasons.

The LBC data imaged two blue objects in the cluster field (highlighted as blue circles in
Fig.~\ref{fig:LBT_015_RGB}) with an apparently bent morphology. The respective spectra of both sources show no clear
emission or absorption feature, rendering no opportunity to determine redshifts. A clear similarity between both
spectral shapes is obvious, on the other hand, substantiating the hypothesis that both objects are lensed images of
a source with a redshift in the redshift desert between \(1.4 < z < 2.0\).

X-ray source photons of 2XMMp~J123759.3+180332 were extracted within a radius of \(\sim0.95~\rm Mpc\).        
The low number of photons from the source (63/102/73 counts for MOS1/MOS2/PN) resulted in large error margins for
the ICM temperature: \(k_{B}T=5.0^{+1.9}_{-1.1}~\textnormal{keV}\). The extracted flux was derived to be
\(F_{0.5-2}(300\rm kpc)=(4.75\pm0.47)\times10^{-14}~\rm erg~cm^{-2}~s^{-1}\), and using the cluster redshift from
optical spectroscopy, we calculated the X-ray luminosity to be \(L_{0.5-2}(300\rm kpc)=(134\pm13)\times10^{42}~\rm erg~s^{-1}\).

\subsubsection{2XMMp~J133853.9+482033}
\label{sec:LBT_018}
The LBT imaging shows the field of 2XMMp~J133853.9+482033 as a rich cluster of galaxies. However, the red sequence
(Fig.~\ref{fig:LBT_018_colormag}) is not very pronounced.
We identified the brightest galaxy near the X-ray centre as BCG (magenta square in Fig.~\ref{fig:LBT_018_LBC}), and
the initial sample of tentative cluster galaxies comprises 13 galaxies within \(0.749 \lesssim z \lesssim 0.756\).
Our iterative member clipping, however, divides the sample into two groups, indicated by red and magenta bars in the
redshift histogram in Fig.~\ref{fig:LBT_018_histo}. The method finally excludes the inital BCG and surrounding
group of galaxies (magenta) from  the sample of cluster member galaxies (red) during the iteration. We therefore note
that the following estimates are most possibly biased by insufficient member statistics. With to the
procedure we applied for the whole spectroscopic sample in this paper, we calculate a mean cluster redshift based on eight
galaxies of \(z=0.74969\pm0.00030\) and a velocity dispersion of \(\sigma=\left(122^{+32}_{-26}\right)\rm km~s^{-1}\).
The low-velocity dispersion shifts 2XMMp~J133853.9+482033 more into the regime of galaxy groups than clusters,
and results in a dynamical mass estimate of only \(M_{200}=1.3^{+1.4}_{-0.6}\times10^{12}M_{\odot}\).
We summarize all properties in Table~\ref{tab:resulttab}.

We were unable to constrain the gas temperature of 2XMMp~J133853.9+482033 based on the publicly available X-ray data.
The X-ray flux and luminosity were measured to be \(F_{0.5-2}(300\rm kpc)=(1.1\pm0.16)\times10^{-14}~\rm erg~cm^{-2}~s^{-1}\)
and \(L_{0.5-2}(300\rm kpc)=(18.8\pm2.8)\times10^{42}~\rm erg~s^{-1}\).

\subsection{Results: galaxy clusters with redshifts from the literature}
\label{sec:results_known}
This section summarizes results from the subsample of X-ray selected cluster candidates, which are known to be clusters in the literature. We queried the NED for publications on galaxy clusters and
took advantage of their spectroscopic cluster redshifts for our {\tt APEC} fit of the analysis of \emph{XMM-Newton}
data. Our goal was to calculate X-ray cluster properties such
as ICM temperature, flux, and luminosity 
for this sample as well and compare our \emph{XMM-Newton} based properties to the literature values.
We refer to Section~\ref{sec:conclusions}, where we show an \(L-T\) plot of all clusters within this sample.
In Table~D.1 we collect important data on the individual cluster detection, including
available OBSIDs, cleaned exposure times based on our reduction, and the 2XMMp vs. 3XMM-DR6 cross-reference.
The procedure of our X-ray data reduction and analysis is described in Sections~\ref{sec:Xray_reduction} and~\ref{sec:Xray_analysis}, respectively.

\begin{table*}\caption{X-ray properties for clusters with spectroscopic redshift from the literature.}
\label{tab:results_known}
\centering
\begin{tabular}{c c c c c c}
\hline
\hline
name & \(z_{lit}\) & \(F_{0.5-2}(300 \rm kpc)\) & \(L_{bol}(r_{500})\) & \(k_{\rm B}T\) & reference \\
 & & \([10^{-14}~\rm erg~cm^{-2}~s^{-1}]\) & \([10^{42}~\rm erg~s^{-1}]\) & [keV] & \\
\hline
2XMMp~J030212.0-000133 & 1.185 & \(0.884\pm0.097\)& \(162\pm17\) & \(4.7^{+1.5}_{-0.9}\) & \citet{Suhada2011} \\
2XMMp~J084836.4+445345 & 1.273 & \(0.406\pm0.059\) & \(75\pm12\) & \(2.5^{+1.4}_{-0.6}\) & \citet{Stanford1997} \\
2XMMp~J084858.3+445158 & 1.261 & \(0.885\pm0.053\) & \(192\pm12\) & \(4.93^{+0.77}_{-0.62}\) & \citet{Rosati1999} \\
2XMMp~J100451.6+411626 & 0.836 & \(6.65\pm0.17\) & \(664\pm13\) & \(4.84^{+0.31}_{-0.3}\) & \citet{Hoeft2008} \\
2XMMp~J105344.2+573517 & 1.134 & \(1.928\pm0.038\) & \(362.7\pm3.6\) & \(4.52^{+0.24}_{-0.22}\) & \citet{Hashimoto2005} \\
2XMMp~J123113.1+154550 & 0.893 & \(3.89\pm0.3\) & \(609\pm32\) & \(8.9^{+4.1}_{-2.2}\) & Rabitz et al. (in prep.) \\
\hline
\end{tabular}
\tablefoot{The X-ray properties flux (Col.~3) and temperature (Col.~5) were measured within radii of 0.3 Mpc at the redshift
           of the cluster, where we used the spectroscopic redshifts available from the reference noted in Col. 6.
           The bolometric luminosity (Col.~4) was extrapolated to \(r_{500}\) (for details, see Sect.~\ref{sec:Xray_analysis}).
           The X-ray flux is given in units of \(10^{-14}~\rm erg~cm^{-2}~s^{-1}\), and the luminosity in \(10^{42}~\rm erg~s^{-1})\).}
\end{table*}

\subsubsection{2XMMp~J030212.0-000133}
The cluster was discovered in X-rays in the framework of the \emph{XMM-Newton} Distant Cluster Project
\citep[XDCP;][]{Fassbender2011_XDCP}.
From six member galaxy spectra, \citet{Suhada2011} derived a cluster redshift of \(z=1.185\pm0.016\), which
we adopted for our X-ray analysis.

We extracted the X-ray spectrum within 300~kpc in the available \(\sim47\) and \(\sim37\) ksec exposures
of MOS1/MOS2 and PN, respectively. The X-ray flux and the luminosity were calculated to be
\(F_{0.5-2}(300\rm kpc)=(0.884\pm0.097)\times10^{-14}~\rm erg~cm^{-2}~s^{-1}\) and
\(L_{0.5-2}(300\rm kpc)=(35.2\pm3.9)\times10^{42}~\rm erg~s^{-1}\), and the cluster temperature was fitted to
\(k_{B}T=4.7^{+1.5}_{-0.9}~\textnormal{keV}\). Flux and luminosity agree within \(3\sigma\) with those
derived by \citet{Suhada2011}, which are based on an extraction radius of \(55\arcsec\).
The ICM temperature is higher in our case but is based on the spectral fits, while \citet{Suhada2011}
estimated their \(T_{500}\) according to scaling relations. Our X-ray properties are summarized in
Table~\ref{tab:results_known}.

\subsubsection{2XMMp~J084836.4+445345 and 2XMMp~J084858.3+445158}
We now describe together the properties of two single extended X-ray detections.

Numerous publications in addition to the initial discovery papers \citep{Stanford1997,Rosati1999} are available
for these two clusters, for which more recent work suggested the terminus supercluster because of their close neighbourhood
in on-sky projection and in redshift space \citep[compare ][]{Mei2012}. Since they were initially
found as extended sources in the RDCS, first X-ray fluxes and luminosities are based on ROSAT data. Observations with
\emph{Chandra} \citep[i.e.][]{Stanford2001} helped to reject point-sources from the data analysis because of their increased
spatial resolution, and this improved the quality of the parameters derived for the clusters. On the other hand, \emph{Chandra}
results show a large spread, at least partly caused by the different calibration of the low-energy QE degradation
as summarized and detailed by \citet{Jee2006}.

Our X-ray analysis is based on \emph{XMM-Newton} data and not directly comparable to previous results because of cross-calibration
issues between \emph{XMM-Newton} and \emph{Chandra}.
We used all observations included in Table~D.1, covering the fields of 2XMMp~J084836.4+445345 (Lynx-W) and
2XMMp~J084858.3+445158 (Lynx-E). For Lynx-W, we extracted source photons within 0.3~Mpc for the MOS1/MOS2
and PN instruments, leading to \(F_{0.5-2}(300\rm kpc)=(0.406\pm0.059)\times10^{-14}~\rm erg~cm^{-2}~s^{-1}\) and
\(L_{0.5-2}(300\rm kpc)=(30.8\pm4.4)\times10^{42}~\rm erg~s^{-1}\). The spectral model fits returned a best-fitting temperature of
\(k_{B}T=2.5^{+1.4}_{-0.6}~\textnormal{keV}\), where the large uncertainties reflect the low number of photons received from
this cluster in X-rays. \citet{Stanford2001} calculated flux and luminosity for this cluster from a smaller extraction radius,
and thus their results are not directly comparable to ours. However, our cluster temperature is comparable to their work
(\(k_{B}T=1.6^{+0.8}_{-0.6}~\textnormal{keV}\)), as well as to an analysis of \emph{Chandra} data by \citet{Jee2006}
(\(k_{B}T=1.7^{+0.7}_{-0.4}~\textnormal{keV}\)).

The \emph{XMM-Newton} spectral analysis for the cluster Lynx-E resulted in fluxes and luminosities of
\(F_{0.5-2}=(0.885\pm0.053)\times10^{-14}~\rm erg~cm^{-2}~s^{-1}\) and \(L_{0.5-2}=(50.2\pm3)\times10^{42}~\rm erg~s^{-1}\).
The flux is in agreement with measurements from \citet{Stanford2001}, where their aperture flux was extrapolated.
Our best-fitting parameter for the ICM temperature is \(k_{B}T=4.93^{+0.77}_{-0.62}~\rm keV,\) which is also comparable with the
results by \citet{Stanford2001} (\(k_{B}T=5.8^{+2.8}_{-1.7}~\rm keV\)); a comparison with additional results based on
\emph{Chandra} data is summarized in \citet{Jee2006}. 

\subsubsection{2XMMp~J100451.6+411626}
\citet{Hoeft2008} found the extended source 2XMMp~J100451.6+411626 in the field of a lensed quasar
(SDSS J1004+4112), and identified the cluster in deep SUBARU-imaging of the field.
They noted clear visibility of extended X-ray emission in all EPIC instruments and a faint optical
counter-part in the centre of the emission in the SDSS \emph{i}-band, indicating a distant galaxy
cluster. The cluster redshift they provided in their analysis (\(z=0.82\pm0.02\)) was derived from
the X-ray spectrum itself.

We extracted the source within 300 kpc, corresponding to \(35.4\arcsec\), in contrast to \citet{Hoeft2008} (\(50\arcsec\)).
No redshift based on spectroscopic data was available, but the quality of the available X-ray data allowed
for a precise determination of the galaxy clusters redshift. We measured a redshift
of \(z=0.836^{+0.031}_{-0.041}\), which agrees very well with the value given by \citet{Hoeft2008} (\(z=0.82\)).
The resulting X-ray flux and luminosity are \(F_{0.5-2}(300\rm kpc)=(6.66\pm0.17)\times10^{-14}~\rm erg~cm^{-2}~s^{-1}\) and
\(L_{0.5-2}(300\rm kpc)=(171.1\pm4.4)\times10^{42}~\rm erg~s^{-1}\), respectively.
The temperature derived in our {\tt APEC} fit, \(k_{B}T=4.84^{+0.31}_{-0.30}~\rm keV\), is also in within
the 1\(\sigma\) interval of the findings by \citet{Hoeft2008} (\(k_{B}T=4.2\pm0.4~\rm keV\)).

\begin{figure*}\centering
 \begin{subfigure}{0.33\textwidth}\centering
  \includegraphics[width=5.75cm]{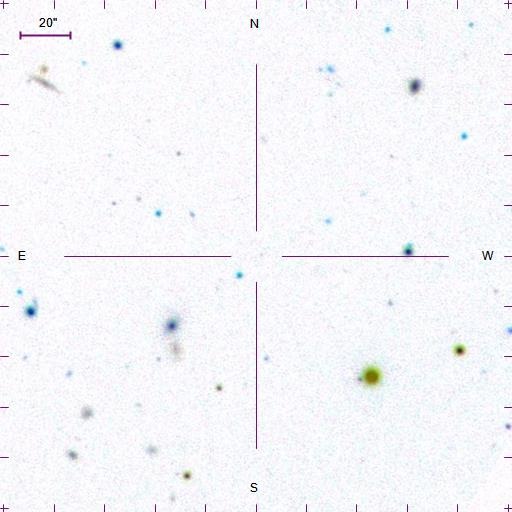}
  \caption{}
  \label{fig:LBT_006_SDSS}
 \end{subfigure}\hfill
 \begin{subfigure}{0.33\textwidth}\centering
  \includegraphics[width=5.75cm]{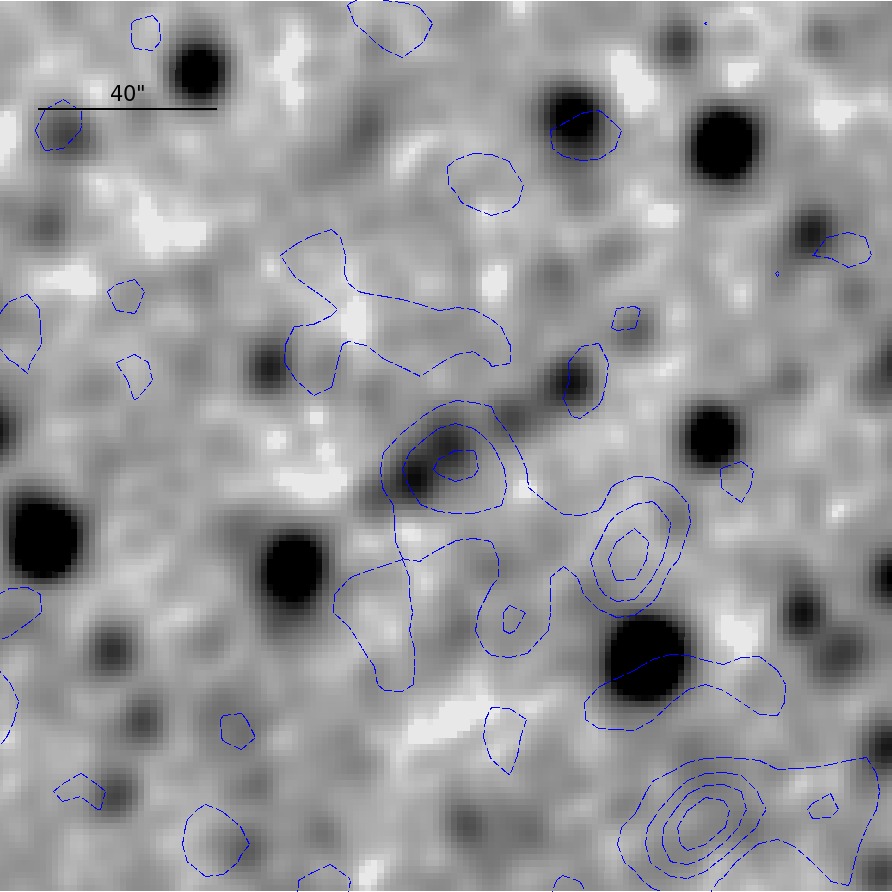}
  \caption{}
  \label{fig:LBT_006_WISE}
 \end{subfigure}\hfill
 \begin{subfigure}{0.33\textwidth}\centering
  \includegraphics[width=5.75cm]{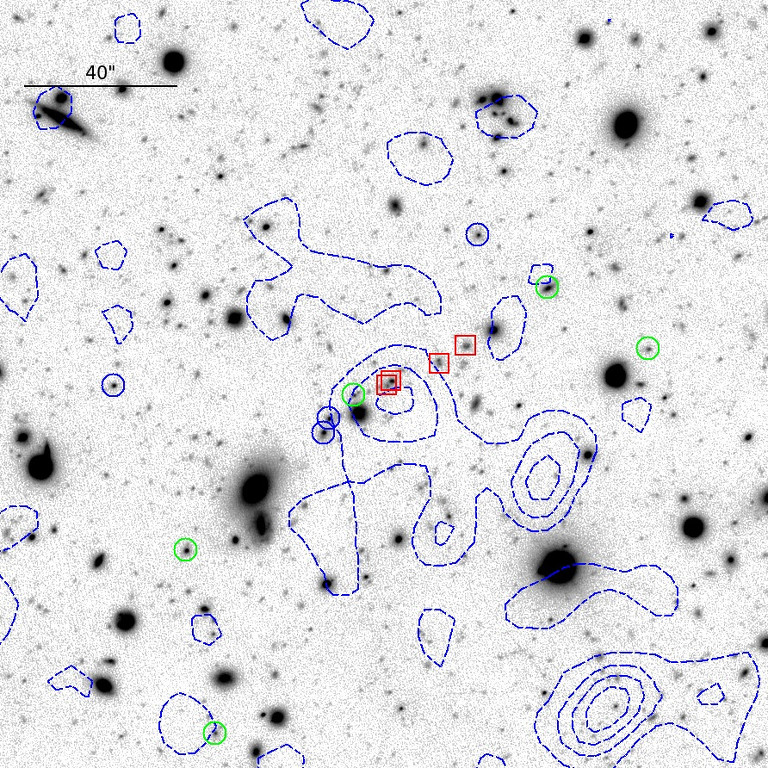}
  \caption{}
  \label{fig:LBT_006_LBC}
 \end{subfigure}\hfill
\caption{\label{fig:LBT_006_cutouts}
         Optical and near-IR images of the field of 2XMMp~J092120.2+371735.
         (\subref{fig:LBT_006_SDSS}): SDSS DR6 image. (\subref{fig:LBT_006_WISE}): W1-band (3.4\(\mu m\)) image from the AllWISE survey.
         (\subref{fig:LBT_006_LBC}): LBC image (mean of the r- and z-SLOAN filters). North is up, east is left, and all images cover
         the same region. Blue dashed lines (middle and right images) indicate contours from the X-ray flux. Red squares overplotted
         on the LBC image refer to the \(z\sim1.2\) group, and
green circles to spectroscopic sources outside our membership criteria.
         Blue cycles show galaxies from the foreground population (\(z\sim1\)).}
\end{figure*}

\subsubsection{2XMMp~J105344.2+573517}
The initial detection of galaxy cluster 2XMMp~J105344.2+573517 was named RX~J1053.7+5735 and is based on deep
(\(\sim1.3\rm Ms\)) \emph{ROSAT} observations \citep{Hasinger1998a} in a field with very low column density of
the neutral galactic hydrogen \citep{Lockman1986}. The structure of the source in the so-called Lockman-Hole
was found to be double-lobed in the X-ray data.
Follow-up imaging in the optical and NIR uncovered potential cluster galaxies for both sides of the lobe and the
peculiarity of a bright lensing arc at redshift \(z=2.57\), lensed by the BCG of the eastern lobe
\citep{Thompson2001}.
Using the Deep Imaging Multi-Object Spectrograph \citep[DEIMOS;][]{Faber2003} at the Keck II telescope,
\citet{Hashimoto2005} confirmed six galaxies at concordant redshifts in both lobes of RX~J1053.7+5735, giving a mean
cluster redshift of \(z=1.134\) --- in good agreement with \(z\sim1.16^{+0.02}_{-0.03}\) and
\(z\sim1.14^{0.02}_{-0.01}\) estimated from the X-ray Fe-K line for the eastern and western lobe and based on
\emph{XMM-Newton} data \citep{Hashimoto2004}. The coinciding redshift measurements of both lobes with independent
methods underlines the possibility that RX~J1053.7+5735 is in fact a merging cluster system.

For the X-ray analysis in this work, we used all observations from the XSA noted in Table~D.1.
After cleaning for times of high background events, the summed exposure times are \(\sim135\), \(\sim147,\) and
\(\sim114\)~ksec for MOS1, MOS2, and PN. Our {\tt APEC} model fits give a flux and rest frame luminosity of 
\(F_{0.5-2}(300\rm kpc)=(1.928\pm0.038)\times10^{-14}~\rm erg~cm^{-2}~s^{-1}\) and
\(L_{0.5-2}(300\rm kpc)=(106.4\pm2.1)\times10^{42}~\rm erg~s^{-1}\). Owing to the differences in the extraction region, these
results are not directly comparable to values in the literature.
The best-fitting ICM temperature according to our model, \(k_{B}T=4.52^{+0.24}_{-0.22}~\rm keV\), is in agreement
with \citet{Hashimoto2004} and their results for the eastern and western lobe (\(3.4^{+0.2}_{-0.1}\) and
\(4.4^{+0.3}_{-0.3}\)), but also with earlier work \citep[\(k_{B}T=4.9^{+1.5}_{-0.9}\rm keV\), see][]{Hashimoto2002}. 

\subsubsection{2XMMp~J123113.1+154550}
The source 2XMMp~J123113.1+154550 has previously been selected by the XDCP, and received deep imaging and spectroscopic
follow-up with VLT/FORS2.
We will present further details on the optical analysis for this cluster as part of a larger sample in an upcoming
paper (Rabitz et al. in prep.). However, we use the spectroscopic mean cluster redshift of \(z=0.893\) here to reduce
the free parameter for the {\tt APEC} model fit.
We extracted photons of the cluster within \(0.3 \rm~Mpc\) around the source position of 2XMMp~J123113.1+154550.
The best-fit parameter for the cluster temperature is \(k_{B}T=8.9^{+4.1}_{-2.2}\rm keV\), and
therefore larger than for the highly luminous cluster 2XMMp~J083026.2+524133 (compare Sect.~\ref{sec:LBT_003}), but with
significant uncertainties. Furthermore, flux and luminosity were computed to
be \(F_{0.5-2}(300\rm kpc)=(3.89\pm0.3)\times10^{-14}~\rm erg~cm^{-2}~s^{-1}\) and
\(L_{0.5-2}(300\rm kpc)=(99.3\pm7.8)\times10^{42}~\rm erg~s^{-1}\), which is almost a factor of two smaller than the
\(F_{0.5-2}(300\rm kpc)\) and \(L_{0.5-2}(300\rm kpc)\) of the
latter highly luminous cluster. We summarize all X-ray properties of our analysis in
Table~\ref{tab:results_known}.

\subsection{Results: rejected and as yet unclassified fields}
\label{sec:results_noclusters}

\subsubsection{2XMMp~J092120.2+371735}
\label{sec:LBT_006}

\begin{figure}
  \centering
  \includegraphics[width=9cm]{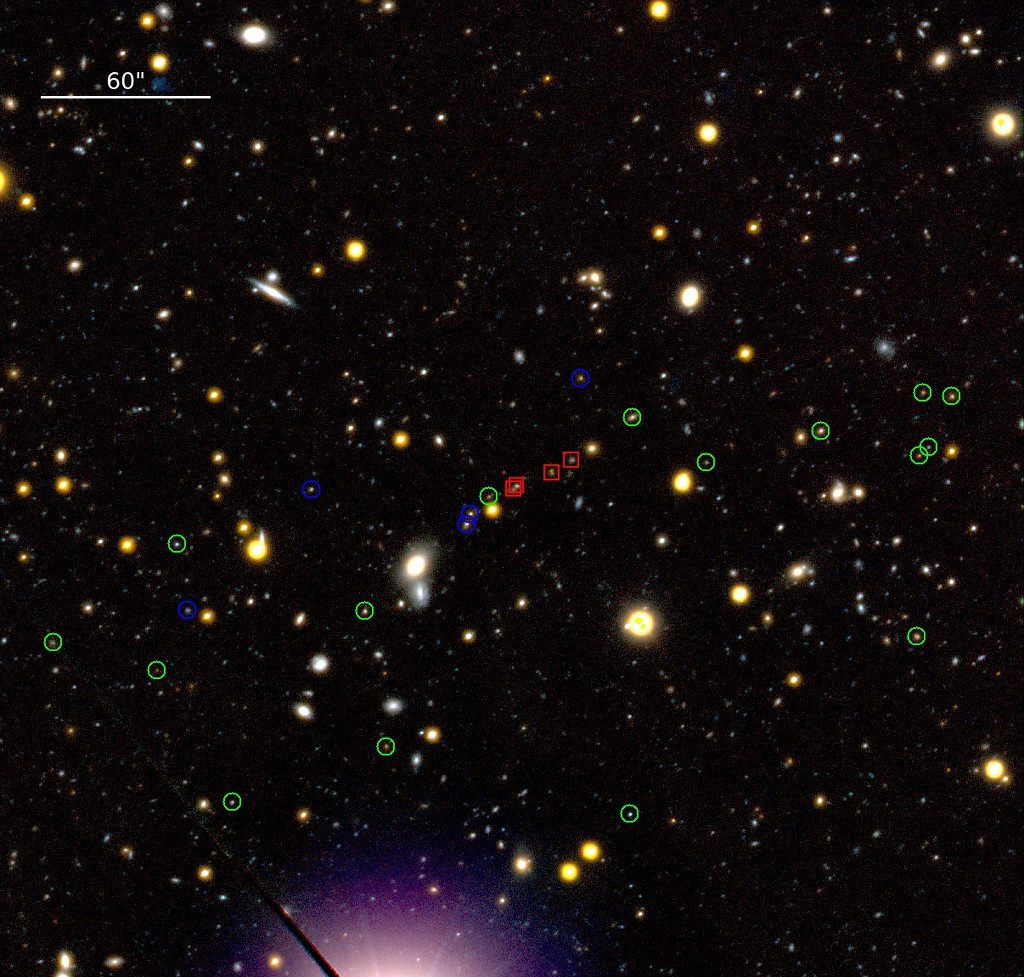}
\caption{\label{fig:LBT_006_RGB}
         LBC colour image of 2XMMp~J092120.2+371735, the high-redshift group at \(z\sim1.21\). We assigned
         r-SLOAN and z-SLOAN images from the LBC to the blue and red channel of the colour image. The green channel
         was created from the mean of both bands. Overplotted red squares refer to the \(z\sim1.2\) group, and green
         circles show spectroscopic sources outside our membership criteria, while blue cycles indicate galaxies of a
         foreground population (see Sect.~\ref{sec:LBT_006}).
         North is up, east is left, and the image is centred on the X-ray position.}
\end{figure}

\begin{figure}
  \begin{subfigure}{0.5\textwidth}
    \includegraphics[width=8.5cm]{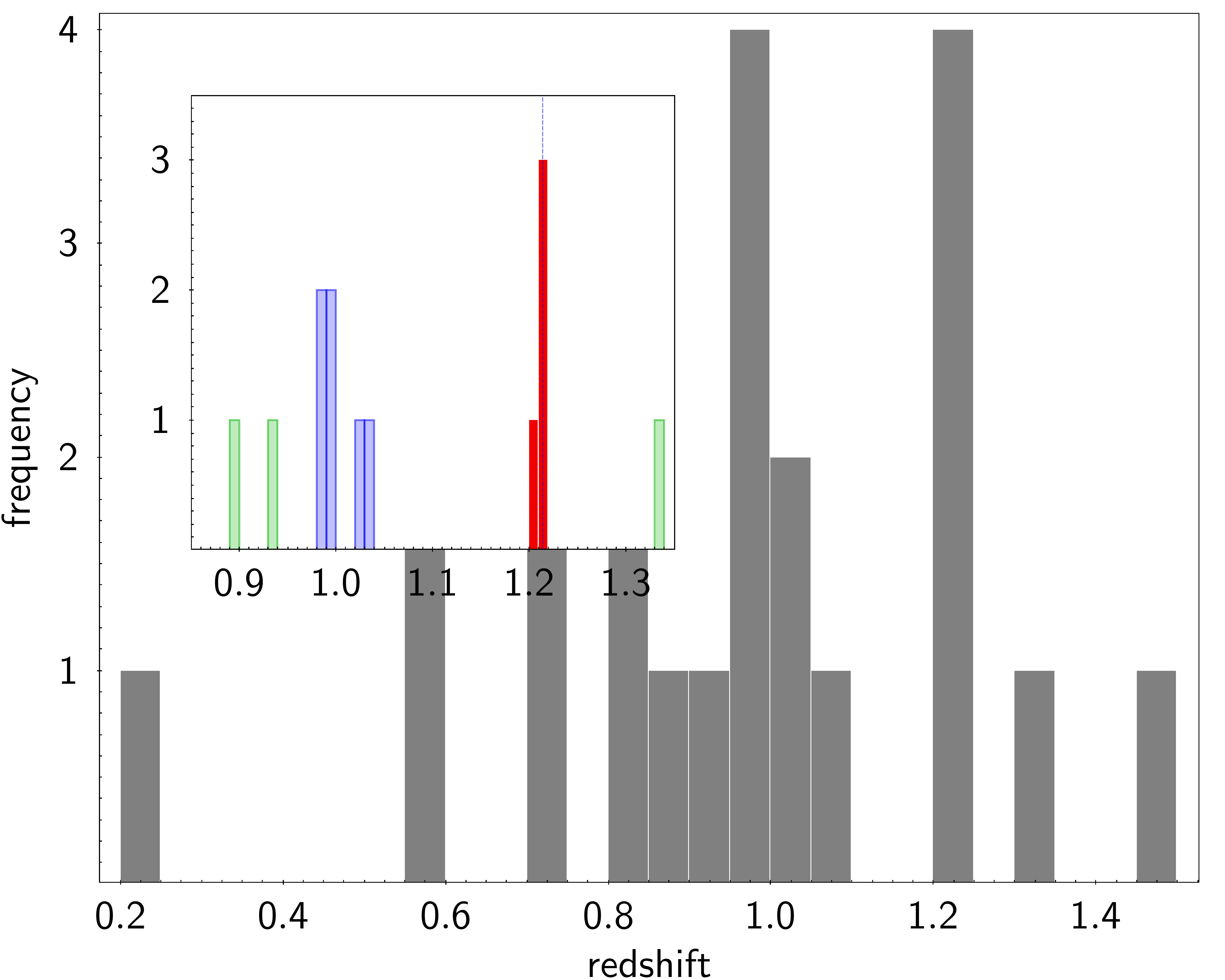}
    \subcaption{colour-magnitude diagram}
    \label{fig:LBT_006_histo}   
  \end{subfigure}

  \begin{subfigure}{0.5\textwidth}
    \includegraphics[width=8.5cm]{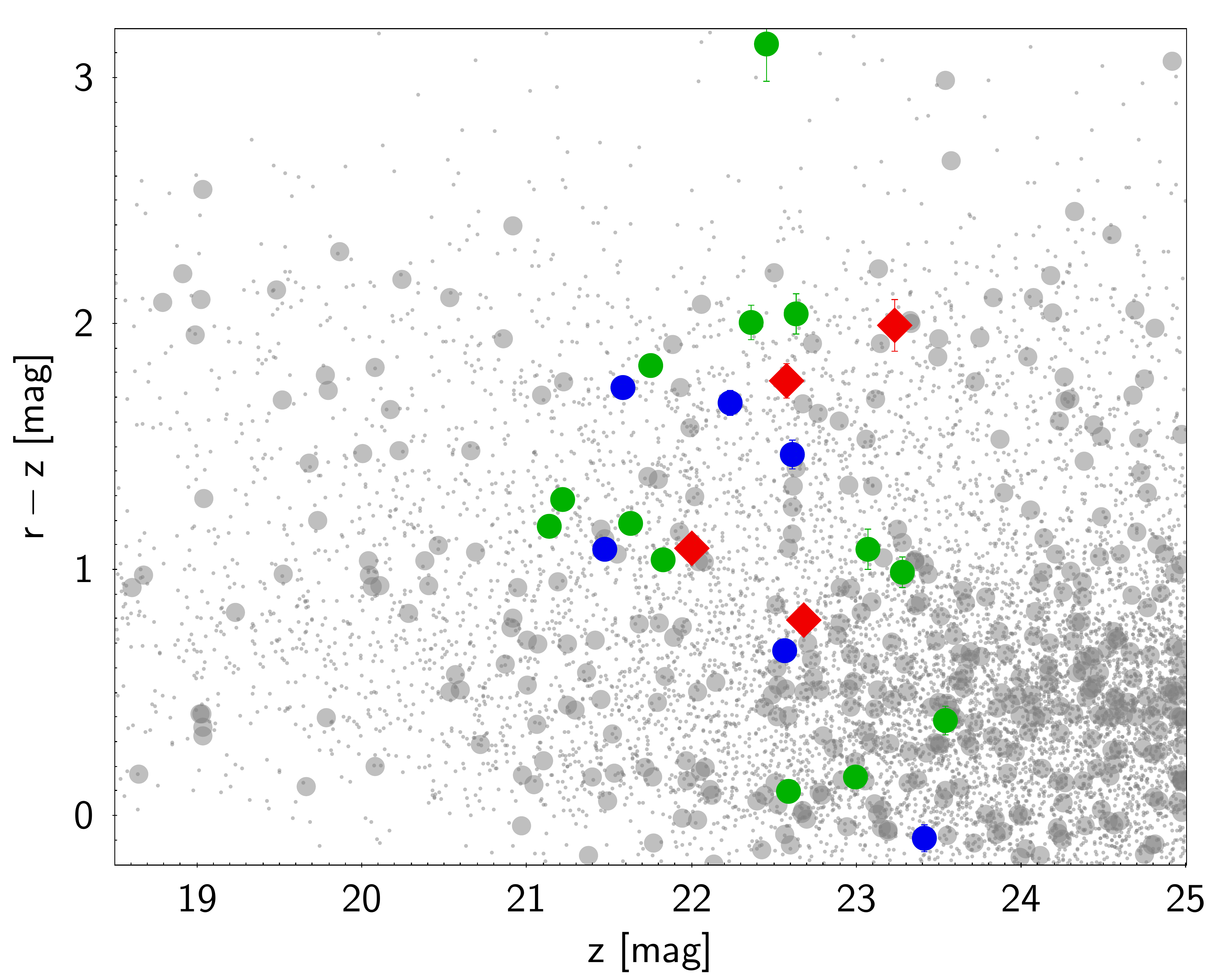}
    \subcaption{colour-magnitude diagram}
    \label{fig:LBT_006_colormag}
  \end{subfigure}
  \caption{\label{fig:LBT_006_colorhist}
           Redshift histogram (top panel) and colour-magnitude diagram (bottom) of 2XMMp~J092120.2+371735.
           Objects with spectra are marked green, while the members of the high-redshift group are shown in red.
           Blue refers to the foreground population of galaxies, which is unlikely to be responsible for the X-ray emission. 
           Error bars plotted in panel (\subref{fig:LBT_006_colormag}) indicate the photometric uncertainties and are visible when they exceed the symbol size.}
\end{figure}
The SDSS field of this source is empty, while WISE data show a clear signal in the central region of the X-ray emission
(compare cutouts in Fig.~\ref{fig:LBT_006_SDSS}-\ref{fig:LBT_006_LBC}). LBC pre-imaging (Fig.~\ref{fig:LBT_006_RGB})
reveals a field that has no obvious overdensity of red galaxies, but is dominated by very faint sources. However, the
possibility exists that the 4000\AA~break is already redshifted out of the z-SLOAN band, causing no red ridge line
to be present in our given filter set, which is also reflected by the colour-magnitude diagram
(Fig.~\ref{fig:LBT_006_colormag}). Since the source was considered to contain a potential high-redshift (\(z>1\)) cluster,
an attempt to select possible member galaxies was made for the spectroscopic follow-up.
Spectra of the brightest extragalactic sources near the centre of the extended X-ray emission, however, yielded two
redshift overdensities, at \(z\sim1.0\) and \(z\sim1.2\) (corresponding to the blue and red
colours; see Figs.~\ref{fig:LBT_006_LBC} and~\ref{fig:LBT_006_colorhist}).

The possible low-redshift cluster, with galaxies in the range of \(0.984 \lesssim z \lesssim 1.033\), would
result in a velocity dispersion that is too high. These galaxies are therefore not considered as a bound system.
The four galaxies within the range of \(1.209 \lesssim z \lesssim 1.214\) are, however,  much closer in redshift space,
with their two brightest members in the very X-ray centre (Fig.~\ref{fig:LBT_006_LBC}). We furthermore note that in the
WISE IR imaging in Fig.~\ref{fig:LBT_006_WISE}, we clearly see counterparts to the high-redshift galaxies.
The four members of the high mean redshift population (\(z=1.2134 \pm 0.0014\)) do not form a typical red sequence
(Fig.~\ref{fig:LBT_006_colormag}), but were found to be active and are hence relatively blue galaxies according to their
[OII] emission line -- an indication for ongoing star formation.
The X-ray emission from 2XMMp~J092120.2+371735 is clearly extended (\(\sim13\arcsec\)~in 3XMM-DR6; compare
Table~D.1), and a meaningful {\tt tbabs} fit of the extracted spectra was not possible, given the quality
of the archival data. Because only four galaxies with coinciding redshifts are confirmed and we lack a clear detection of a
red-sequence or even a passive BCG, we cannot confirm the status of a galaxy cluster for this group of \(z\sim1.21\) galaxies.

\subsubsection{2XMMp~J093607.2+613245}
The X-ray source 2XMMp~J093607.2+613245 is catalogued as an extended source in 2XMMp, and its extent is also
confirmed by the 3XMM-DR6 catalogue. Deep imaging with the LBC shows only faint blue galaxies within the
central X-ray emission region, with a WISE counterpart in the W1 band (see upper panel in Fig.~C.2),
while survey data from the SDSS do not detect foreground sources. Spectroscopic follow-up
with MODS revealed three galaxies within a relatively wide redshift range (\(z=0.81~..~0.84\)), including the galaxy
closest to the X-ray position, an OII emitter at \(z=0.811\). An object at 12\arcsec~angular distance from the X-ray position
was identified as a QSO at \(z=1.55\).
A re-inspection of the X-ray source showed that the 2XMMp source is a blend of two sources to which
the QSO contributes part of its flux. Since it remains unclear whether the galaxies at \(z\sim 0.83\) form a cluster
that could give rise to the remaining X-ray emission, we did not add this source to the sample of identified clusters.

\subsubsection{2XMMp~J120735.1+250538}
The X-ray emission related to 2XMMp~J120735.1+250538 is of low extent (6.6\arcsec) and blended with a point source.
Because of the uncertain X-ray extent and the lack of obvious member galaxies, spectroscopic follow-up was not
executed.

\subsubsection{2XMMp~J133038.6-013832}
The X-ray emission of 2XMMp~J133038.6-013832 is of relatively low extent (\(\sim6.8\arcsec\); see Table~D.1)
and well isolated. In the centre, LBC imaging detects a concentration of red galaxies with counterparts in WISE data
(see Figs.~C.2i and C.2h).
We skipped further follow-up for this cluster candidate, since its BCG is already visible in newer releases of the SDSS,
where it is listed with a photometric redshift of \(z\sim 0.75\) and has a magnitude of 19.92 in the z-band of the LBC pre-imaging.
This renders a high-redshift nature unlikely.

\subsubsection{2XMMp~J144854.8+085400}
This X-ray source is clearly extended in the softer XMM-EPIC images [0.2--2] keV.
In the harder bands, at [2.0--12] keV, a point source is apparent near the centre of the soft extended emission.
No obvious galaxy cluster can be found in the LBC imaging. However, the hard point source is identified with a galaxy that is also detected in the SDSS with $m_{\rm r}=22.64$ and has a photometric redshift $z_{\rm ph}=0.75$.
A search in radio catalogues of the field revealed two 1.4 GHz NVSS \citep{Condon1998} sources  $\simeq 1$ arcmin east
and west of the hard X-ray source, roughly corresponding to the elongation of the extended X-ray emission.
The 5 GHz FIRST survey \citep{Becker1994} also shows several compact and extended sources, and the low-frequency GLEAM
survey \citep{Hurley-Walker2017} detected one source near the position of the eastern NVSS source.

We therefore conclude that the extended X-ray emission is more likely associated with  emission from
the jets of a radio-loud AGN in the galaxy at \(z_{\rm ph}=0.75\). 
The core of the AGN must be highly obscured, since it is not visible at soft X-rays below 2 keV.

\subsubsection{2XMMp~J145220.8+165458}
In the 2XMMp catalogue, 2XMMp~J145220.8+165458 is listed as an
extended source, but it has lost the extent flag in the more
recent 3XMM-DR6 catalogue. The central part of the X-ray emission harbours a galaxy that was also visible in the SDSS.
Deep LBC imaging and IR data from WISE do not indicate an additional background overdensity of galaxies (compare
Fig.~C.1e and the middle panel of Fig.~C.3).
Therefore the source 2XMMp~J145220.8+165458 was rejected as high-redshift galaxy cluster and excluded from the
follow-up program.

\subsubsection{2XMMp~J151716.8+001302}
The source of 2XMMp~J151716.8+001302 is extended in both 2XMMp and 3XMM-DR6, considering the OBSID 0103860601.
A following observation (0201902001), only present in the more recent catalogue, lists no extent for this
source, hence it remains uncertain whether the source is extended at all.
The deep imaging with the LBC and survey data of WISE do not indicate a population of high-redshift galaxies either.
Therefore this source was removed from the list of tentative high-redshift galaxy clusters.

\subsection{Results: \(L-T\) relation}
\label{sec:results_L-T}
\begin{figure*}
 \includegraphics[width=\textwidth]{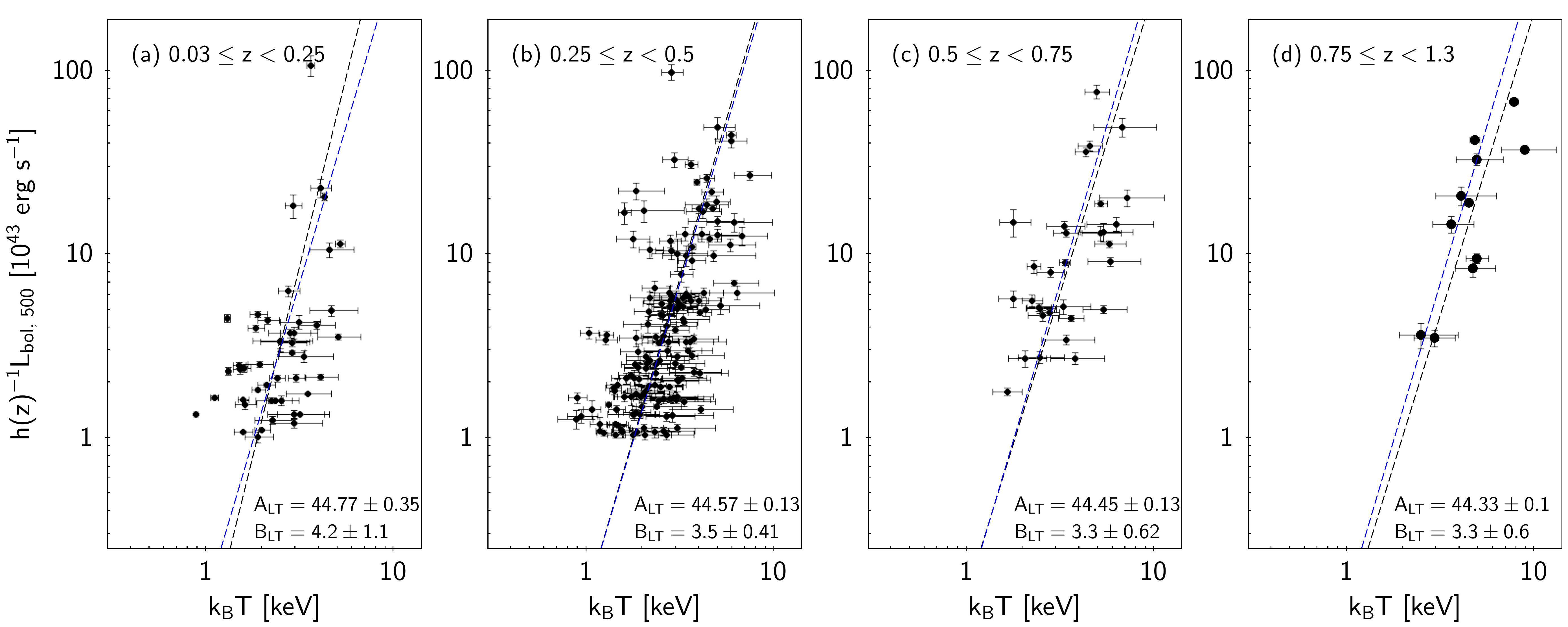}
 \caption{Cluster temperature plotted against the bolometric luminosity of two samples of galaxy clusters.
  Panels (a), (b), and (c) show data from \citet{Takey2013}, restricted to \(L_{bol}h(z)^{-1}>10^{43}~\rm erg~s^{-1}\),
  and present their three redshift bins (see individual panel). The clusters from this work are plotted
  in panel (d). In all panels we highlight the best-fitting solution of Eq.~\ref{eq:LT} for the
  combined sample of \citet{Takey2013} and our high-redshift extension, including the luminosity cut, as a blue
  dashed line (parameters are listed in Table~\ref{tab:BCESfit}).}
 \label{fig:L-T}
\end{figure*}
Based on the X-ray properties derived for the confirmed clusters of galaxies, we analysed the relation between
\(L_{bol}(r_{500})\) and \(k_{B}T\).
The bolometric luminosities within the radius \(r_{500}\) were calculated using an iterative method \citep{Takey2013},
and the cluster temperatures were derived from spectral model fits to the innermost 300~kpc region of the cluster, which
we described in Section~\ref{sec:Xray_analysis}. We note that the spatial resolution of \emph{XMM-Newton} does not allow
us to exclude the core of clusters at the high redshifts of our sample, and therefore ICM temperatures might
be biased by a cool cluster core.
When we analysed our combined sample of 11 clusters (Tables~\ref{tab:resulttab} and~\ref{tab:results_known}) using the
BCES orthogonal regression method \citep{Akritas1996} on the logarithm of luminosity and temperature, we found as the best-fitting linear relation
\begin{equation}
\label{eq:LT}
 \textnormal{log}\left(\frac{L_{bol}(r_{500})}{h(z)}\right)=A_{LT}+B_{LT}\cdot\textnormal{log}\left(\frac{k_{B}T}{5}\right),
\end{equation}
 with the intersect \(A_{LT}=44.33\pm0.1\), the slope \(B_{LT}=3.3\pm0.6,\) and the Hubble parameter \(h(z)\) (compare right
 panel of Fig.~\ref{fig:L-T}).

Slope and intercept of the \(L-T\) relation are in good agreement with \citet{Takey2013}. Larger cluster statistics enabled
\citet{Takey2013} to analyse three redshift bins of their sample. Although our slope indeed matches the value derived
for their highest redshift bin (\(0.5 \leq z \leq 0.7\)) best, it does not contradict their intermediate bin
(\(0.25 < z < 0.5\)) either, taking into account the uncertainties of the fits.
We restricted the data from their work to \(L_{bol}h(z)^{-1}>10^{43}~\rm erg~s^{-1}\), rendering the lowest luminosity of
the sample of \citet{Takey2013} at similar values as our high-redshift sample.
The slopes and intersects were again derived using the BCES method, as described above. The derived fitting parameters are
displayed in Fig.~\ref{fig:L-T} with the refitted sample of \citet{Takey2013} in panels a, b, and c for the lowest, medium,
and highest of their redshift bins. In panel d we plot the data from our current sample, which only consists of clusters with
\(z>0.75\) from either LBT/MODS spectroscopy or spectra from the literature, as black dots.
We generated a combined sample, including the data from \citet{Takey2013} and our high-redshift sample of confirmed clusters in
this work, and applied the \(L_{bol}h(z)^{-1}>10^{43}~\rm erg~s^{-1}\) cut. The best-fitting parameters for this new sample
are \(A_{LT}=44.535\pm0.083\) and \(B_{LT}=3.45\pm0.29\), the solution is plotted in Fig.~\ref{fig:L-T} as a dashed blue line.
It is apparent that using the cut in bolometric luminosity, the parameters of the \(L-T\) relation are in agreement across all
analysed redshift bins. Hence, we see no evolution in the \(L-T\) relation with redshifts using
the present cluster statistics.

Additionally, we analysed the intrinsic scatter of the data with respect to the derived best-fitting \(L-T\) relation for
each sample using the procedure described by \citet{Pratt2009}.
We first computed the raw scatter of our data points as the error-weighted orthogonal distances to the regression line.
The quadratic difference between the raw scatter and the statistical uncertainties now give the intrinsic scatter,
\(\sigma_{log}L_{bol}\), of our samples.

We summarize the fitting parameters for the different samples in Table~\ref{tab:BCESfit}.
\begin{table*}\caption{Best-fitting parameters of \(L-T\) relation from the BCES method.}
\label{tab:BCESfit}
\centering
\begin{tabular}{c c c c c c}
\hline
\hline
sample & n & redshift range & \(B_{LT}\) &  \(A_{LT}\) & \(\sigma_{\rm log}L_{bol}\) \\
 \hline
\citet{Takey2013} & 47 & \(0.03\leq z<0.25\) & \(4.2\pm1.1\) & \(44.77\pm0.77\) & \(1.05\pm0.15\) \\
\citet{Takey2013} & 157 & \(0.25\leq z<0.5\) & \(3.5\pm0.41\) & \(44.57\pm0.13\) & \(0.543\pm0.043\) \\
\citet{Takey2013} & 31 & \(0.5\leq z<0.75\) & \(3.3\pm0.62\) & \(44.45\pm0.13\) & \(0.406\pm0.073\) \\
current work & 11 & \(0.75\leq z<1.27\) & \(3.3\pm0.6\) & \(44.33\pm0.1\) & \(0.232\pm0.07\) \\
combined & 244 & \(0.03\leq z<1.27\) & \(3.45\pm0.29\) & \(44.535\pm0.083\) & \(0.676\pm0.043\) \\
\hline
\hline
\end{tabular}
\tablefoot{Summary of the BCES fit on the sample with \(L_{bol}h(z)^{-1}>10^{43}~\rm erg~s^{-1}\).
          The first three columns give the origin of the data, their sample size, and the redshift range of the particular
          clusters. The resulting best-fitting values for slope and intersect of Eq.~\ref{eq:LT}, their
          respective errors, and the intrinsic scatter are given in the last three column.}
\end{table*}

\section{Conclusions}
\label{sec:conclusions}

In the input sample of 19 extended X-ray sources  with empty SDSS fields,  we found 13 distant clusters of galaxies 
with redshifts in the range \(z=0.75-1.27\). While 6 of these clusters are (X-ray selected) objects with identifications and 
spectroscopic redshifts that have previously been published in the literature, we identified 6 new clusters, measured their redshifts, and also detected a group of four galaxies at the redshift of \(z\sim1.21\),  using the LBT.
The high fraction of confirmed high-redshift clusters demonstrates the efficiency of this selection method, which is based solely
on public archival data, for the discovery of these cosmologically interesting objects.
The high-redshift objects complement the samples published by \citet{Takey2013}, which comprise
extended 2XMM sources that have been confirmed as galaxy clusters using SDSS imaging and spectra.

Our spectroscopic LBT observations confirmed the redshift (z=0.99) of the cluster 2XMMp~083026.2+524133, 
which was previously determined from its X-ray spectrum \citep{Lamer2008}.
The analysis of new \emph{XMM-Newton} observations of this object also confirmed its high X-ray temperature
(\(k_{B}T=7.82^{+0.4}_{-0.39}~\rm{keV}\)) and luminosity (\(L_{bol}(r_{500}) = (1168 \pm 4)\times10^{42}~\rm erg~s^{-1} \)).
However, this cluster optically appears surprisingly inconspicuous, and only eight of the 
galaxies targeted spectroscopically were found to be cluster members. The resulting constraints to its dynamical mass 
(\(M_{200}=5.1_{-2.9}^{+6.7}\times10^{14}M_{\odot}\)) are not accurate enough to favour either the 
high-mass estimates based on the X-ray temperature \citep[\(M_{500}=5.6\times10^{14}M_{\odot}\);][]{Lamer2008}
or the lower estimates from  SZ measurements \citep[\(M_{200}=3.6\times10^{14}M_{\odot}~/~4.7\times10^{14}M_{\odot}\), depending
on the model;][]{Schammel2013}.

We were able to determine meaningful X-ray temperatures and bolometric luminosities for 11 clusters in the new sample.
The sample shows a tight  \(L-T\) relation of the form
\begin{equation}
\label{eq:LT}
 \textnormal{log}\left(\frac{L_{bol}(r_{500})}{h(z)}\right)=\left(44.33\pm0.1\right)+\left(3.3\pm0.6\right)\textnormal{log}\left(\frac{k_{B}T}{5}\right).
\end{equation}
We have re-analysed the \(L-T\) relation in the sample of \citet{Takey2013} using only clusters with
\( L_{bol}(r_{500})h(z)^{-1} > 10^{43} \rm erg~s^{-1} \) and find that the \(L-T\) relation in our high-redshift
sample and the \(L-T\) relations in all redshift bins of the
sample of \citet{Takey2013} are consistent with each other.
When we combine our sample with the sample of \citet{Takey2013}, the  best-fitting \(L-T\) relation is
\begin{equation}
\label{eq:LT}
 \textnormal{log}\left(\frac{L_{bol}(r_{500})}{h(z)}\right)=\left(44.54\pm0.1\right)+\left(3.45\pm0.29\right)\textnormal{log}\left(\frac{k_{B}T}{5}\right).\end{equation}

When it is compared with the BCES fit to the XMM XXL data \citep{Giles2016} without  bias correction, the slope
for our combined 2XMM sample is somewhat steeper, but still consistent within the errors.
It is worth noting that with an intrinsic scatter of \(\sigma_{log}L_{bol} = 0.232 \pm 0.07,\)  the \(L-T\)
relation in the high-z sample is significantly tighter than those of the \citet{Takey2013} and XMM XXL samples.
The low scatter might be an indication of a lower fraction of cool-core clusters in the high-z sample.
A decline of the cool-core fraction at high redshifts can be expected because of the cosmological evolution of clusters.
On the other hand, at increasing distance, cool-core clusters might also become too compact to be detected as extended X-ray sources.

Finally, we note that for 5 of the 12 clusters we measured X-ray fluxes \(F_{0.5-2}\) above \(3\cdot 10^{-14}~\rm erg~cm^{-2}~s^{-1}\), 
which is the approximate limit for the detection of extended  sources in  the eROSITA all-sky survey \citep{Merloni2012}. 
Given the solid angle of \(60~\rm deg^{2}\) covered by our survey, we expect several thousand clusters with redshifts
\(z>0.8\) to be discovered by eROSITA.  

\begin{acknowledgements}
  We thank the anonymous referee for the constructive report that helped us to improve the readability
  of the paper. The author acknowledges the use of the TOPCAT\footnote{http://www.starlink.ac.uk/topcat/} and
  STILTS\footnote{http://www.starlink.ac.uk/stilts/} software packages written by Mark Taylor.
  This paper uses data taken with the MODS spectrographs built with funding from NFS grant AST-9987045 and
  the NSF Telescope System Instrumentation Program (TSIP), with additional funds from the Ohio Board of
  Regents and the Ohio State University Office of Research.
  We acknowledge financial support from the ARCHES project (7th Framework of the European Union, n 313146).
  Funding for the SDSS and SDSS-II has been provided by the Alfred P. Sloan Foundation, the Participating
  Institutions, the National Science Foundation, the U.S. Department of Energy, the National Aeronautics and
  Space Administration, the Japanese Monbukagakusho, the Max Planck Society, and the Higher Education Funding
  Council for England. The SDSS Web Site is http://www.sdss.org/.
  The SDSS is managed by the Astrophysical Research Consortium for the Participating Institutions. The
  Participating Institutions are the American Museum of Natural History, Astrophysical Institute Potsdam,
  University of Basel, University of Cambridge, Case Western Reserve University, University of Chicago, Drexel
  University, Fermilab, the Institute for Advanced Study, the Japan Participation Group, Johns Hopkins University,
  the Joint Institute for Nuclear Astrophysics, the Kavli Institute for Particle Astrophysics and Cosmology,
  the Korean Scientist Group, the Chinese Academy of Sciences (LAMOST), Los Alamos National Laboratory, the
  Max-Planck-Institute for Astronomy (MPIA), the Max-Planck-Institute for Astrophysics (MPA), New Mexico State
  University, Ohio State University, University of Pittsburgh, University of Portsmouth, Princeton University,
  the United States Naval Observatory, and the University of Washington.
  Based on observations obtained with XMM-Newton, an ESA science mission with instruments and contributions directly
  funded by ESA Member States and NASA.
  This research has made use of data obtained from the 3XMM XMM-Newton serendipitous source catalogue compiled by the
  10 institutes of the XMM-Newton Survey Science Centre selected by ESA.
\end{acknowledgements}

\nocite{Taylor2005}
\nocite{Taylor2006}
\bibliographystyle{aa.bst}
\bibliography{literature.bib}
 
\end{document}